# Konzeption und Realisierung eines skalierbaren Simulators für die Magnetresonanz-Tomographie

Von der Fakultät für Elektrotechnik und Informationstechnik
der Rheinisch-Westfälischen Technischen Hochschule Aachen
zur Erlangung des akademischen Grades eines
Doktors der Ingenieurwissenschaften genehmigte Dissertation

vorgelegt von

Diplom-Ingenieur

Jürgen Kürsch

aus Köln

Diese Dissertation ist auf den Internet-Seiten der Hochschulbibliothek online verfügbar.

# Vorwort

Die vorliegende Arbeit entstand während meiner Tätigkeit als wissenschaftlicher Mitarbeiter am Lehrstuhl für Allgemeine Elektrotechnik und Datenverarbeitungssysteme der RWTH Aachen.

Dem Leiter des Instituts, Herrn Prof. Dr.-Ing. Tobias G. Noll, danke ich für die hervorragende, fachlich herausfordernde Betreuung der Arbeit sowie für fruchtbare Diskussionen über ihren Inhalt; beides hat die Qualität der Arbeit maßgeblich beeinflußt. Herrn Prof. Dr. rer. nat. Günter Rau danke ich für die freundliche Übernahme des Korreferats und für viele Hinweise, die die Ausarbeitung haben klarer werden lassen.

Desweiteren danke ich Herrn Dr.-Ing. Stephan Kannengießer für lange Diskussionen zum Thema K-t-Formalismus und für einen steten Strom von Hinweisen zu dem entstehenden Simulationswerkzeug. Herr Dr.-Ing. Erik Penner hat vor vielen Jahren mein Interesse an der Magnetresonanz-Tomographie geweckt und so für die Initialzündung dieser Arbeit gesorgt.

Gedankt sei weiterhin den Herren Dipl.-phys. Matthias Drobnitzky und Dr.-Ing. Arndt Glowinski, die während ihrer Tätigkeit an der Klinik für Radiologische Diagnostik des Klinikums der RWTH Aachen die praktische Umsetzung so mancher Idee auf den klinischen Magnetresonanz-Tomographen ihrer Einrichtung nach Kräften unterstützt haben.

Meiner Frau Claudia Kürsch und meinem Sohn Martin danke ich herzlich für ihre unendliche Geduld und ihre Hilfe bei der Entstehung und Überarbeitung der Niederschrift, ohne die die Fertigstellung in den Sternen gestanden hätte.

Gedacht sein soll meiner Mutter, Frau Elfriede Kürsch, die meine Ausbildung trotz vieler Widrigkeiten ermöglicht und nach Kräften unterstützt hat.

# Inhalt

# Einheiten und Formelzeichen

## Einheiten

| | |
|---|---|
| $1$ s | Sekunde, Zeit |
| $1$ m | Meter, Länge |
| $1$ Hz $= 1$ s$^{-1}$ | Hertz, Frequenz |
| $1$ V | Volt, elektrische Spannung |
| $1$ A | Ampère, elektrischer Strom |
| $1$ T $= 1$ Vs m$^{-2}$ | Tesla, magnetische Flußdichte |

## Notation

| | |
|---|---|
| $\boldsymbol{A}$ | Matrix |
| $\boldsymbol{A}^T$ | Transponierte Matrix |
| $\boldsymbol{A}^{-1}$ | Inverse Matrix |
| $\vec{a} = (a_x, a_y, a_z)^T$ | Vektor |
| $\vec{a} \cdot \vec{b}$ | Skalarprodukt |
| $\vec{a} \times \vec{b}$ | Vektorprodukt, Kreuzprodukt |
| $j = \sqrt{-1}$ | Imaginäre Einheit |
| $\underline{a}$ | Komplexe Zahl |
| $\underline{a}^*$ | Konjugiert komplexe Zahl |
| $\boldsymbol{A}'$, $\vec{a}'$, $a'$ | Größen im rotierenden Bezugssystem |
| $\forall$ | Für alle |
| $\nabla$ | Nabla-Operator |

## Größen (griechische Symbole)

| | |
|---|---|
| $\alpha$ | Flipwinkel eines hochfrequenten Magnetfeldpulses, $[1°]$ |
| $\gamma$ | Gyromagnetisches Verhältnis, $[1$ s$^{-1}$ T$^{-1}]$ |
| $\varepsilon$ | Maschinengenauigkeit (dimensionslos) |
| $\eta$ | Spindurchsatz, $[1$ s$^{-1}]$ |
| $\theta$ | Azimutwinkel, $[1°]$ |
| $\mu$ | Magnetisches Moment, $[1$ A m$^2]$ |
| $\rho$ | Spindichte $[1$ m$^{-3}]$, relative Spindichte (dimensionslos) |
| $\tau$ | Zeitdauer, $[1$ s$]$ |
| $\phi$ | Phasenwinkel eines hochfrequenten Pulses, $[1°]$ |
| $\Delta\varphi$ | Phasenänderung der Magnetisierung, $[1°]$ |
| $\chi$ | Magnetische Suszeptibilität (dimensionslos) |
| $\psi$ | Deklinationswinkel, $[\,1°\,]$ |
| $\omega$ | Kreisfrequenz, $[\,1\ \mathrm{s}^{-1}\,]$ |
| $\omega_{max}$ | Maximal mögliche Kreisfrequenz am Rand des FOV |
| $\omega_0$ | Kreisfrequenz im statischen Magnetfeld, Larmor-Frequenz |
| $\omega_{HF}$ | Trägerfrequenz des HF-Felds, Kreisfrequenz des rotierenden Bezugssystems |
| $\Delta\omega$ | Abweichung der Kreisfrequenz, Bandbreite eines HF-Pulses |
| $\Delta\omega_{CS}$ | Kreisfrequenzabweichung durch Chemical Shift |
| $\Delta\omega_0$ | Kreisfrequenzabweichung durch Inhomogenitäten von $B_0$ |
| $\Delta\omega_{susz}$ | Kreisfrequenzabweichung an Suszeptibilitätsgrenzflächen |

## Größen (lateinische Symbole)

| | |
|---|---|
| $\boldsymbol{A}$ | Koeffizientenmatrix des Anfangswertproblems, $[\,1\ \mathrm{s}^{-1}\,]$ |
| $\underline{a}_i$ | Population der transversalen Konfiguration, Ordnung $i$, $[\,1\ \mathrm{A}\ \mathrm{m}^{-1}\,]$ |
| $\underline{a}^{(i)}$ | Population der $i$-ten Konfiguration, $[\,1\ \mathrm{A}\ \mathrm{m}^{-1}\,]$ |
| $\underline{b}_i$ | Population der longitudinalen Konfiguration, Ordnung $i$, $[\,1\ \mathrm{A}\ \mathrm{m}^{-1}\,]$ |
| $\vec{B} = (B_x, B_y, B_z)^T$ | Magnetische Flußdichte, $[\,1\ \mathrm{T}\,]$ |
| $\vec{\underline{B}}$ | Magnetische Flußdichte nach Koordinatentransformation |
| $\vec{B}_0$ | Magnetische Flußdichte des statischen Magnetfelds |
| $\vec{B}_{grad}$ | Magnetische Flußdichte des Gradientenmagnetfelds |
| $\vec{B}_1$ | Magnetische Flußdichte des hochfrequenten Magnetfelds |
| $\vec{B}_{eff}$ | Effektive magnetische Flußdichte im rotierenden Bezugssystem |
| $\Delta\vec{B}$ | Abweichung der magnetischen Flußdichte |
| $C$ | Skalierungsfaktor für Magnetfeldinhomogenität, $[\,1\ \mathrm{T}\,]$ |
| $D$ | Durchmesser Empfangsspule, $[\,1\ \mathrm{m}\,]$ |
| $\boldsymbol{E}$ | Relaxationsmatrix in der Blochschen Gleichung, $[\,1\ \mathrm{s}^{-1}\,]$ |
| $\vec{E}_0$ | Relaxationsvektor der longitudinalen Magnetisierung, $[\,1\,\mathrm{A}\,\mathrm{m}^{-1}\mathrm{s}^{-1}\,]$ |
| $\vec{e}_x, \vec{e}_y, \vec{e}_z$ | Einheitsvektoren im kartesischen Koordinatensystem (dimensionslos) |
| $\Delta E_{stör}$ | Störabstand, $[\,1\ \mathrm{dB}\,]$ |
| $\underline{E}_i$ | Transversale Konfiguration der Ordnung $i$, $[\,1\ \mathrm{A}\ \mathrm{m}^{-1}\,]$ |
| $\vec{f}(t)$ | Inhomogener Anteil des Anfangswertproblems, $[\,1\,\mathrm{A}\,\mathrm{m}^{-1}\mathrm{s}^{-1}\,]$ |
| $\vec{G} = \nabla \cdot B_{\text{grad}, z}$ | Gradient des Gradientenmagnetfelds ohne Transversalkomponenten, $[1\ \text{T m}^{-1}]$ |
| $G_x, G_y, G_z$ | Komponenten von $\vec{G}$ |
| $I$ | Pixelintensität (dimensionslos) |
| $i$ | Index der Elementarsequenz, Ordnung der Konfiguration (dimensionslos) |
| $\vec{k} = (k_x, k_y, k_z)^T$ | Koordinaten im K-Raum, $[1\ \text{m}^{-1}]$ |
| $\Delta\vec{k}$ | Einheitsschritt im K-Raum |
| $\Delta K_x, \Delta K_y, \Delta K_z,$ | Periode wiederholter Ortsfrequenzfunktionen |
| $K_{x, \max}, K_{y, \max}, K_{z, \max}$ | Maximum der K-Raum-Koordinaten aller Konfigurationen |
| $k_{x, \max}, k_{y, \max}, k_{z, \max}$ | Maximum der K-Raum-Koordinaten, das nach dem Abtasttheorem zur korrekten Auflösung des FOV erforderlich ist |
| $\vec{M} = (M_x, M_y, M_z)^T$ | Magnetisierung, $[1\ \text{A m}^{-1}]$ |
| $\underset{\sim}{\vec{M}}$ | Magnetisierung nach Koordinatentransformation |
| $M_0$ | Magnetisierung im thermischen Gleichgewicht, Ruhemagnetisierung |
| $\underline{\vec{M}} = (\underline{M}_{xy}, \underline{M}^*_{xy}, M_z)^T$ | Komplexe Magnetisierung |
| $\underline{M}_{xy}$ | Transversalkomponente der komplexen Magnetisierung |
| $M_{\text{ref}}$ | Magnetisierung des Referenzabtastwerts |
| $M_{\text{test}}$ | Magnetisierung des Testabtastwerts |
| $m, n$ | Ordnungen einer Konfiguration in $y$- und $z$-Richtung |
| $N$ | Anzahl Abtastwerte pro Echo (dimensionslos) |
| $N_{\text{spin}}$ | Anzahl Spins |
| $P_{12}$ | Legendre-Polynom zwölfter Ordnung (dimensionslos) |
| $p$ | Quantenmechanische Wahrscheinlichkeitsfunktion (dimensionslos) |
| $q$ | Ordnungsinkrement einer Konfiguration (dimensionslos) |
| $\boldsymbol{R}_0$ | Koordinatentransformationsmatrix (dimensionslos) |
| $r = |\vec{x}|$ | Abstand vom Koordinatenursprung, $[1\ \text{m}]$ |
| $R$ | Radius der Homogenitätskugel eines MR-Systems, $[1\ \text{m}]$ |
| $s(t)$ | Magnetresonanz-Signal im Zeitbereich (dimensionslos) |
| $S(\vec{k}), S(t)$ | Ortsfrequenzfunktion (dimensionslos) |
| $\vec{S}_{\text{HF}}$ | Sensitivität der Empfangsspule, $[1\ \text{Vs A}^{-1}\ \text{m}^{-2}]$ |
| $T_1$ | Spin-Gitter-Relaxation, longitudinale Relaxation $[1\ \text{s}]$ |
| $T_2$ | Spin-Spin-Relaxation, transversale Relaxation $[1\ \text{s}]$ |
| $t$ | Zeit, [1 s] |
| $t_0$ | Anfangszeitpunkt, [1 s] |
| $T_E$ | Echozeit, [1 s] |
| $T_R$ | Repetitionszeit, [1 s] |
| $TF$ | Turbo-Faktor bei TSE-Sequenzen (dimensionslos) |
| $\boldsymbol{T}_{\text{Relax}}$ | Relaxationsoperator der Lösung der Blochgleichung (dimensionslos) |
| $\vec{T}_{\text{Relax, z}}$ | Relaxationsvektor für die longitudinale Magnetisierung, [1 A m$^{-1}$] |
| $\boldsymbol{T}_0$, $\boldsymbol{T}'_0$ | Rotationsoperator für das statische Feld (dimensionslos) |
| $\boldsymbol{T}_{\text{Grad}}$, $\boldsymbol{T}'_{\text{Grad}}$ | Rotationsoperator für das Gradientenmagnetfeld (dimensionslos) |
| $\boldsymbol{T}'_{\text{HF}}$ | Rotationsoperator für das HF-Feld (dimensionslos) |
| $\boldsymbol{U}$ | Rotationsoperator des rotierenden Bezugssystems (dimensionslos) |
| $u(t)$ | Elektrische Spannung an einer Empfangsspule, [1 V] |
| $u_r(t)$, $u_i(t)$ | Quadraturkomponenten von $u(t)$ |
| $V$, d$V$ | Volumen, infinitesimales Volumenelement, [1 m$^3$] |
| $\vec{x} = (x, y, z)^T$ | Ortsvektor, [1 m] |
| $\vec{x}' = (x', y', z')^T$ | Ortsvektor im rotierenden Bezugssystem |
| $\Delta x$, $\Delta y$, $\Delta z$ | Abtastabstand, [1 m] |
| $\mathcal{Z}_i$ | Longitudinale Konfiguration der Ordnung $i$, [1 A m$^{-1}$] |

## Abkürzungen und Namen

| | |
|---|---|
| CS | Chemical Shift |
| EPI | Echo Planar Imaging |
| FOV | Field of View |
| HF | Hochfrequenz |
| MR | Magnetresonanz |
| NMR | Nuclear Magnetic Resonance |
| Oufis | Optimized Ultra-Fast Imaging Sequence |
| ParSpin | Parallelized, Spin-based MR Simulator |
| PVM | Parallel Virtual Machine |
| Quest | Quick Echo Split imaging Technique |
| SE | Spin-Echo |
| Steam | Stimulated Echo Acquisition Mode |
| TSE | Turbo-Spin-Echo |

# Einleitung

Die Magnetresonanz-Tomographie, auch Kernspintomographie genannt, ist neben der Röntgen- und Ultraschallbildgebung eine der bedeutendsten Bildgebungsmodalitäten in der Medizin. Aufgrund ihrer hervorragenden diagnostischen Aussagekraft hat sie sich einen festen Platz im Repertoire der Radiologie erobert.

Die, verglichen mit anderen Modalitäten, junge MR-Bildgebung (MR: Magnet-Resonanz) besitzt ein hohes Entwicklungspotential, das sie weltweit zum Gegenstand zahlreicher Forschungsaktivitäten macht. Die Komplexität der zur Bildgebung genutzten Physik und der für ein MR-Experiment benötigten Hardware erschwert es in vielen Forschungsarbeiten, neue experimentelle Ansätze erfolgreich auf realen MR-Tomographen zu entwickeln und zu evaluieren. Detaillierte Einblicke lassen sich oft erst in der Kombination von realer Messung mit der numerischen Simulation des Bildgebungsvorgangs auf Digitalrechnern gewinnen.

Losgelöst von der Hardware macht es die numerische Simulation möglich, flexibel neue Bildgebungsexperimente zu entwerfen, zu testen und zu optimieren. In der Realität fest vorgegebene Grenz- und Randbedingungen eines Gerätes können in der Simulation in weiten Grenzen variiert werden. Somit wird es möglich, nicht nur die Machbarkeit eines Experiments auf einer gegebenen Hardware zu analysieren, sondern auch umgekehrt Anforderungen an die Eigenschaften eines Geräts zu formulieren, mit dem bestimmte Experimente möglich sein sollen.

Die numerische Simulation ist ein ausgezeichnetes Mittel, die Abläufe auf allen in der Realität oftmals unzugänglichen Ebenen des Bildgebungsverfahrens nachvollziehen zu können. Daneben erlaubt sie, gezielt nach Ursachen für Artefakte, d.h. systematischen Störungen der Bildqualität, zu suchen, indem mögliche Verursacher separat von anderen Einflüssen aktiviert oder deaktiviert werden können.

Geräte für die MR-Tomographie, mit denen große Volumina, etwa das eines Menschen, zwei- und dreidimensional untersucht werden können, haben hohe Anschaffungs- und laufende Betriebskosten. Aufgrund dessen ist vielen Forschergruppen ein uneingeschränkter Zugang zu einem MR-Tomographen nicht jederzeit möglich. Hier hilft die numerische Simulation durch Verkürzung der Sitzungen an einem MR-Tomographen und durch Unterstützung bei der Vor- und Nachbereitung von realen Experimenten.

In der Forschung werden meist Simulatoren entwickelt und genutzt, die eigens auf eine bestimmte Fragestellung zugeschnitten sind. Flexible Simulationswerkzeuge mit einem uni-versellen Einsatzbereich sind in der MR-Forschung selten. Gründe hierfür sind sowohl der hohe für einen universellen Simulator zu bemessende Entwicklungsaufwand, als auch die hohe Laufzeit eines Simulators mit einem umfangreichen physikalischen Modell.

Diese unbefriedigende Situation machte die Entwicklung des neuartigen Simulators PARSPIN erforderlich, dem Gegenstand der vorliegenden Arbeit. Primäres Ziel war die Unterstützung der Forschungsaktivitäten des Lehrstuhls für Allgemeine Elektrotechnik und Datenverarbeitungssysteme der RWTH Aachen im Bereich der Magnetresonanz-Bildgebung durch ein möglichst flexibles, schnelles und bezüglich der vorhandenen Rechenleistung skalierbares Simulationswerkzeug. Von besonderem Interesse war hier der Anwendungsbereich der medizinischen Bildgebung in Verbindung mit der Simulierbarkeit von MR-Tomographiesystemen mit idealen und nichtidealen Eigenschaften.

Die Beteiligung des Lehrstuhls an der interdisziplinären, von der Deutschen Forschungsgemeinschaft geförderten Forschergruppe „NMR-Oberflächentomographie", die sich mit der Entwicklung eines Oberflächen-MR-Tomographen kleiner und preiswerter Bauart beschäftigt, führte zu einer Intensivierung der Anstrengungen, nichtideale Geräteeigenschaften simulierbar zu machen.

Die vorliegende Arbeit beschreibt alle für die Realisierung relevanten Aspekte des Simulators. In den folgenden Kapiteln wird zunächst das Anforderungsprofil definiert und ein Überblick über Simulationsansätze anderer Gruppen gegeben. Nachfolgend legen eine Beschreibung der mathematischen Grundlagen und die Ableitung einer im Rahmen der Arbeit entwickelten Analysemethode für MR-Experimente, der K-t-Formalismus, ein solides theoretisches Fundament.

Mögliche Konzepte für einen MR-Simulator werden anschließend diskutiert und auf ihre Eignung hinsichtlich der gegebenen Anforderungen untersucht. Erstmalig wird die präzise mathematische Herleitung einer Vorschrift für die Ortsraumdiskretisierung bei sogenannten spinbasierten Simulatoren gegeben.

Im Anschluß hieran werden die wichtigsten Überlegungen bei der Umsetzung des am besten geeigneten Konzepts erörtert. In diesem Teil der Arbeit liegt besonderes Gewicht auf Ansätzen, die Gesamtlaufzeit eines rechenintensiven Simulators durch eine skalierbare, parallelisierte Programmstruktur wesentlich zu verkürzen. Mögliche Ansätze werden dargestellt und hinsichtlich ihrer Leistungsfähigkeit bewertet.

Die Arbeit wird durch reichhaltige Simulationsbeispiele und einen Vergleich ausgewählter, realer MR-Experimente mit ihrer simulierten Entsprechung abgerundet.

# 1 Motivation und Zielsetzung

In den späten dreißiger und den vierziger Jahren des zwanzigsten Jahrhunderts fanden mehrere Forschergruppen den Effekt der magnetischen Resonanz, mit dessen Hilfe eine Vielzahl von magnetischen Eigenschaften der Atomkerne verschiedener Nuklide untersucht werden können. Damit war es erstmals möglich, sehr präzise Aussagen über das magnetische Verhalten von Elementarteilchen zu treffen, aus denen sich neue Erkenntnisse ableiten ließen. Praktische Anwendungen fanden sich lange Zeit hauptsächlich in den Materialwissenschaften und der Werkstoffprüfung.

Von methodischen und technologischen Weiterentwicklungen des Meßverfahrens begleitet, begann in den achtziger Jahren die breite Nutzung der magnetischen Resonanz in der Medizin. Die medizinische Magnetresonanz-Bildgebung fand rasche Verbreitung in der Radiologie, da sie zum einen ohne ionisierende, den Patienten belastende Strahlung auskommt, zum anderen neue Arten von Bildkontrasten, insbesondere im Bereich anatomisch weichen Gewebes, hervorbrachte.

Der Effekt der magnetischen Resonanz läßt sich am präzisesten mit der Theorie der Quantenmechanik beschreiben, die Informationen über die quantenmechanische Wahrscheinlichkeitsfunktion $p$ bzw. das magnetische Moment $\vec{\mu}$ eines Nukleons in Abhängigkeit von den es umgebenden magnetischen Feldern liefert. Es zeigt sich, daß in vielen Fällen anstelle eines quantentheoretischen ein klassischer Ansatz zur Modellierung von Spin-Systemen herangezogen werden kann [1].

Im Zusammenhang mit der magnetischen Resonanz werden die betrachteten Nukleonen verallgemeinernd häufig als *Spin* bezeichnet; dieser Name wird in der vorliegenden Arbeit übernommen. Anstelle des magnetischen Moments $\vec{\mu}$ wird die Magnetisierung $\vec{M}$ eines Spin-Ensembles mit einer großen Anzahl von Spins betrachtet. $\vec{M}$ ist hier die resultierende Magnetisierung des Spin-Ensembles, hervorgerufen durch die magnetischen Momente der Einzelspins.

Kontext dieser Arbeit bildet die auf $^1$H-Nukleonen, d.h. auf Atomkernen des Wasserstoff-Atoms, basierende MR-Bildgebung in medizinischen Anwendungen. Andere Nukleonen, z.B. Phosphor, sind ebenfalls für Bildgebung und Spektroskopie nutzbar, weisen jedoch in medizinisch relevantem Gewebe eine geringere Konzentration als $^1$H-Nukleonen auf und sind daher in der Praxis weniger bedeutsam.

[1]H-Nukleonen interagieren gewöhnlich mit ihrer Umgebung, z. B. mit benachbarten Spins oder, verursacht durch thermische Bewegung, mit der Materie in ihrer Nähe. Bei der Bildgebung in Flüssigkeiten oder flüssigkeitsähnlichen Substanzen sind diese Kopplungen schwach, man spricht dann von quasi-freien Spins. Diese Kopplungen können phänomenologisch durch zwei zeitlich exponentiell abklingende Relaxationsvorgänge beschrieben werden, die *Spin-Spin-Relaxation* und die *Spin-Gitter-Relaxation* [6]. Diese Relaxationsarten besitzen charakteristische, materialabhängige Zeitkonstanten $T_1$ (Spin-Gitter-Relaxation) und $T_2$ (Spin-Spin-Relaxation).

Die zeitliche Änderung der Magnetisierung $\vec{M}$ kann von außen durch zeitkonstante und zeitabhängige Magnetfelder beeinflußt werden. Bei geeigneter Wahl dieser Felder können durch Messung der zeitlichen Veränderung der Magnetisierung Rückschlüsse auf Strukturen und Eigenschaften des untersuchten Objekts gezogen werden. Für die Erzeugung eines Tomogramms sind drei Typen von äußeren Magnetfeldern erforderlich:

- ein *statisches* Magnetfeld mit einer sehr hohen magnetischen Flußdichte im Tesla- oder Zehntel-Teslabereich,

- ein relativ langsam veränderliches *Gradienten*magnetfeld mit charakteristischen Zeitkonstanten im Bereich von einigen zehn Mikrosekunden und einer idealerweise linearen Abhängigkeit der magnetischen Flußdichte von den drei Raumkoordinaten, sowie

- ein *hochfrequentes* Magnetfeld mit Flußdichten im Bereich einiger Mikrotesla und Frequenzen im Bereich der Kurz- und Ultrakurzwellen.

Diese drei Feldtypen werden simultan mit stromdurchflossenen Spulen erzeugt. Weitere, im Hochfrequenzbereich operierende Spulenanordnungen dienen der meßtechnischen Aufnahme der zeitlichen Änderung der Magnetisierung $\vec{M}$. Die Anforderungen an die magnetische Entkopplung der Spulen und an die Flexibilität ihrer Ansteuerung sind sehr hoch. Ein kommerzielles Meßsystem zur Nutzung der magnetischen Resonanz für die Bildgebung ist aus diesen Gründen ein hochkomplexes und teures System einer Vielzahl von technischen Komponenten.

Ähnlich wie die Relaxationskonstanten ist die Anzahl der Spins pro Volumeneinheit eine materialabhängige Größe. Diese *Spindichte* $\rho(\vec{x})$ ist von außen nicht unmittelbar meßbar. In einem Volumenelement stellt sich jedoch in Verbindung mit dem äußeren statischen Magnetfeld eine *Ruhemagnetisierung* $M_0(\vec{x})$ ein, die zur Spindichte proportional ist. Die Ruhemagnetisierung kann mit geeigneten Verfahren meßtechnisch erfaßt werden.

Spins besitzen daneben im gesamten Objekt eine zur Flußdichte des statischen Magnetfelds proportionale, vom Nukleon abhängige nominale *Resonanzfrequenz* $\omega_0$. Auf sie muß das hochfrequente Magnetfeld abgestimmt sein, um – aus quantenmechanischer Sicht – das Energieniveau der Spins verändern zu können, bzw. – aus klassischer Sicht – die Magnetisierungs-vektoren kohärent über alle Spins verändern zu können. Aufgrund von weiteren Materialeigenschaften kann die Resonanzfrequenz eines Spins vom Nominalwert $\omega_0$ abweichen.

## 1.1 Vorzüge numerischer Simulation

Bei der Entwicklung von MR-Experimenten werden i. a. Antworten auf eine oder mehrere der folgenden Fragestellungen gesucht:

- Wie ist die zeitliche Abfolge des Experiments, die sogenannte *Sequenz*, d.h. wie werden die zeitveränderlichen Magnetfelder angesteuert ?

- Wie sind die Parameter der Sequenz einzustellen, so daß das Ergebnis im Hinblick auf anwendungsspezifische Kriterien optimal ist ?

- Wie muß eine für ein MR-Experiment geeignete Hardware beschaffen sein, und wie lauten die optimalen Parameter einer gegebenen Hardware-Konfiguration ?

- Wie läßt sich die Qualität des aus den Meßwerten gewonnenen Ergebnisses, z. B. eines Abbilds des Objekts, durch Signalverarbeitung oder angepaßte Bildrekonstruktionsverfahren verbessern ?

Ihre Beantwortung ist mit realen Experimenten oftmals nur eingeschränkt möglich. Insbesondere bei der Suche nach einer optimalen Hardware können in der Praxis nur wenige Alternativen untersucht werden. Aber auch die Identifikation der Ursache für eine schlechte Ergebnisqualität einer Sequenz ist nach einem realen Experiment nicht immer möglich, weil inhärent eine Vielzahl von möglichen Einflußgrößen präsent ist.

Die numerische Simulation des Bildgebungsvorgangs auf Digitalrechnern eröffnet hier Auswege, da sie frei von den prinzipiellen Einschränkungen einer realen Messung ist. So eröffnet sie die Möglichkeit, Einflußgrößen gezielt zu aktivieren oder zu deaktivieren. Damit kann unmittelbar und zielgerichtet ermittelt werden, unter welchen Voraussetzungen die Ergebnisqualität einer Sequenz den Anforderungen genügt.

Neuartige Hardware-Designs können auf ihre Eignung für einen bestimmten Zweck untersucht werden, noch bevor ein Prototyp realisiert wird. Darüber hinaus können unter Berücksichtigung der gesamten Meßkette die Hardware-Parameter, wie z. B. Abmessungen oder Materialeigenschaften, im Hinblick auf die Qualität des Ergebnisses numerisch optimiert werden.

Die numerische Simulation eines MR-Bildgebungssystems bietet, zusammengefaßt, einen Mehrwert, der über das Einsparen teurer Meßzeit weit hinausgeht, und ist daher unerläßlicher Bestandteil von Forschungsvorhaben.

Ein weiterer Aspekt soll nicht unerwähnt bleiben: die Magnetresonanz-Tomographie ist ein technisch und methodisch anspruchsvolles Bildgebungsverfahren, das im Gegensatz zur Rönt-

gen- oder Ultraschallbildgebung keinen einfachen, anschaulichen Zugang bietet. Einerseits sind die Vorgänge im Inneren des untersuchten Objekts nicht unmittelbar beobachtbar, der experimentelle Aufbau ist technisch hochkompliziert und das gemessene Signal als eine Transformierte des gesuchten Bildes nur wenig aussagekräftig. Andererseits liefert die häufig in der Literatur vorzufindende Beschreibung der Dynamik einzelner Spins nicht den nötigen Überblick über ein Experiment, um es als ganzes verstehen zu können.

Die numerische Simulation bietet hier als Bindeglied zwischen Experiment und Theorie die Möglichkeit, die Vorgänge in einem Objekt und die Entstehung des gemessenen Signals unmittelbar beobachten zu können.

## 1.2    Bedarf an numerischer Simulation

Am Lehrstuhl für Allgemeine Elektrotechnik und Datenverarbeitungssysteme der RWTH Aachen wurde im Rahmen von Forschungsarbeiten, hauptsächlich zur medizinischen Kernspintomographie, stets auf rechnergestützte Simulationswerkzeuge zurückgegriffen. Als störend erwies sich wiederholt, daß vorhandene bzw. verfügbare Werkzeuge nicht den Anforderungen genügten, und ihre Nachbesserung, so überhaupt möglich, nicht zu einem allgemein und flexibel einsetzbaren Simulator führte.

Forschungsschwerpunkte im Bereich der Bildgebungsmethoden waren die Eigenschaften sehr komplexer Sequenzen (ultraschnelle Methoden wie PREVIEW [22], QUEST [39], BURST [43], OUFIS [106] oder Verfahren mit stochastischer Anregung [7]) in Abhängigkeit von dem analysierten Objekt und der Systemhardware [16] [58]. Darüber hinaus wurde untersucht, welche Auswirkungen stark inhomogene Magnetfelder auf den Bildgebungsprozeß haben [33] und wie die Qualität gestörter Bilder durch Anpassungen des Experiments und Erweiterungen der Signal- und Bildverarbeitung verbessert werden kann [15] [50] [51].

Ultraschnelle oder stochastische Sequenzen unterscheiden sich strukturell stark von herkömmlichen Methoden und lassen sich auf kommerziellen medizinischen MR-Systemen oft nur unvollkommen implementieren. Die Erforschung des Potentials dieser Bildgebungsansätze ist auf die Verfügbarkeit eines flexiblen Simulationswerkzeugs zwingend angewiesen.

Nicht zuletzt ist im Bereich der universitären Lehre Erkenntnisgewinn willkommen, z. B. bei der Einarbeitung und der Ausbildung studentischer Mitarbeiter (Hilfskräfte, Diplomanden) oder bei Lehrveranstaltungen für Studierende höherer Semester. Mit den weitergehenden Einblicken, die eine numerische Simulation bietet, und unabhängig von der knappen Ressource „Meßzeit" auf einem realen MR-System, kann ein größerer Personenkreis in die Thematik eingeführt werden.

Mit der Beteiligung des Lehrstuhls in der interdisziplinären Forschergruppe „NMR Oberflächentomographie" der Deutschen Forschungsgesellschaft (DFG) erweiterten sich die Anforderungen an ein Simulationswerkzeug insbesondere im Bereich der Nachbildung der Systemhardware. Die hier betrachteten Geometrien des Bildgebungssystems zeichnen sich dadurch aus, daß das System das zu untersuchende Objekt nicht umschließt, sondern lediglich auf die Oberfläche des Objekts aufgelegt wird.

Im Vergleich zu üblichen medizinischen MR-Systemen weisen die von einer solchen Anordnung erzeugten Magnetfelder eine wesentlich höhere Inhomogenität auf. Daneben besitzen die Felder i. a. keine Vorzugsrichtung im kartesischen Koordinatensystem, sondern sind örtlich variabel und wechselnd orientiert. Eine Reihe von bei der mathematischen Modellierung oft eingesetzten Vereinfachungen läßt sich hierdurch nicht verwenden.

## 1.3  Anforderungen an einen MR-Simulator

Zur Deckung des vorhandenen Simulationsbedarfs ist ein Simulationswerkzeug erforderlich, das einer Reihe von Anforderungen genügt. Das Anforderungsprofil umfaßt vier Bereiche, die die Simulierbarkeit bestimmter Fragestellungen oder die Bedienung des Werkzeugs betreffen. Allen voran steht die Modellierung des Objekts, dessen physikalisches Model wesentlich die Bandbreite der möglichen virtuellen Experimente bestimmt. Weniger grundlegend, aber dennoch wichtig ist die Abbildung von Sequenz und System-Hardware auf Programmstrukturen des Simulators.

### Anforderungen an das Objektmodell

Grundlegende Anforderung im Bereich des Objekts ist die Modellierung der Gesamtmagnetisierung $\vec{M}_{\text{ges}}(t)$ des Objekts als Volumenintegral über alle Einzelmagnetisierungen $\vec{M}(\vec{x}, t)$. Für weitergehende Einblicke in einen Bildgebungsvorgang ist es nützlich, zusätzlich auf die Einzelmagnetisierungen zugreifen zu können. Daneben müssen die Relaxationskonstanten $T_1(\vec{x})$ und $T_2(\vec{x})$ in Abhängigkeit vom Ort definiert und berücksichtigt werden können; gleiches gilt für die zur Spindichte proportionale Ruhemagnetisierung $\vec{M}_0(\vec{x})$. Für realitätsnahe Simulationen müssen sich außerdem Objektausdehnungen in alle drei Raumrichtungen spezifizieren lassen.

Neben diesen grundlegenden Anforderungen an das Objekt, ohne deren Berücksichtigung ein MR-Simulationswerkzeug nicht auskommt, müssen für ein verfeinertes Modell weitere Größen herangezogen werden. So kann es aufgrund der molekularen Umgebung eines Spins zu einer Abweichung der Resonanzfrequenz vom Nominalwert $\omega_0$ kommen. Diese Verschiebung wird *Chemical Shift* (CS) genannt; sie ist ebenfalls gewebeabhängig und kann in einer Simulation in Form einer Kreisfrequenzabweichung $\Delta\omega_{\text{CS}}(\vec{x})$ berücksichtigt werden.

Weiterhin sind die magnetischen Eigenschaften des Materials in der Bildgebung bedeutsam, allen voran die magnetische Suszeptibilität $\chi(\vec{x})$. In der Nähe von Grenzflächen zwischen Materialien mit verschiedenen Suszeptibilitätskonstanten kommt es zu Verzerrungen des magnetischen Felds in Betrag und Richtung. Hiervon besonders betroffen ist das statische Magnetfeld. Im einfachsten Fall der Abweichung des Betrags des statischen Magnetfelds vom Nennwert lassen sich die Effekte an Suszeptibilitätsgrenzflächen mit einer Verschiebung $\Delta\omega_{susz}(\vec{x})$ der Resonanzfrequenz modellieren. Doch auch Abweichungen der Richtung des Magnetfelds von der Nennrichtung müssen berücksichtigt werden können.

Die Feldverzerrung an Suszeptibilitätsgrenzflächen kann mit Hilfe der Maxwellschen Gleichungen der Elektrodynamik ermittelt werden. Sie ist i. a. nur mit numerischen Mitteln zu finden, deren Diskussion den Rahmen dieser Arbeit übersteigt. Ein MR-Simulator muß aber zumindest die mit anderen Werkzeugen berechnete Feldverzerrungsfunktion verarbeiten und in die Simulation einbeziehen können.

### Anforderungen an das Sequenzmodell

Für die MR-Bildgebung gibt es eine Vielzahl von Bildgebungssequenzen, die sich in ihrer Struktur teilweise sehr stark voneinander unterscheiden. Ein Simulationswerkzeug muß bei der Gestaltung von Sequenzen größtmögliche Flexibilität bieten, um auch neue Sequenzansätze unterstützen zu können. Konkret bedeutet dies, daß die Zeitfunktionen von Gradienten- und HF-Magnetfeldern über das gesamte Experiment nahezu beliebig spezifiziert werden können müssen, ohne daß ein Simulationswerkzeug an Grenzen stößt.

Sequenzelemente wie gepulste HF-Magnetfelder (*HF-Pulse*) mit einer Amplituden- und Phasenmodulation müssen ebenso simulierbar sein wie Gradientenmagnetfelder mit speziellen Zeitabhängigkeiten. Sehr wichtig ist, daß in einer Simulation alle drei Raumrichtungen berücksichtigt werden können.

### Anforderungen an das Systemmodell

Ein reales Bildgebungssystem besteht aus einer Vielzahl von Komponenten, z. B. den Spulen für statische, Gradienten- und hochfrequente Magnetfelder, den für ihren Betrieb erforderlichen Steuereinheiten und der Bedienkonsole, um nur die wichtigsten zu nennen. Jede Komponente besitzt spezifische physikalische Eigenschaften, die das Meßergebnis mittelbar oder unmittelbar beeinflussen.

Für das zeitliche Verhalten der Magnetisierung eines Spins sind ausschließlich seine MR-spezifischen physikalischen Eigenschaften und das ihn umgebende magnetische Feld relevant. Hinzu kommt der Vorgang der Messung des MR-Signals, bei dem die Magnetisierung des Objekts mit einer HF-Spule erfaßt und anschließend aufgezeichnet wird.

In einem idealen MR-System sind das statische und das hochfrequente Magnetfeld jeweils homogen in $\vec{x}$. Die magnetische Flußdichte der Gradientenmagnetfelder ist linear von den drei Raumkoordinaten $x$, $y$ und $z$ abhängig. Die Feldlinien des hochfrequenten Magnetfelds stehen senkrecht auf denen des statischen Magnetfelds, während die des Gradientenmagnetfelds parallel zu denen des statischen Magnetfelds liegen. In einem nichtidealen System sind diese Bedingungen verletzt. Im Hinblick auf den beschriebenen Bedarf müssen sowohl ein ideales System als auch ein nichtideales System abbildbar sein.

**Sekundäre Anforderungen**

Weitere Anforderungen ergeben sich aus dem Einsatz im Forschungsbereich. Ein Simulationswerkzeug muß in diesem Kontext schnell, effizient und verfügbar sein und die primäre Aufgabe des Benutzers unterstützen. Auch komplexe Fragestellungen müssen simuliert werden können; dazu muß ein Simulator so skalierbar sein, daß er durch Hinzuführen von Rechenkapazität auf einfache Weise leistungsfähiger wird.

Nicht zuletzt ist eine gründliche Validierung des Simulationswerkzeugs von großer Bedeutung, um Zuverlässigkeit und Aussagekraft der Simulationsergebnisse sicherstellen zu können.

Bevor auf die Umsetzung der Anforderungen näher eingegangen wird, gibt das folgende Kapitel einen Überblick über den Stand der Technik der Simulation von MR-Experimenten und die einschlägige Literatur zu diesem Thema.

# 2 Computergestützte Simulation von Magnetresonanz-Experimenten

Die numerische, rechnerbasierte Simulation des Bildgebungssystems spielt seit Beginn der achtziger Jahre des zwanzigsten Jahrhunderts weltweit eine immer wichtigere Rolle in Forschungseinrichtungen. Beginnend mit Realisierungen auf Großrechnern, danach verstärkt übergehend zu Arbeitsplatzrechnern, wurden von verschiedenen Gruppen entsprechende Arbeiten in der Literatur dokumentiert. Die verwendeten Ansätze zur Simulation des Bildgebungssystems lassen sich in vier Klassen unterteilen:

1. Spinbasierte Ansätze, bei denen das zu simulierende Objekt in Elemente unterteilt wird, die dem physikalischen Modell der Nukleonen sehr nahekommen (z.B. [5] [90]);

2. Ansätze, die auf dem von Hennig vorgestellten erweiterten Phasengraph-Algorithmus [41] [42] basieren oder mit ihm verwandt sind (z.B. [78] [79]);

3. Ansätze, die die Ortsfrequenzfunktion eines gegebenen Objekts aufgrund der Abbildungseigenschaften eines MR-Systems verzerren (z.B. [4] [72]);

4. Ansätze, die ausschließlich das Objekt im Ortsbereich betrachten und nach der simulierten Sequenz verfremden (z.B. [87] [101]).

Neben den in der Literatur dokumentierten Simulationswerkzeugen werden in vielen Forschungseinrichtungen auch Werkzeuge erstellt und verwendet, die lediglich den zur Erfüllung der jeweils aktuellen Aufgabenstellung erforderlichen Funktionsumfang besitzen. Die wenigen verfügbaren Informationen über derartige Simulatoren lassen eine Klassifizierung und Bewertung nicht zu.

Im folgenden werden die in der Literatur veröffentlichten Realisierungen von MR-Simulatoren dokumentiert. Abschließend werden sie hinsichtlich ihrer Eignung bewertet, die im vorhergehenden Kapitel dargestellten Anforderungen erfüllen zu können.

## 2.1 Spinbasierte Simulatoren

Bei spinbasierten Simulatoren wird versucht, die Vorgänge bei der Entstehung eines MR-Signals auf der Ebene von Spins numerisch nachzubilden. Die Detailliertheit der Umsetzung reicht von der Berechnung und Lösung der quantenmechanischen Wellenfunktion eines Spin-systems bis hin zu makroskopischen, klassischen, also nicht quantentheoretischen Ansätzen. Letztere basieren meist auf einem von Bloch 1946 vorgestellten Differentialgleichungssystem [6] für die Magnetisierung $\vec{M}$, der später nach ihm benannten *Blochschen Gleichung*. Spinbasierte Simulatoren auf Basis der Blochschen Gleichung sind im Kontext der medizinischen MR-Bildgebung am häufigsten anzutreffen.

Mit ersten Anwendungen der MR-Tomographie in der Medizin entstanden auf Großrechnern ablaufende Simulationwerkzeuge. Hierzu gehören Arbeiten von Siebold [90], Bittoun et al. [5], John et al. [48] und Machin et al. [64].

Der von Siebold 1982 knapp beschriebene Simulator integrierte eine Reihe von nichtidealen Eigenschaften der System-Hardware, die am Anfang der kommerziellen medizinischen Nutzung bedeutend waren. Dazu gehören neben der Inhomogenität der Felder auch Aspekte wie die zeitliche Stabilität der magnetischen Felder oder das Phasenverhalten des Empfängers für die hochfrequente Antwort des untersuchten Objekts.

Der 1984 erschienene Aufsatz von Bittoun et al. beschreibt detailliert ein weiteres spinbasiertes Simulationswerkzeug. Es unterstützt allerdings nur eindimensionale Experimente unter idealen Bedingungen, d.h. ohne die Berücksichtigung von Magnetfeldinhomogenitäten oder nichtlinearen Gradientenmagnetfeldern.

Der von John et al. 1984 verfaßte Aufsatz beschreibt Theorie und Resultate eines Simulators für die Spektroskopie. Das vorgestellte Werkzeug liefert, ähnlich einem Formelmanipulationssystem, analytische Ausdrücke für die Reaktion eines Systems stark miteinander gekoppelter Spins auf eine Sequenz von hochfrequenten Pulsen. Für die numerische Simulation von Bildgebungsexperimenten ist es nicht geeignet.

Das von Machin et al. 1986 entwickelte Simulationswerkzeug nutzt im Gegensatz zu den meisten anderen Bloch- bzw. spinbasierten Vertretern nicht analytische Lösungen der Blochschen Gleichung, sondern numerische Lösungsverfahren. Diese Veröffentlichung ist sehr knapp, insbesondere hinsichtlich des Funktionsumfangs.

In der Folgezeit entstanden mehr und mehr Simulationswerkzeuge auf Mini-Computern verschiedenster Bauart. In den letzten zehn Jahren wurden auch Arbeitsplatzrechner so leistungsfähig, daß sie als Simulationsplattform genutzt wurden.

Ein interessantes Programm dieser Reihe von Summers et al. [94], erschienen 1986, bietet ein detailliertes physikalisches Modell nebst einer grafikunterstützten Menüführung, siehe Bild 2.1. Die mögliche Komplexität der untersuchten Probleme, hier vor allem die Anzahl der Spins, ist gering. Durch eine stark hardwareabhängige Implementierung ist das Programm nur mit sehr großem Aufwand auf andere Plattformen übertragbar. Damit ist die Verwendung in aktuellen Forschungsvorhaben nicht sinnvoll.

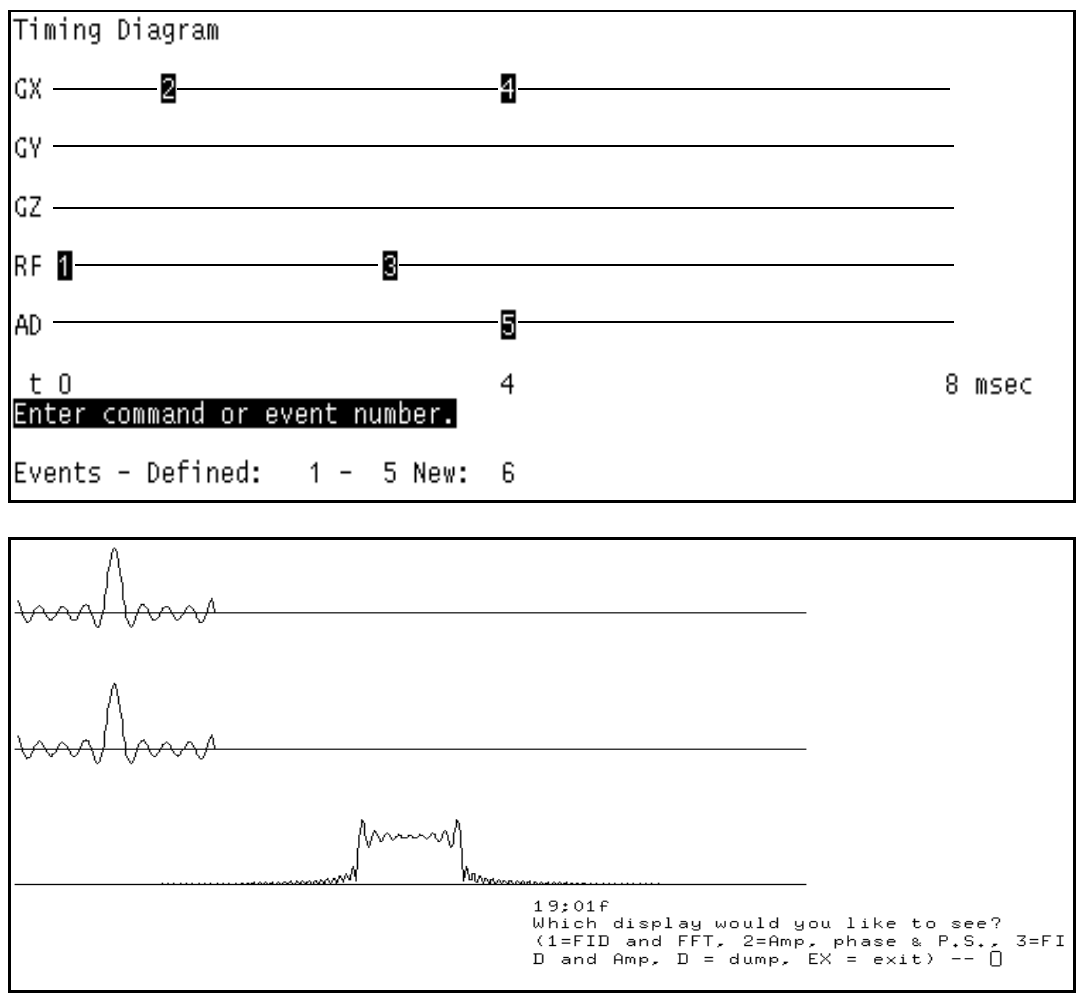

Bild 2.1: Bedienoberfläche des VAX-basierten MR-Simulators nach Summers et. al [94]: Sequenzeditor (oben) und Simulationsergebnis (unten)

Roberts skizzierte 1991 in [85] das Funktionsprinzip einiger kleiner Simulationswerkzeuge, die jeweils auf einen Anwendungsfalls zugeschnitten waren. Die gewünschte allgemeine Einsetzbarkeit eines Simulators kann mit ihnen nicht erreicht werden.

Eine sehr umfangreiche Simulationsumgebung ist das von Smith et al. 1994 vorgestellte Paket GAMMA [91], das Simulationen spektroskopischer Experimente bei mehrfach gekoppelten Spinsystemen unterstützt. Sein mathematisches Modell basiert auf einer quantenmechanischen Beschreibung des Spinsystems, daher sind gute theoretische Kenntnisse der Quantenmechanik Voraussetzung für die Benutzung. Das Paket besteht aus einer objektorientierten Programmierbibliothek zur Beschreibung und zur Handhabung von Spinsystemen. Der Benutzer muß das zu simulierende Experiment prozedural beschreiben. Wegen des mathematisch aufwendigen Modells, das viele verschiedene Kopplungsarten nachbildet, lassen sich mit GAMMA nur wenige Spins gleichzeitig simulieren. Für Simulationen von Bildgebungsexperimenten ist das Paket unzureichend.

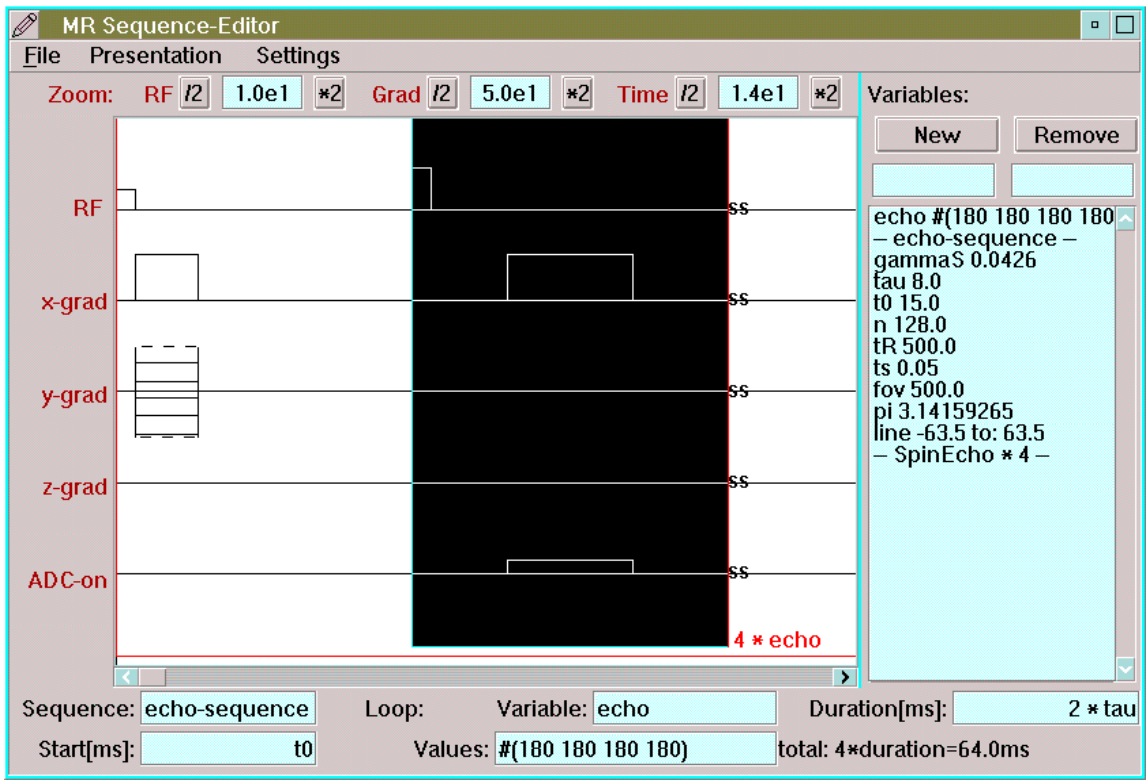

Bild 2.2: Sequenzeditor eines am Lehrstuhl erstellten objektorientierten Simulationspro-
gramms unter dem Betriebssystem OS/2 und der Programmiersprache Smalltalk.

1995 erschien die Veröffentlichung von Olsson et al. [74]. Sie baut auf der Arbeit von Bittoun
et al. [5] auf und erweitert den dort beschriebenen Funktionsumfang um die Berücksichtigung
von Feldinhomogenitäten in der Nähe von Grenzflächen mit unterschiedlichen magnetischen
Suszeptibilitäten. Besonderes Gewicht wird auf die Elimination fehlerhafter Signale gelegt, die
bei einer zu geringen Abtastung des simulierten Objekts entstehen (vgl. auch Abschnitt 5.2
dieser Arbeit). Das Simulationswerkzeug ist für Arbeitsplatzrechner konzipiert. Ähnlich wie
das GAMMA-Paket erfordert es vom Benutzer Programmierkenntnisse, da die gewünschte
Sequenz und die zu simulierenden Ereignisse prozedural spezifiziert werden müssen.

Taniguchi et al. legten in [96] besonderes Augenmerk auf eine effiziente Methode, amplitu-
den- und phasenmodulierte hochfrequente Magnetfelder zu simulieren. Eine starke Rechen-
zeitreduktion ergibt sich, wenn ein spezifischer zeitlicher Verlauf mehrfach in einem Experi-
ment auftritt. Inhomogenitäten von Magnetfeldern bleiben in dem Aufsatz unberücksichtigt.

Im Rahmen einer Diplomarbeit [107] wurde in der Programmiersprache Smalltalk der Proto-
typ eines Simulators mit einer komfortablen Bedienoberfläche fertiggestellt, siehe Bild 2.2.
Haupthindernis beim Einsatz dieses Simulators in Forschungsaktivitäten war eine sehr geringe
Performance. Hinzu kam ein sehr großer Speicher- und Prozessorleistungsbedarf des umfang-
reichen, aus Betriebssystem und einer Smalltalk-Laufzeitumgebung bestehenden Gesamtsy-
stems.

Parallel zu der vorliegenden Arbeit entstand an der Katholischen Universität Leuven, Belgien, ein objektorientiert programmiertes Simulationswerkzeug, das ein Cluster von vernetzten UNIX-Workstations für eine parallelisierte Simulation verwendet. Die Programmierung erfolgte in der Sprache C++. Die 1996 verfaßte Beschreibung in [69] läßt darauf schließen, daß es sich bei diesem Projekt um eine Technologiestudie über Vorteile objektorientierter Herangehensweisen handelt. Der Funktionsumfang des Simulators ist prinzipiell stark erweiterbar, aber im vorhandenen Ausbau klein.

Mit Ausnahme des Werkzeugs von Michiels et al. besitzt keines der vorgestellten Werkzeuge das Potential, allgemein einsetzbar zu sein und gleichzeitig eine geringe Rechenzeit auch bei sehr komplexen Experimenten zu erreichen. Leider veröffentlichte die Gruppe ihren Aufsatz nur als internen Bericht an der Universität Leuven, so daß sie für die vorliegende Arbeit erst sehr spät in Betracht gezogen werden konnte. Bis zum jetzigen Zeitpunkt hat es keine Publikation in der Fachliteratur gegeben.

## 2.2 Simulatoren auf Basis des erweiterten Phasengraph-Algorithmus'

Der erweiterte Phasengraph-Algorithmus ist eine häufig für komplexe Experimentabläufe genutzte, auf der Blochschen Gleichung basierende Analysemethode, vgl. [39] [40] [43]. Ausgehend von einer Reihenentwicklung der Magnetisierung des untersuchten Objekts kann mit mathematisch einfachen Mitteln das zeitliche Auftreten und die Amplitude von MR-Signalen bestimmt werden, die bei einem gegebenen Experimentablauf an einem Objekt gemessen werden können.

Der Phasengraph-Algorithmus umfaßt neben einer mathematischen auch eine graphische Komponente, die das Resultat der Reihenentwicklung anschaulich darzustellen vermag (vgl. Kapitel 4 dieser Arbeit). Die graphische Komponente, sie führt zu den *Phasengraphen*, läßt sich auch ohne die mathematische einsetzen. Für eine numerische Simulation ist nur die mathematische Komponente des Phasengraph-Algorithmus' von Interesse, da nur sie quantitative Aussagen über Zeitpunkt, Hüllkurve und Amplitude eines MR-Signals zuläßt.

Petersson et al. stellten 1993 in [78] erstmalig eine Simulationsmethode vor, die eine mit dem erweiterten Phasengraph-Algorithmus verwandte Methode für eine Simulation nutzt. Sie bietet gegenüber einem spinbasierten Ansatz mit gleichem Funktionsumfang einen erheblichen Laufzeitvorteil, da die Simulation zunächst für Objektbereiche mit gleichen MR-Eigenschaften, z. B. gleichen $T_1$-$T_2$-Kombinationen, durchgeführt wird. Erst anschließend wird mit einfachen Operationen der Systemtheorie die Gestalt des Objekts in das Ergebnis einbezogen. Magnetfeldinhomogenitäten können zunächst nicht berücksichtigt werden. Eigene Arbeiten, vgl. [54] [56] [57], bestätigten die Eignung der Vorgehensweise für komplexe Sequenzen und das ausgezeichnete Laufzeitverhalten.

In [79] erweiterten Petersson et al. 1997 ihren Ansatz aus [78] um die Berücksichtigung von Inhomogenitäten des statischen Magnetfelds, ohne das Laufzeitverhalten wesentlich zu verschlechtern. Die hierfür nötigen Erweiterungen der theoretischen Basis waren jedoch umfangreich und sind nicht ohne weiteres auf Inhomogenitäten von Gradienten- und hochfrequenten Magnetfeldern übertragbar. Die bis zum gegenwärtigen Zeitpunkt einzige Möglichkeit, Inhomogenitäten beliebiger Art flexibel berücksichtigen zu können, besteht in der Aufgliederung des Objekts in Quader und der Durchführung des Algorithmus' für jeden Quader getrennt. Hierdurch geht der Laufzeitvorteil des Ansatzes gegenüber einer spinbasierten Lösung vollständig verloren, ohne daß deren Anpassungsfähigkeit erreicht wird.

## 2.3    Ortsfrequenz-Simulatoren

In der MR-Bildgebung entsprechen die aufgezeichneten Signale in erster Näherung der Ortsfrequenzfunktion der gesuchten Spindichte-Ortsfunktion, verzerrt oder gestört durch diverse Einflüsse, z. B. durch Rauschen oder durch Abweichungen der eingesetzten magnetischen Felder von denen des idealen MR-Systems, vgl. Abschnitt 1.3.

Ortsfrequenz-Simulatoren gehen von der ungestörten Ortsfrequenzfunktion eines Objekts aus und verzerren sie anschließend mit einer Störungsfunktion, die aufgrund einer Einflußgröße zu erwarten ist. Ein durch inverse Fouriertransformation aus dem gestörten Datensatz gewonnenes Bild weist Artefakte auf, die mit experimentell beobachtbaren Bildartefakten oft gut übereinstimmen.

Die Simulationsmethode wird in der Literatur meist zusammen mit Studien zu einzelnen Aspekten von Bildgebungsexperimenten vorgestellt. Dementsprechend sind die erstellten Simulationswerkzeuge in ihrem Funktionsumfang auf die Fragestellung zugeschnitten, also nicht allgemein einsetzbar.

Einige Beispiele für Simulationen im Ortsfrequenzbereich sind in der Literatur dokumentiert. Bakker et al. veröffentlichten 1994 in [4] eine Simulationsstudie über Bildartefakte, die an Grenzflächen zwischen Materialien mit unterschiedlichen magnetischen Suszeptibilitätskoeffizienten auftreten. Nishimura et al. beschäftigten sich 1995 in [72] mit den Effekten von sich bewegenden Materialien auf das entstehende Bild.

## 2.4    Ortsraum-Simulatoren

Ortsraum-Simulatoren ermitteln anhand einstellbarer Parameter einer kleinen Menge von Experimentabläufen, welche Einflüsse auf das Bild zu erwarten sind. Hierzu zählt insbeson-dere der Einfluß des Experiments auf Bildkontraste zwischen unterschiedlichen Objektregio-
nen mit verschiedenen MR-spezifischen Eigenschaften.

Weis et al. diskutierten 1990 in [101] den Einfluß von Inhomogenitäten des statischen und des
Gradientenmagnetfelds und ihre Korrekturmöglichkeiten. Für eine Simulation wird mit Matri-
zen, die die durch Inhomogenitäten entstehenden Intensitätsänderungen und geometrischen
Verzerrungen beschreiben, ein ungestörtes Bild modifiziert. Nach einer Interpolation auf das
ursprüngliche Pixelraster ergibt sich das gewünschte Simulationsergebnis.

Rundle et al. stellten 1990 in [87] ein für die Ausbildung von Radiologen entwickeltes Werk-
zeug vor. Es basiert auf einem Pool von mit verschiedenen Sequenzen und Parametern auf
einem realen MR-System aufgenommenen Bildern. Abhängig von den Sequenzparametern,
die der Benutzer für eine Simulation vorgibt, wird aus einer Datenbank das zugehörige Bild
entnommen und dargestellt, falls es vorhanden ist. Das Programm ist eher als Multimedia-
Anwendung denn als Simulator einzustufen, da es vorhandene Bilder entweder nur darstellt
oder in einem speziellen Betriebsmodus ein Bild aus anderen durch Gewichtung und Überlage-
rung synthetisiert.

Einen ähnlichen Ansatz zeigten Torheim et al. 1994 in [97]. Anhand von realen MR-Bildern
werden pixelweise die MR-Eigenschaften Spindichte, $T_1$- und $T_2$-Relaxationszeiten ermittelt.
Mit einfacher Arithmetik wird unter Berücksichtigung der gegebenen Parameter ein Bild
berechnet. Die Datenbank besteht hier aus Bildern verschiedener anatomischer Regionen mit
künstlich eingefügten pathologischen Veränderungen.

Crum et al. veröffentlichten 1997 in [23] einen speziellen Simulator für *Tagging*-Sequenzen,
die biologisches Gewebe mit einem örtlich regelmäßigen Muster anregen. Die Veränderungen
des Musters durch Bewegung des Gewebes lassen Rückschlüsse auf krankhafte Veränderun-
gen zu. Das präsentierte Werkzeug ermittelt für jeden Bildpunkt die aufgrund der eingesetzten
Sequenz zu erwartende Intensität. Das von der Sequenz erzeugte Tagging-Muster wird nach-
folgend durch ortsabhängige Gewichtung in das Bild integriert. Zwei Typen von Tagging-
Sequenzen werden unterstützt.

Alle so erstellten Simulatoren sind, ähnlich wie Ansätze im Ortsfrequenzbereich, auf eine spe-
zifische Fragestellung zugeschnitten und nicht allgemein einsetzbar. Sie erfordern spezielles
Vorwissen über die Wirkung von Einflußgrößen auf das Bild, das bei unbekannten oder modi-
fizierten Experimenten nicht verfügbar ist. Somit sind auch solche Ansätze zur Abdeckung des
Anforderungsspektrums ungeeignet.

# 3 Grundlegende Modellierung

Ein klassisches Modell für zeitliche Veränderung der Magnetisierung $\vec{M}$ eines Spins erleichtert, verglichen mit einer quantenmechanischen Behandlung, die mathematische Beschreibung und das anschauliche Verständnis des zeitlichen Verhaltens von Spins in einem äußeren magnetischen Feld erheblich. Bloch leitete in [6] eine klassische Gleichung für die zeitliche Änderung der Magnetisierung ab.

Die *Blochsche Gleichung* ist ein gewöhnliches Differentialgleichungssystem erster Ordnung. Sie besitzt einen weiten Gültigkeitsbereich für *quasi-freie* Spins, die nur schwach mit benachbarten Spins und dem umgebenden Atomgitter interagieren. Insbesondere für den in der Medizin wichtigen Fall des Experiments mit Wasserstoff-Nukleonen bildet die Blochsche Gleichung eine hervorragend geeignete Basis für die mathematische Modellierung.

Das vorliegende Kapitel beschreibt, wie zum einen die Entstehung und zum anderen die Messung von MR-Signalen mathematisch erfaßt werden kann. Die mathematische Modellierung ihrer *Entstehung* ist eng mit der Blochschen Gleichung verknüpft. Diese muß für eine Vielzahl von Randbedingungen jeweils separat betrachtet werden. Nicht minder wichtig, aber mathematisch einfacher zu fassen, ist die am Ende des Kapitels behandelte *Messung* eines MR-Signals.

## 3.1 Definitionen

Bei der mathematischen Modellierung des Magnetresonanzeffekts bei quasi-freien Spins interessiert die zeitliche Entwicklung der Magnetisierung $\vec{M}$ in einem Einheitsvolumen unter dem Einfluß äußerer magnetischer Felder mit der Flußdichte $\vec{B}$. Magnetisierung und Flußdichte sind i. a. vom Ort $\vec{x} = (x, y, z)^T$ und von der Zeit $t$ abhängig, und sie besitzen eine beliebige Orientierung. In der vorliegenden Arbeit gelten folgende grundlegende Definitionen:

$$\vec{M} = \vec{M}(x, y, z, t) = \begin{pmatrix} M_x(x, y, z, t) \\ M_y(x, y, z, t) \\ M_z(x, y, z, t) \end{pmatrix} \qquad (3.1)$$

für die Magnetisierung und

$$\vec{B} \;=\; \vec{B}(x, y, z, t) \;=\; \begin{pmatrix} B_x(x, y, z, t) \\ B_y(x, y, z, t) \\ B_z(x, y, z, t) \end{pmatrix} \tag{3.2}$$

für die magnetische Flußdichte. Die magnetische Flußdichte des äußeren magnetischen Feldes ist von der Magnetisierung unabhängig und kann frei gewählt werden, wohingegen die Magnetisierung vom zeitlichen Verlauf der Flußdichte, den Anfangswerten sowie einigen MR-spezifischen Materialkonstanten abhängt.

## 3.2    Die Blochsche Gleichung

Die zeitliche Änderung der Spinmagnetisierung $\vec{M}$ hängt vom Momentanwert der Magnetisierung und der magnetischen Flußdichte ab. Die Magnetisierung, ihre zeitliche Änderung und die magnetische Flußdichte sind über ein Differentialgleichungssystem miteinander verknüpft, und es gilt

$$\frac{\mathrm{d}\vec{M}}{\mathrm{d}t} \;=\; \gamma \cdot (\vec{M} \times \vec{B})\,. \tag{3.3}$$

Die Größe $\gamma$ ist hier eine Materialkonstante, das *gyromagnetische Verhältnis*[1].

Gleichung (3.3) ist für die Beschreibung der Magnetischen Resonanz nur bedingt geeignet. Sie vernachlässigt die in der Praxis wichtigen, *schwachen* quantenmechanischen Kopplungen zwischen nicht weit voneinander entfernten Spins sowie zwischen Spins und der umgebenden Materie. Abhilfe schafft die Blochsche Gleichung, die durch Erweiterung des Differentialgleichungssystems (3.3) um die Spin-Gitter- ($T_1$) und die Spin-Spin-Relaxation ($T_2$) entsteht (vgl. Kapitel 1).

Die magnetische Flußdichte $\vec{B}$ wird von bis zu drei Quellen unterschiedlichen Charakters hervorgerufen, die statische, quasi-stationäre und hochfrequente Magnetfelder erzeugen. Das statische Magnetfeld $\vec{B}_0$ hat bei der Betrachtung der Blochschen Gleichung eine Sonderrolle: seine Orientierung gibt die Richtung vor, in der sich ein *thermisches Gleichgewicht* ausbildet.

Das thermische Gleichgewicht ist der Zustand eines Spin-Ensembles, der sich einstellt, wenn sich das Ensemble sehr lange in einem statischen Magnetfeld aufhält. Wird ein Spin-Ensemble gestört, so stellt sich durch einen Ausgleichsvorgang stets wieder das thermische Gleichgewicht ein. Dabei konvergiert die Longitudinalmagnetisierung zeitlich exponentiell mit der charakteristischen Zeitkonstanten $T_1$ gegen ihren Wert im thermischen Gleichgewicht.

---

[1] Für Protonen hat das gyromagnetische Verhältnis den Zahlenwert $2\pi \cdot 42{,}6 \cdot 10^6 \; (\mathrm{s} \cdot \mathrm{T})^{-1}$.

Zwischen der longitudinalen Relaxationszeit $T_1$ und der Orientierung des statischen Magnetfelds existiert demnach ein Zusammenhang. Für eine mathematisch effiziente Formulierung der Blochschen Gleichung ist es deshalb hilfreich, eine Achse des Bezugskoordinatensystems entlang der Richtung des statischen Magnetfelds zu wählen. Es ergeben sich Komponenten von Magnetisierung und magnetischer Flußdichte *transversal* und *longitudinal* zu dieser Achse des Bezugssystems.

In der Literatur ist es gebräuchlich, die *z*-Achse des Bezugssystem in die Richtung des statischen Magnetfelds zu definieren. Die Blochsche Gleichung lautet für diese spezielle Wahl der Orientierung von $\vec{B}_0$

$$\frac{\mathrm{d}\vec{M}}{\mathrm{d}t} = \gamma \cdot (\vec{M} \times \vec{B}) + \begin{pmatrix} -\dfrac{1}{T_2} & 0 & 0 \\[2mm] 0 & -\dfrac{1}{T_2} & 0 \\[2mm] 0 & 0 & -\dfrac{1}{T_1} \end{pmatrix} \cdot \vec{M} + \begin{pmatrix} 0 \\[1mm] 0 \\[1mm] \dfrac{M_0}{T_1} \end{pmatrix} \tag{3.4}$$

$$= \gamma \cdot (\vec{M} \times \vec{B}) + \boldsymbol{E} \cdot \vec{M} + \vec{E}_0.$$

In dieser Arbeit stellt die Annahme einer im gesamten Ortsraum gleichen Orientierung des statischen Magnetfeldvektors eine unzulässige Beschränkung dar. Weicht die Orientierung des statischen Magnetfelds an einem Ort $\vec{x}$ von der für Gleichung (3.4) gewählten ab, kann vor der Anwendung der Blochschen Gleichung eine lineare Koordinatentransformation der Vektoren $\vec{M}$ und $\vec{B}$ durchgeführt werden. Dazu wird das Bezugssystem lokal in $\vec{x}$ gegenüber dem Laborsystem so gedreht, daß eine der Achsen des gedrehten Systems in Richtung des statischen Magnetfelds weist.

In dem so gedrehten Bezugssystem ist eine Formulierung der Blochschen Gleichung analog zu Gleichung (3.4) möglich, und der Magnetisierungvektor kann ermittelt werden. Durch Inversion der Koordinatentransformation ergibt sich abschließend der gesuchte Magnetisierungsvektor im Laborsystem.

## 3.3   Ansätze zur Lösung der Blochschen Gleichung

Die Gleichungen (3.3) und (3.4) stellen zusammen mit den zugehörigen Anfangswerten $\vec{M}(t = t_0)$ jeweils ein inhomogenes, gewöhnliches Differentialgleichungssystem (DGL-System) erster Ordnung dar. Für die Lösung dieses DGL-Systems existieren zwei grundsätzlich verschiedene Herangehensweisen: zum ersten kann versucht werden, eine analytische Lösung für das DGL-System zu finden, zum zweiten können DGL-Systeme mit numerischen Verfahren gelöst werden.

Numerische Methoden erlauben das Lösen des DGL-Systems auf sehr allgemeine Art, d.h. für nahezu beliebige Randbedingungen und zeitliche Abhängigkeiten der äußeren Magnetfelder. Einschränkungen der Variabilität von Randbedingungen und Zeitabhängigkeiten ergeben sich, wenn anstelle solcher numerischer Lösungen ein Satz analytischer, möglichst allgemein verwendbarer Lösungen der Blochschen Gleichung zur Anwendung kommt.

Bei der Verwendung analytischer Lösungen muß die Blochsche Gleichung nicht während der Simulation für jeden Spin erneut gelöst werden. Vielmehr werden unter Beachtung des jeweiligen Gültigkeitsbereichs Randbedingungen und Zeitabhängigkeiten in eine der vorhandenen Lösungen eingesetzt und numerisch ausgewertet. Die Verwendung analytischer Lösungen bietet meist einen erheblichen Laufzeitvorteil gegenüber iterativen numerischen Lösungsverfahren. Auch virtuell lang andauernde Vorgänge können mit einem einzelnen arithmetischen Schritt berechnet werden, solange der Gültigkeitsbereich einer Lösung nicht verlassen wird.

Numerische Lösungsmethoden bieten sich an, wenn für eine spezielle Konstellation keine analytische Lösung auffindbar ist. Vor allem zeitabhängige HF-Felder stellen bei der Suche nach analytischen Lösungen der Blochschen Gleichung oft ein Problem dar. In der vorliegenden Arbeit konnte auf numerische Lösungsmethoden verzichtet werden, da die verwendeten analytischen Lösungen ein sehr großes Anwendungsspektrum abdecken. Die bei der Behandlung von zeitabhängigen HF-Feldern genutzten analytischen Ansätze sind weiter unten beschrieben (Abschnitt 3.4.4).

Das Auffinden des analytischen Lösungsvektors ist im Falle von Systemen mit linearen Differentialgleichungen erster Ordnung gut schematisiert [17]. Rechnerbasierte Formelmanipulationssysteme, beispielsweise Maple [53], vereinfachen auf effiziente Weise den Lösungsvorgang auch bei komplexen Randbedingungen.

Obwohl die Blochsche Gleichung theoretisch für jeden zeitlich stetigen Verlauf von $\vec{B}$ eine Lösung besitzt [34][38][100], kann diese analytisch i. a. nicht ohne die Kenntnis einer speziellen zeitlichen Abhängigkeit von $\vec{B}$ angegeben werden. In einigen Fällen lassen sich dennoch Lösungen finden, ohne eine spezielle Zeitfunktion festzulegen. Derartige Lösungen werden im folgenden Abschnitt präsentiert. Lösungen der Blochschen Gleichung sind in der Literatur häufig und für verschiedene Einsatzbereiche dokumentiert, z. B. in [1], [37], [47], [65] oder [99]. Für diese Arbeit wurde die Notation vereinheitlicht, die Gültigkeit einiger Lösungsvektoren wurde verifiziert, und einige für ein Simulationswerkzeug notwendige Ergänzungen wurden hinzugefügt.

# 3.4   Analytische Lösungen der Blochschen Gleichung

Im folgenden werden die für das Abdecken des Anforderungsprofils relevanten analytischen Lösungen der Blochschen Gleichung angegeben. Ergänzungen zu einzelnen Lösungen finden sich in Anhang A.

## 3.4.1   Lösung für statische Magnetfelder

Nach Einsetzen der Flußdichte $\vec{B}_0 = (0, 0, B_{0_z})^T$ eines statischen Magnetfelds in Gleichung (3.4) und Lösung des DGL-Systems ergibt sich als Lösungsvektor, ausgehend von einem Anfangswert $\vec{M}(t_0)$, nach einer Zeit $\Delta t = t - t_0$ der neue Magnetisierungsvektor $\vec{M}(t)$

$$\vec{M}(t) = \begin{pmatrix} \exp(-\Delta t/T_2) & 0 & 0 \\ 0 & \exp(-\Delta t/T_2) & 0 \\ 0 & 0 & \exp(-\Delta t/T_1) \end{pmatrix} \cdot$$

$$\begin{pmatrix} \cos(\gamma \cdot B_{0_z} \cdot \Delta t) & \sin(\gamma \cdot B_{0_z} \cdot \Delta t) & 0 \\ -\sin(\gamma \cdot B_{0_z} \cdot \Delta t) & \cos(\gamma \cdot B_{0_z} \cdot \Delta t) & 0 \\ 0 & 0 & 1 \end{pmatrix} \cdot \vec{M}(t_0) + \qquad (3.5)$$

$$\begin{pmatrix} 0 \\ 0 \\ M_0 \cdot [1 - \exp(-\Delta t/T_1)] \end{pmatrix}.$$

Durch Einführen der Größen $T_{\text{Relax}}$, $T_0$ und $\vec{T}_{\text{Relax,z}}$ ergibt sich in Matrixschreibweise

$$\vec{M}(t) = T_{\text{Relax}} \cdot T_0 \cdot \vec{M}(t_0) + \vec{T}_{\text{Relax,z}}. \qquad (3.6)$$

Gleichung (3.5) beschreibt eine zeitlich abklingende Präzessionsbewegung der Magnetisierung um die $z$-Achse. Die Transversalmagnetisierung rotiert mit der Kreisfrequenz $\gamma \cdot B_{0_z}$ in mathematisch negativer Richtung um die $z$-Achse. Für die Präzessionskreisfrequenz ergibt sich

$$\vec{\omega}_0(\vec{x}) = -\gamma \cdot \vec{B}_0(\vec{x}), \qquad (3.7)$$

die *Larmor-Frequenz*. Durch Spin-Spin-Relaxation vermindert sich die Transversalmagnetisierung exponentiell mit der charakteristischen Zeitkonstanten $T_2$. Die longitudinale Magnetisierung strebt exponentiell mit der Zeitkonstanten $T_1$ die Ruhemagnetisierung $M_0$, das thermische Gleichgewicht, an.

Bei der Analyse der Bewegung von Magnetisierungsvektoren ist es hilfreich, ein mit der Kreisfrequenz $\omega_{\mathrm{HF}}$ in mathematisch negativem Sinn um die $z$-Achse rotierendes Bezugssystem zu definieren. Gleichung (3.6) geht in diesem Fall über in

$$\vec{M}'(t) \,=\, \boldsymbol{T}_{\mathrm{Relax}} \cdot \boldsymbol{T}'_{0} \cdot \vec{M}'(t_0) + \vec{T}_{\mathrm{Relax,z}}\,; \tag{3.8}$$

hierbei bezeichnen gestrichene Terme Größen im rotierenden Bezugssystem. Für die Matrix $\boldsymbol{T}'_{0}$ gilt

$$\boldsymbol{T}'_{0} = \begin{pmatrix} \cos\left[(\gamma \cdot B_{0z} - \omega_{\mathrm{HF}})\Delta t\right] & \sin\left[(\gamma \cdot B_{0z} - \omega_{\mathrm{HF}})\Delta t\right] & 0 \\ -\sin\left[(\gamma \cdot B_{0z} - \omega_{\mathrm{HF}})\Delta t\right] & \cos\left[(\gamma \cdot B_{0z} - \omega_{\mathrm{HF}})\Delta t\right] & 0 \\ 0 & 0 & 1 \end{pmatrix}. \tag{3.9}$$

Bei kleinen Differenzen $\gamma \cdot B_{0z} - \omega_{\mathrm{HF}} \approx 0$ ist $\boldsymbol{T}'_{0}$ näherungsweise identisch mit der Einheitsmatrix und kann in Gleichung (3.8) vernachlässigt werden.

### 3.4.2   Chemical Shift

Die Präzessionskreisfrequenz $\omega(\vec{x})$ eines Spins, der Chemical Shift unterliegt, ist die Summe aus der vom äußeren Magnetfeld abhängigen Larmor-Frequenz und der Verschiebung durch Chemical Shift. Mit einer Larmorfrequenzverschiebung $\Delta\omega_{\mathrm{CS}}(\vec{x})$ wird die Präzessionskreisfrequenz eines Spins zu

$$\omega(\vec{x}) \,=\, \gamma \cdot B_{0z}(\vec{x}) + \Delta\omega_{\mathrm{CS}}(\vec{x})\,. \tag{3.10}$$

Chemical Shift äußert sich durch eine Erweiterung der Matrixelemente aus Gleichung (3.9), und es gilt für eine sowohl das statische Magnetfeld als auch Chemical Shift berücksichtigende Rotationsmatrix

$$\boldsymbol{T}'_{0,\,\mathrm{CS}} = \begin{pmatrix} \cos\left[(\gamma \cdot B_{0z} - \omega_{\mathrm{HF}} + \Delta\omega_{\mathrm{CS}})\Delta t\right] & \sin\left[(\gamma \cdot B_{0z} - \omega_{\mathrm{HF}} + \Delta\omega_{\mathrm{CS}})\Delta t\right] & 0 \\ -\sin\left[(\gamma \cdot B_{0z} - \omega_{\mathrm{HF}} + \Delta\omega_{\mathrm{CS}})\Delta t\right] & \cos\left[(\gamma \cdot B_{0z} - \omega_{\mathrm{HF}} + \Delta\omega_{\mathrm{CS}})\Delta t\right] & 0 \\ 0 & 0 & 1 \end{pmatrix}. \tag{3.11}$$

Mit der Näherung $\gamma \cdot B_{0z} - \omega_{\mathrm{HF}} \approx 0$ folgt

$$\boldsymbol{T}'_{0,\,\mathrm{CS}} \approx \begin{pmatrix} \cos(\Delta\omega_{\mathrm{CS}}\Delta t) & \sin(\Delta\omega_{\mathrm{CS}}\Delta t) & 0 \\ -\sin(\Delta\omega_{\mathrm{CS}}\Delta t) & \cos(\Delta\omega_{\mathrm{CS}}\Delta t) & 0 \\ 0 & 0 & 1 \end{pmatrix}. \tag{3.12}$$

### 3.4.3 Lösungen für Gradientenmagnetfelder

Gradientenmagnetfelder mit der Flußdichte $B_{\mathrm{grad}}(\vec{x}, t)$ sind langsam zeitabhängige Magnetfelder mit einer definierten Ortsabhängigkeit, die bei der MR-Bildgebung für die *Ortskodierung* benötigt werden. Im Idealfall hat die Flußdichte des Gradientenmagnetfelds zwei besondere Eigenschaften. Zum einen besitzt sie nur eine Komponente $B_{\mathrm{grad},z}$ parallel zum statischen Magnetfeld, zum anderen ist der zeitabhängige Gradientenvektor $\vec{G}(t) = \nabla B_{\mathrm{grad},z}(\vec{x}, t)$ der verbleibenden Komponente örtlich konstant, das resultierende Magnetfeld $\vec{B}_{\mathrm{grad},z} = (\vec{G} \cdot \vec{x}) \cdot \vec{e}_z$ also linear vom Ort $\vec{x}$ abhängig.

#### 3.4.3.1 Gradientenmagnetfelder parallel zum statischen Feld

Es gilt für die magnetische Flußdichte von Gradientenmagnetfeldern, die parallel zum statischen Magnetfeld gerichtet sind,

$$\vec{B}_{\mathrm{grad}}(\vec{x}, t) = [0, 0, B_{\mathrm{grad},z}(\vec{x}, t)]^T. \tag{3.13}$$

Für diesen Fall läßt sich eine Lösung für beliebige, zeitlich stetige Funktionen $B_{\mathrm{grad},z}(\vec{x}, t)$ angeben. Für den Lösungsvektor $\vec{M}$ nach Einwirkung eines Gradientenmagnetfelds mit der Zeitdauer $\Delta t = t - t_0$ ergibt sich

$$\vec{M}(t) = \begin{pmatrix} \exp(-\Delta t/T_2) & 0 & 0 \\ 0 & \exp(-\Delta t/T_2) & 0 \\ 0 & 0 & \exp(-\Delta t/T_1) \end{pmatrix} \cdot$$

$$\begin{pmatrix} \cos\left(\gamma \cdot \int\limits_{t_0}^{t} B_{\mathrm{grad},z}(\tau)\mathrm{d}\tau\right) & \sin\left(\gamma \cdot \int\limits_{t_0}^{t} B_{\mathrm{grad},z}(\tau)\mathrm{d}\tau\right) & 0 \\ -\sin\left(\gamma \cdot \int\limits_{t_0}^{t} B_{\mathrm{grad},z}(\tau)\mathrm{d}\tau\right) & \cos\left(\gamma \cdot \int\limits_{t_0}^{t} B_{\mathrm{grad},z}(\tau)\mathrm{d}\tau\right) & 0 \\ 0 & 0 & 1 \end{pmatrix} \cdot \vec{M}(t_0) +$$

$$\begin{pmatrix} 0 \\ 0 \\ M_0 \cdot [1 - \exp(-\Delta t/T_1)] \end{pmatrix}. \tag{3.14}$$

bzw.

$$\vec{M}(t) = \boldsymbol{T}_{\mathrm{Relax}} \cdot \boldsymbol{T}_{\mathrm{grad}} \cdot \vec{M}(t_0) + \vec{T}_{\mathrm{Relax},z}. \tag{3.15}$$

Für die Integral-Terme in Gleichung (3.14) wird oft ein Größe $k$ mit der Dimension einer Orts-frequenz eingeführt:

$$\vec{k}(t) - \vec{k}(t_0) = \frac{\gamma}{2\pi} \cdot \int_{t_0}^{t} \vec{G}(\tau) \, d\tau. \tag{3.16}$$

Die von einem idealen Gradientenmagnetfeld mit der Flußdichte $\vec{B}_{\text{grad}}(\vec{x}, t) = [\vec{G}(t) \cdot \vec{x}] \cdot \vec{e}_z$ hervorgerufene Phasenänderung der transversalen Magnetisierung in der $x$-$y$-Ebene ist

$$\Delta\varphi(\vec{x}, t) = -\gamma \cdot \int_{t_0}^{t} \vec{G}(\tau) \cdot \vec{x} \, d\tau. \tag{3.17}$$

### 3.4.3.2  Gradientenmagnetfelder mit transversalen Komponenten

Besitzt das Gradientenmagnetfeld zum Hauptmagnetfeld transversale Komponenten, liegt die Vorzugsrichtung für $T_1$-Relaxation nicht mehr parallel zum Hauptmagnetfeld, sondern äqui-valent zur resultierenden Richtung des umgebenden Magnetfelds, bestehend aus Haupt- und Gradientenmagnetfeldern.

Nach [24] kann die Blochsche Gleichung zur Modellierung eingesetzt werden, wenn man die Änderung der Vorzugsrichtung als eine im systemtheoretischen Sinn sprungförmige Rich-tungsänderung des Magnetfelds auffaßt. Vorausgesetzt wird, daß $\vec{B}_{\text{grad}}$ zeitlich konstant im betrachteten Zeitraum $[t_0, t]$ ist. In diesem Fall folgt die Magnetisierung durch $T_1$-Relaxation der resultierenden Richtung des umgebenden Magnetfelds. Gradientenmagnetfelder mit trans-versalen Komponenten lassen sich so auf eine Richtungsänderung des Hauptmagnetfeldes zurückführen. Die Gleichungen (3.14) und (3.15) gelten dann in dem nach Abschnitt 3.2 rotier-ten Bezugssystem.

Für kleine, parasitäre transversale Komponenten der Gradientenmagnetfelder, die sogenannten *Concomitant Gradients*, wird in [73] für einige Bildgebungssequenzen gezeigt, daß ein Einfluß auf das MR-Signal nur in Fällen beobachtbar ist, in denen der über alle Komponenten ermit-telte Betrag der Flußdichte des Gradientenmagnetfelds in der Größenordnung der des stati-schen Magnetfelds ist.

## 3.4.4  Lösungen für hochfrequente Magnetfelder

Hochfrequente Magnetfelder dienen der Anregung eines Spin-Ensembles, also der Zufuhr von Energie bzw. der Störung des thermischen Gleichgewichts. Makroskopisch läßt sich mit hoch-frequenten Magnetfeldern eine Richtungsänderung der Magnetisierung erreichen, die zu einer von außen beobachtbaren Präzessionsbewegung des Magnetisierungsvektors führt. Um die Energieeinkopplung zu maximieren, muß die Trägerfrequenz gleich der Larmorfrequenz des Spin-Ensembles oder ihr zumindest nahe sein. Durch eine zeitliche Modulation des hochfrequenten Magnetfelds lassen sich seine anregenden Eigenschaften verändern, z. B. hinsichtlich der Einschränkung des Ortsbereichs, in dem es anregend wirkt.

### 3.4.4.1  HF-Felder mit zeitkonstanter Hüllkurve

In diesem Abschnitt werden hochfrequente Magnetfelder $\vec{B}_1$ diskutiert, die zirkular polarisiert sind. Ihr Feldvektor rotiert in einer Ebene transversal zum Hauptmagnetfeld mit einer Kreisfrequenz $\omega_{HF}$. Die Hüllkurve der Flußdichte ist zeitlich konstant. Für die Flußdichte eines hochfrequenten Magnetfelds, das in mathematisch negativer Richtung rotiert, gilt somit

$$\vec{B}_1 = \begin{pmatrix} B_1 \cdot \cos(\omega_{HF}t - \phi) \\ -B_1 \cdot \sin(\omega_{HF}t - \phi) \\ 0 \end{pmatrix}. \tag{3.18}$$

Die Ermittlung eines Lösungsvektors der Blochschen Gleichung für ein derartiges Magnetfeld vereinfacht sich, wenn das Spinsystem in einem um die $z$-Achse rotierenden Bezugssystem betrachtet wird, das sich synchron mit $\vec{B}_1$ dreht (vgl. Anhang A.2.1.1). Das hochfrequente Magnetfeld ist in diesem Bezugssystem zeitunabhängig. Für gegenüber den Relaxationszeiten $T_1$ und $T_2$ vernachlässigbare Einwirkdauern $\Delta t = t - t_0$ ist der Lösungsvektor im rotierenden Bezugssystem

$$\vec{M}'(t) = \begin{pmatrix} \cos^2\phi + \sin^2\phi\cos\alpha & \sin\phi\cos\phi(1-\cos\alpha) & -\sin\phi\sin\alpha \\ \sin\phi\cos\phi(1-\cos\alpha) & \sin^2\phi + \cos^2\phi\cos\alpha & \cos\phi\sin\alpha \\ \sin\phi\sin\alpha & -\cos\phi\sin\alpha & \cos\alpha \end{pmatrix} \cdot \vec{M}'(t_0). \tag{3.19}$$

Der Winkel $\alpha$ heißt *Flipwinkel*, und für ihn gilt $\alpha = \gamma \cdot B_1 \cdot \Delta t$. Das hochfrequente Magnetfeld dreht den Magnetisierungsvektor um den Flipwinkel $\alpha$ und um eine Achse in der $x'$-$y'$-Ebene, die mit der $x'$-Achse den Phasenwinkel $\phi$ einschließt. Analog zu den Gleichungen (3.6) und (3.15) gilt dann

$$\vec{M}'(t) = \mathbf{T}'_{HF} \cdot \vec{M}'(t_0). \tag{3.20}$$

### 3.4.4.2  Zeitveränderliche Hüllkurven — Näherungslösung

Ortsselektive Pulse werden üblicherweise durch eine Amplituden- und Phasenmodulation eines HF-Pulses realisiert, d.h. mittels einer zeitveränderlichen komplexen Hüllkurve. Der HF-Puls erhält dadurch ein charakteristisches Frequenzspektrum, das nur Spins anregt, deren Präzessionskreisfrequenz innerhalb des HF-Spektrums liegt. Mit der bisher ermittelten Lösung der Blochschen Gleichung für hochfrequente Magnetfelder können ortsselektive HF-Pulse nicht modelliert werden. Ein HF-Puls mit einer Hüllkurve $B_{1x}(t)$, $B_{1y}(t)$ und einem Träger nach Gleichung (3.18) besitzt eine magnetische Flußdichte

$$\vec{B}_1 = \begin{pmatrix} B_{1x}(t) \cdot \cos(\omega_{HF} t - \phi) \\ -B_{1y}(t) \cdot \sin(\omega_{HF} t - \phi) \\ 0 \end{pmatrix}. \tag{3.21}$$

Mit ihr ist die Blochsche Gleichung nicht mehr allgemein lösbar. Grund ist, daß auch in einem rotierenden Bezugssystem die magnetische Flußdichte noch zeitabhängig ist. Damit kann ohne explizite Kenntnis von $B_{1x}(t)$ und $B_{1y}(t)$ eine analytische Lösung nicht ermittelt werden. Ein Ansatz zu einer allgemeinen Näherungslösung besteht in der Annahme kleiner Flußdichten derart, daß während des HF-Pulses die Longitudinalmagnetisierung unverändert bleibt [99]. Das Blochsche Differentialgleichungssystem vereinfacht sich damit deutlich.

Für eine übersichtliche Darstellung des Lösungsvektors empfiehlt sich die Einführung der komplexen Transversalmagnetisierung $\underline{M}'_{xy} = M'_x + j \cdot M'_y$ und der komplexen transversalen Flußdichte $\underline{B}'_1 = B'_{1x} + j \cdot B'_{1y}$. Bei gleichzeitigem Vorhandensein einer zeitlich konstanten Flußdichte $B_z$ in $z$-Richtung ergibt sich als Näherungslösung

$$\underline{M}'_{xy}(t) \approx \underline{M}'_{xy}(t_0) \cdot \exp[-j\gamma \cdot B_z \cdot (t - t_0)] +$$
$$j\gamma \cdot M_z(t_0) \cdot \exp(-j\gamma \cdot B_z \cdot t) \int_{t_0}^{t} \underline{B}'_1(\tau) \cdot \exp(j\gamma \cdot B_z \cdot \tau) \ d\tau. \tag{3.22}$$

Die vor dem HF-Puls vorhandene transversale Magnetisierung erfährt für dessen Dauer eine von $B_z$ abhängige Phasenverschiebung. Zusätzliche transversale Beiträge entstehen aus longitudinaler Magnetisierung $M_z(t)$, die vor dem Puls vorhanden war. Sie werden im wesentlichen von der Fouriertransformierten der komplexen Hüllkurve des HF-Pulses bestimmt, ergänzt um eine Phasenverschiebung. Die Näherungslösung (3.22) ist nach [75] in der Praxis für Flipwinkel von bis zu 30° benutzbar.

### 3.4.4.3  Zeitveränderliche Hüllkurven — Lösung durch Diskretisierung

Die Lösungen (3.19) und (3.22) haben den Nachteil, daß für ihre Herleitung Relaxationseffekte vernachlässigt wurden. In der Praxis liegt hingegen die Dauer von modulierten HF-Pulsen oft in der Größenordnung der Relaxationskonstanten $T_1$ und $T_2$, so daß Relaxation i.a. nicht vernachlässigt werden darf. Darüber hinaus muß in (3.22) eine Beschränkung des Gesamtflipwinkels hingenommen werden.

Ein Ansatz zur Modellierung von zeitlich ausgedehnten, modulierten HF-Pulsen der Länge $T$ unter Verwendung analytischer Lösungen der Blochschen Gleichung ist die *Hard-Pulse-*

*Approximation* [76]. Hierbei wird die Hüllkurve eines HF-Pulses diskretisiert, also in $N_{HF}$ Teilpulse zerlegt (vgl. Anhang B.2).

Stimmt die Winkelgeschwindigkeit des rotierenden Bezugssystems mit der Larmorfrequenz überein, können die Einflüsse des Gradientenmagnetfelds von denen des HF-Felds separiert werden, und der Einfluß des statischen Magnetfelds entfällt. Der hochfrequente Teilpuls wird als sehr kurzer, „harter" HF-Puls aufgefaßt. Für Teilpuls $i$ gilt dann

$$\vec{M}'(t_i + \tau) \; = \; \boldsymbol{T}_{Relax} \cdot \boldsymbol{T}_{grad,\,i} \cdot \boldsymbol{T}'_{HF,\,i} \cdot \vec{M}'(t_i) + \vec{T}_{Relax,z} \tag{3.23}$$

mit den Definitionen aus Gleichung (3.15) und (3.19).

# 3.5  Messung des Magnetresonanzsignals

In den vorangegangenen Abschnitten wurde der Einfluß verschiedener Arten von äußeren magnetischen Feldern auf die Magnetisierung an einem Ort $\vec{x}$ diskutiert. Hier nun wird erläutert, wie die Magnetisierung eines Objekts meßtechnisch erfaßt werden kann.

Von außen läßt sich nur die Gesamtmagnetisierung eines Objekts messen, d.h. die Summe über alle Einzelmagnetisierungen $\vec{M}(\vec{x}, t)$. Für ihre Messung werden in der Nähe des Objekts Spulen angebracht, in denen aufgrund von magnetischer Induktion eine Spannung $u(t)$ erzeugt wird.

## 3.5.1  Signal bei einer inhomogenen Empfangssensitivität

Nach dem Reziprozitätsprinzip [20] lassen sich Eigenschaften einer Antenne im Sendefall auf den Empfangsfall übertragen. Ist also die von einer Spule erzeugte, von einem Strom $i(t)$ durch die Spule hervorgerufene magnetische Flußdichte $\vec{B}_1(\vec{x}, t)$ bekannt, ist auch ihre passive Sensitivität bekannt. Eine geeignete, zeitunabhängige Größe zur Charakterisierung der Spulensensitivität ist der Quotient aus der Flußdichte und dem sie erzeugenden Strom

$$\vec{S}_{HF}(\vec{x}) \; = \; \frac{\vec{B}_1(\vec{x}, t)}{i(t)} \,. \tag{3.24}$$

Die Empfangssensitivität hängt nicht nur von der Geometrie der Spule ab. In ihr ist auch auch die Anordnung der Spule relativ zum Objekt abgebildet, d.h. die Größe $\vec{x}$ enthält absolute Koordinaten im verwendeten Laborsystem.

Das von einem Spin an einem Ort $\vec{x}$ hervorgerufene Magnetfeld wird vor dem Empfang mit der Empfangssensitivität gewichtet. Die in einer Empfangsspule induzierte Spannung ergibt sich anschließend aus der linearen Superposition der gewichteten Magnetisierungen aller Spins im Volumen $V$ des Objekts. Nach [44] gilt für die induzierte Leerlaufspannung

$$u(t) = -\int_V \frac{\mathrm{d}}{\mathrm{d}t}[\vec{S}_{\mathrm{HF}}(\vec{x}) \cdot \vec{M}(\vec{x}, t)] \; \mathrm{d}V. \tag{3.25}$$

## 3.5.2   Signal bei einer homogenen Empfangssensitivität

Gleichung (3.25) vereinfacht sich, wenn Haupt- und Gradientenmagnetfelder parallel zur $z$-Achse gerichtet sind, $\vec{S}_{\mathrm{HF}}(\vec{x})$ über dem Volumen homogen ist und ausschließlich transversale Komponenten besitzt. In diesem Falle präzedieren die Magnetisierungsvektoren $\vec{M}(\vec{x}, t)$ mit der Kreisfrequenz $\omega(\vec{x}) = \gamma \cdot B_z(\vec{x})$ um die z-Achse, und nur die Transversalkomponenten der Magnetisierung sind relevant.

Unter Vernachlässigung von Relaxationseffekten wird Gleichung (3.5) zu

$$\vec{M}(\vec{x}, t) = \begin{pmatrix} M_x(\vec{x}, t_0) \cdot \cos[\omega(\vec{x})\Delta t] + M_y(\vec{x}, t_0) \cdot \sin[\omega(\vec{x})\Delta t] \\ -M_x(\vec{x}, t_0) \cdot \sin[\omega(\vec{x})\Delta t] + M_y(\vec{x}, t_0) \cdot \cos[\omega(\vec{x})\Delta t] \\ M_z(\vec{x}, t_0) \end{pmatrix}. \tag{3.26}$$

Im mit der Kreisfrequenz $\omega_{\mathrm{HF}}$ rotierenden Bezugssystem (vgl. Abschnitt 3.4.4.1) ergibt sich daraus die Magnetisierung

$$\vec{M}'(\vec{x}, t) = \begin{pmatrix} M_x(\vec{x}, t_0) \cdot \cos(\Delta\omega\Delta t) + M_y(\vec{x}, t_0) \cdot \sin(\Delta\omega\Delta t) \\ -M_x(\vec{x}, t_0) \cdot \sin(\Delta\omega\Delta t) + M_y(\vec{x}, t_0) \cdot \cos(\Delta\omega\Delta t) \\ M_z(\vec{x}, t_0) \end{pmatrix} \tag{3.27}$$

mit $\Delta\omega = \omega(\vec{x}) - \omega_{\mathrm{HF}}$.

Mit einer homogenen Empfangssensitivität $\vec{S}_{\mathrm{HF}}(\vec{x}) = (S_{\mathrm{HF}}, 0, 0)^T$ ergibt sich die in der Empfangsspule induzierte Spannung nach Gleichung (3.25) zu

$$u(t) = S_{\mathrm{HF}} \cdot \int_V \omega(\vec{x}) \cdot [M_x(\vec{x}, t_0) \cdot \sin(\omega(\vec{x})\Delta t) - M_y(\vec{x}, t_0) \cdot \cos(\omega(\vec{x})\Delta t)] \; \mathrm{d}V. \tag{3.28}$$

Bei klinischen Geräten ist die Varianz von $\omega(\vec{x})$, die von Hauptfeldinhomogenitäten und Gradientenmagnetfeldern hervorgerufen wird, klein gegenüber $\omega$. Näherungsweise kann daher $\omega(\vec{x})$ durch $\omega_0 = \omega(\vec{x}_0)$ nach Gleichung (3.7) ersetzt werden, und es gilt

$$u(t) \approx \omega_0 \cdot S_{\mathrm{HF}} \cdot \int_V [M_x(\vec{x}, t_0) \cdot \sin\omega(\vec{x})\Delta t - M_y(\vec{x}, t_0) \cdot \cos\omega(\vec{x})\Delta t] \; \mathrm{d}V. \tag{3.29}$$

Dieses Signal wird in der Praxis meist einem Quadraturempfänger zugeführt, der mit der Mittenfrequenz $\omega_{HF}$ betrieben wird; dazu wird es mit den beiden Signalen $\sin\omega_{HF}t$ und $-\cos\omega_{HF}t$ multipliziert. Es ergeben sich nach Tiefpaßfilterung die Quadraturkomponenten

$$u_r(t) = \frac{\omega_0 \cdot S_{HF}}{2} \int\limits_V [M_x(\vec{x}, 0) \cdot \cos\Delta\omega t + M_y(\vec{x}, 0) \cdot \sin\Delta\omega t] \ dV \qquad (3.30)$$

und

$$u_i(t) = -\frac{\omega_0 \cdot S_{HF}}{2} \int\limits_V [M_x(\vec{x}, 0) \cdot \sin\Delta\omega t - M_y(\vec{x}, 0) \cdot \cos\Delta\omega t] \ dV. \qquad (3.31)$$

Der Vergleich von (3.27) mit (3.30) und (3.31) zeigt, daß die Quadraturkomponenten proportional zum Ortsintegral über die $x$- und $y$-Komponenten der Magnetisierung im rotierenden Bezugssystem sind. Für viele simulative Studien genügt es daher, lediglich die Magnetisierungsvektoren aller Spins im rotierenden Koordinatensystem zu betrachten und ihre Summe als empfangenes Signal auszuwerten.

## 3.6  Imaging-Gleichung

In diesem Abschnitt wird anhand der Imaging-Gleichung, einem zentralen Konzept der MR-Bildgebung, diskutiert, wie aus dem gemessenen Signal das gewünschte Bild des Objekts rekonstruiert werden kann. Für die Diskussion wird von einer homogenen Empfangssensitivität ausgegangen, und Relaxation und Chemical Shift werden vernachlässigt.

Aus Gleichung (3.30) kann, z.B. nach [99], gefolgert werden, daß das MR-Signal $s(t)$ im rotierenden Bezugssystem proportional zum Volumenintegral über die transversale Magnetisierung des Objekts ist

$$\underline{s}(t) \sim \int\limits_V \underline{M}'_{trans}(x, t) \ dV. \qquad (3.32)$$

Für Zeiten während der Datenakquisition kann dies umgeschrieben werden zu

$$\underline{s}(\vec{k}) \sim \int\limits_V \exp[-2\pi j \cdot \vec{k} \cdot \vec{x}] \cdot \underline{M}'_{trans}(\vec{x}, t_0) \ dV, \qquad (3.33)$$

wobei $\vec{k}(t)$ nach Gleichung (3.16) die Einflüsse des Gradientenmagnetfelds beinhaltet. Gleichung (3.33) ist die Definitionsgleichung der Fourier-Transformation. Sie verknüpft in diesem Fall die Bildinformation $\underline{M}'_{trans}(\vec{x}, t_0)$ im Ortsraum mit dem gemessenen Signal $\underline{s}(\vec{k}(t))$ im Ortsfrequenzraum, der im Kontext der MR-Bildgebung meist als K-Raum bezeichnet wird.

Aus Gleichung (3.33) kann umgekehrt gefolgert werden, daß sich die Ortsinformation aus dem gemessenen Signal durch inverse Fourier-Transformation ermitteln läßt:

$$\underline{M}'(\vec{x}) \sim \int_K \exp(2\pi j \cdot \vec{k} \cdot \vec{x}) \cdot \underline{s}(\vec{k}) \, d\vec{k}. \tag{3.34}$$

Ein Ziel beim Design vieler Sequenztypen für die MR-Bildgebung ist es, mit dem Signal den Ortsfrequenzraum vollständig im Sinne des Abtasttheorems der Systemtheorie abzutasten [60]. Dies gelingt beispielsweise durch eine geeignete Ansteuerung der Gradientenmagnetfelder derart, daß die durch sie festgelegte Trajektorie durch den K-Raum die Abtastbedingung erfüllt (siehe auch Anhang B).

Die Rekonstruktion der Bildinformation aus dem MR-Signal nach Gleichung (3.34) läßt eine Reihe von Einflußgrößen außer acht, beispielsweise die Spin-Spin-Relaxation während der Datenakquisition. Dies führt i.a. zu Rekonstruktionsfehlern im Bild, die vielfältiger Natur sein können [9] [77]. Wegen der guten Handhabbarkeit der Fourier-Transformation auf Digitalrechnern und ihrer sehr guten Stabilität auch bei geringem Signal-Rausch-Abstand ist sie in der medizinischen Praxis dennoch eines der am häufigsten eingesetzten Bildrekonstruktionsverfahren.

# 4 Der K-t-Formalismus

Aus den Lösungen der Blochschen Gleichung läßt sich ein Formalismus ableiten, der das Verständnis von komplexen Bildgebungssequenzen mit einer großen Anzahl von hochfrequenten Pulsen erleichtert und eine numerisch effiziente Simulationsmethode darstellt. Hennig beschrieb in [41] und [42] graphische und arithmetische Komponenten des erweiterten Phasengraph-Formalismus für Spin-Ensembles, die sequentiell HF-Pulsen mit verschiedenen Flip-, aber gleichen Phasenwinkeln ausgesetzt werden. Petersson et al. veröffentlichten in [78] [79] eine auf der Betrachtung der Magnetisierung im K-Raum basierende Simulationsmethode, die dem erweiterten Phasengraph-Formalismus ähnelt, allerdings keine gesonderte graphische Repräsentation aufweist.

Im Rahmen der vorliegenden Arbeit wurde der erweiterte Phasengraph-Formalismus für zwei- und dreidimensionale Probleme und beliebige HF-Phasenwinkel erweitert und von der Betrachtung von Phasen eines Spin-Ensembles entkoppelt. Durch die Betrachtung von Positionen im K-Raum anstelle von Phasen wurde eine ortsunabhängige Beschreibung möglich, der *K-t-Formalismus*. Zusätzlich gelang eine fruchtbare Verknüpfung des K-t-Formalismus' mit dem geläufigen K-Raum-Konzept.

Das vorliegende Kapitel liefert eine detaillierte Beschreibung des K-t-Formalismus und zeigt mögliche Anwendungen auf. Obwohl er eine eigenständige Simulationsmethode ist, liefert er präzise Aussagen über die Ortsdiskretisierung bei einer spinbasierten Simulation und ist damit für die vorliegende Arbeit von unerläßlicher Bedeutung, vgl. Abschnitt 5.2.

## 4.1 Konzept der Elementarsequenz

Eine Bildgebungssequenz setzt sich aus einer Vielzahl von Zeitintervallen zusammen, während denen die aktiven Magnetfelder zeitlich konstant oder zumindest stetig sind. Für eine einheitliche, übersichtliche Beschreibung aller möglichen Zeitintervalle ist eine allgemein verwendbare Strukturierungsmethode nötig.

Für die Ableitung einer geeigneten Struktur wird ein typisches Timing-Diagramm einer einfachen Bildgebungssequenz betrachtet, siehe Bild 4.1. Diese Spin-Echo-Sequenz besteht aus zwei „harten", d.h. sehr kurzen, HF-Pulsen, deren Flipwinkel $\alpha = 90°$ und $\alpha = 180°$ betragen. Die HF-Pulse können verschiedene Phasenwinkel $\phi$ besitzen. Die Wirkung derartiger Pulse auf die Spinmagnetisierung kann z. B. über Gleichung (3.19) ermittelt werden. Zwischen den beiden HF-Pulsen wirkt ein $x$-Gradientenmagnetfeld. Sein Zeitintegral ist halb so groß wie das des $x$-Gradienten nach dem zweiten HF-Puls. Während der konstanten Phase des $x$-Gradienten nach dem zweiten HF-Puls werden die Gesamtmagnetisierung des Objekts bzw. die in der Empfangsspule induzierte Spannung aufgezeichnet; dieser Vorgang wird *Datenakquisition* genannt. Darüber hinaus wirkt zwischen den beiden HF-Pulsen ein $y$-Gradient.

Für ein zweidimensionales Bild wird die Sequenz mehrfach wiederholt. Bei jeder Wiederholung wird der $y$-Gradient $G_y(t)$ verändert, so daß die durch ihn hervorgerufene Phasenänderung $\Delta\varphi$ nach Gleichung (3.17) bei jeder Wiederholung einen anderen Wert hat. Zwischen den Wiederholungen der Sequenz wird eine Pause eingefügt, die dafür sorgt, daß das Spin-Ensemble durch $T_1$-Relaxation wieder sein thermisches Gleichgewicht einnimmt.

Zur Abbildung von Sequenzen auf Software-technische Datenstrukturen wird bei kommerziellen MR-Tomographen meist eine Ereignisliste eingesetzt, bei der die Änderungen der Ansteuerung eines der beteiligten Magnetfelder als Ereignisse aufgefaßt werden. Wegen der Parallelität der Ansteuerungs-Hardware für die Komponenten eines realen Imagers sind Ereignislisten eine effiziente und elegante Lösung. In einer Computer-Simulation müssen die Sequenzelemente jedoch nacheinander auf die Spins des Ensembles angewendet werden. Sinnvoller ist es hier, Zeit*intervalle* zu definieren, während denen sich an den beteiligten Magnetfeldern nichts ändert, also die Zeiträume zwischen einzelnen Ereignissen als Abbildungsgrundlage einzusetzen.

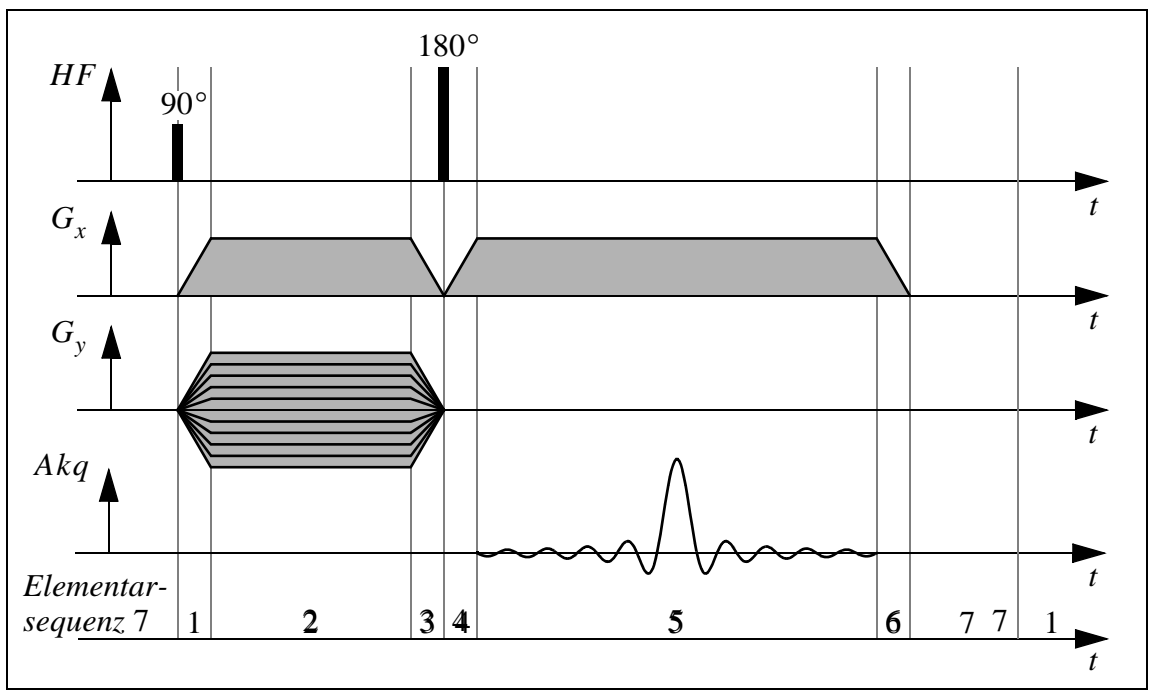

Bild 4.1: Timing-Diagramm einer zweidimensionalen Spin-Echo-Sequenz und Zerlegung der Sequenz in Elementarsequenzen

Die Sequenz aus Bild 4.1 läßt sich in sieben dort gekennzeichnete *Elementarsequenzen* zerlegen, während denen das Gradientenmagnetfeld stetig ist und sich seine funktionale Zeitabhängigkeit nicht ändert. Jeder Elementarsequenz wird ein harter HF-Puls zugeordnet, der am Beginn der Elementarsequenz wirkt. Die siebte Elementarsequenz beschreibt die Wartezeit zwischen zwei Wiederholungen der Sequenz. Anhand dieses Beispiels wird die Definition einer Elementarsequenz deutlich:

- Eine Elementarsequenz ist der allgemeinste Baustein, in den eine Sequenz zerlegt werden kann.

- Eine Elementarsequenz besitzt eine Länge $\Delta t \geq 0$.

- Eine Elementarsequenz besteht immer aus

  - einem harten HF-Puls am Anfang der Elementarsequenz und

  - einem beliebigen Gradientenmagnetfeld, dessen funktionale Abhängigkeit von der Zeit sich während der Dauer $\Delta t$ nicht ändert.

- Während einer Elementarsequenz ist es möglich, Daten zu akquirieren.

- HF- und Gradientenmagnetfelder dürfen in einer Elementarsequenz mit Null identisch sein.

Mathematisch kann die Wirkung einer Elementarsequenz auf die Spinmagnetisierung mit Hilfe von Matrixmultiplikationen modelliert werden. Zur Vereinfachung ist es sinnvoll, im mit $\omega_{\text{HF}}$ rotierenden Bezugssystem zu operieren. Die Gesamtwirkung einer Elementarsequenz der Länge $\Delta t = t - t_0$ auf die Spinmagnetisierung an einem Ort $\vec{x}$ ergibt sich aus den Gleichungen (3.8), (3.15) und (3.20) zu

$$\vec{M}(\vec{x}, t) = \boldsymbol{T}_{\text{Relax}} \cdot \boldsymbol{T}_{\text{grad}} \cdot \boldsymbol{T'}_{\text{HF}} \cdot \vec{M}(\vec{x}, t_0) + \vec{T}_{\text{Relax}, z}. \tag{4.1}$$

Der Vergleich von Gleichung (4.1) mit Gleichung (3.23) zeigt, daß sich Elementarsequenzen darüber hinaus zur Modellierung von HF-Pulsen mit zeitveränderlicher komplexer Hüllkurve eignen. Elementarsequenzen sind daher, auch unabhängig vom K-t-Formalismus, für das in dieser Arbeit beschriebene Simulationswerkzeug von universeller Bedeutung.

## 4.2 Mathematische Formulierung

Mathematische Grundlage des K-t-Formalismus' ist das Zusammenfassen der transversalen Magnetisierungskomponenten $M_x$ und $M_y$ zu einer komplexen Transversalmagnetisierung $\underline{M}_{xy} = M_x + \text{j} \cdot M_y$. Für diesen Fall lauten die Rotationsmatrizen für den Vektor $\underline{\vec{M}} = (\underline{M}_{xy}, \underline{M}^*_{xy}, M_z)^T$, der sogenannten *komplexen Magnetisierung*, für harte HF-Pulse

$$\boldsymbol{T}'_{\text{HF}} = \begin{pmatrix} \cos^2\dfrac{\alpha}{2} & \sin^2\dfrac{\alpha}{2}\,\exp(j2\phi) & \sin\alpha\,\exp\!\left[j\!\left(\phi+\dfrac{\pi}{2}\right)\right] \\[2ex] \sin^2\dfrac{\alpha}{2}\,\exp(-j2\phi) & \cos^2\dfrac{\alpha}{2} & \sin\alpha\,\exp\!\left[-j\!\left(\phi+\dfrac{\pi}{2}\right)\right] \\[2ex] \dfrac{\sin\alpha}{2}\,\exp\!\left[-j\!\left(\phi-\dfrac{\pi}{2}\right)\right] & \dfrac{\sin\alpha}{2}\,\exp\!\left[j\!\left(\phi-\dfrac{\pi}{2}\right)\right] & \cos\alpha \end{pmatrix} \quad (4.2)$$

und für zeitabhängige Gradienten

$$\boldsymbol{T}_{\text{grad}} = \begin{pmatrix} \exp(-j\Delta\vec{k}\cdot\vec{x}) & 0 & 0 \\ 0 & \exp(j\Delta\vec{k}\cdot\vec{x}) & 0 \\ 0 & 0 & 1 \end{pmatrix}. \quad (4.3)$$

Dabei ist

$$\Delta\vec{k} = \gamma\int\limits_{t_0}^{t}\vec{G}(\tau)\,d\tau, \quad (4.4)$$

und die Matrizen und Vektoren aus Gleichung (3.15) zur Berücksichtigung von Relaxation, $\boldsymbol{T}_{\text{Relax}}$ und $\vec{T}_{\text{Relax},\,z}$, bleiben unverändert. Eine Elementarsequenz ist damit für ein homogenes Hauptmagnetfeld und eine der Larmorfrequenz entsprechenden Kreisfrequenz des rotierenden Bezugssystems analog zu Gleichung (4.1) vollständig beschreibbar durch die Matrixgleichung

$$\vec{\underline{M}}(\vec{x},t) = \boldsymbol{T}_{\text{Relax}}\cdot\boldsymbol{T}_{\text{grad}}\cdot\boldsymbol{T}'_{\text{HF}}\cdot\vec{\underline{M}}(\vec{x},t_0) + \vec{T}_{\text{Relax},\,z}. \quad (4.5)$$

## 4.2.1   Zerlegung der Magnetisierung in komplexe Konfigurationen

Betrachtet werde eine einfache Sequenz im eindimensionalen Fall, siehe Bild 4.2. Sie besteht aus zwei sehr kurzen HF-Pulsen in einem zeitlichen Abstand $\Delta t$. Ein zeitlich konstantes Gradientenmagnetfeld mit dem Gradienten $G_x$ wird so geschaltet, daß sein Integral über der Zeit nach den HF-Pulsen proportional zu einer Konstanten $\Delta k_x$ bzw. $2\Delta k_x$ ist.

Der komplexe Magnetisierungvektor läßt sich in Abhängigkeit von der Zeit mit Gleichung (4.5) berechnen. Unter Vernachlässigung von Relaxation ergibt sich beispielsweise zur Zeit $t = 2\Delta t$ die komplexe Magnetisierung zu

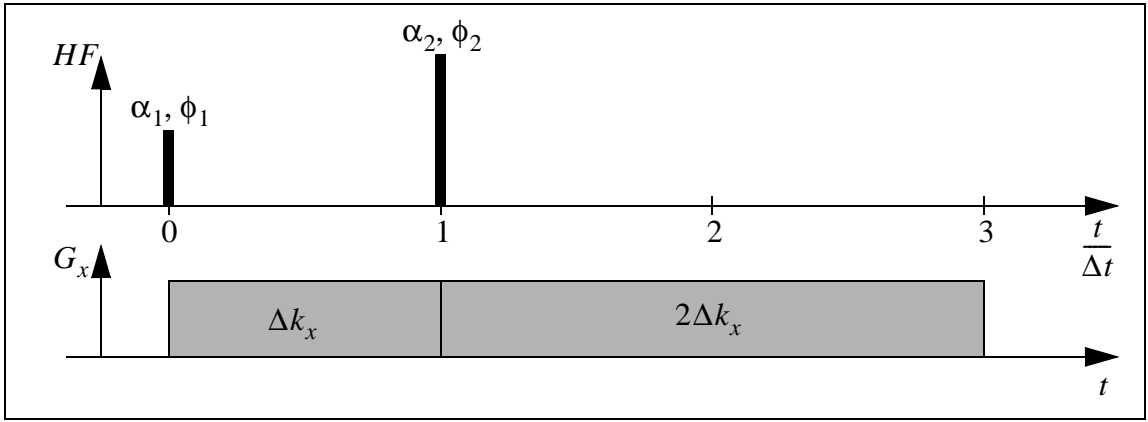

Bild 4.2: Timing-Diagramm einer eindimensionalen Sequenz mit zwei Elementarsequenzen.

$$\frac{\vec{\underline{M}}(\check{x}, 2\Delta t)}{M_0(\check{x})} =$$

$$\begin{pmatrix} -\mathrm{j} \cdot \sin\alpha_1 \sin^2\frac{\alpha_2}{2}\exp 0 + \mathrm{j} \cdot \cos\alpha_1 \sin\alpha_2 \exp(-1 \cdot 2\pi\mathrm{j} \cdot \Delta k_x \cdot x) + \mathrm{j} \cdot \sin\alpha_1 \cos^2\frac{\alpha_2}{2}\exp(-2 \cdot 2\pi\mathrm{j} \cdot \Delta k_x \cdot x) \\ \mathrm{j} \cdot \sin\alpha_1 \sin^2\frac{\alpha_2}{2}\exp 0 - \mathrm{j} \cdot \cos\alpha_1 \sin\alpha_2 \exp(1 \cdot 2\pi\mathrm{j} \cdot \Delta k_x \cdot x) - \mathrm{j} \cdot \sin\alpha_1 \cos^2\frac{\alpha_2}{2}\exp(2 \cdot 2\pi\mathrm{j} \cdot \Delta k_x \cdot x) \\ \cos\alpha_1 \cos\alpha_2 \exp 0 - \frac{1}{2}\sin\alpha_1 \sin\alpha_2 \exp(1 \cdot 2\pi\mathrm{j} \cdot \Delta k_x \cdot x) - \frac{1}{2}\sin\alpha_1 \sin\alpha_2 \exp(-1 \cdot 2\pi\mathrm{j} \cdot \Delta k_x \cdot x) \end{pmatrix}.$$

$$(4.6)$$

Es ist zu erkennen, daß die Komponenten des Magnetisierungsvektors aus Summen von Termen bestehen, die sich aus einem komplexen Gewicht und einem Exponentialterm zusammensetzen. Das Gewicht ist dabei ausschließlich von den Parametern der HF-Pulse abhängig. Das Argument der Exponentialterme ist jeweils ein ganzzahliges Vielfaches der Konstanten $2\pi\mathrm{j} \cdot \Delta k_x \cdot x$ und nur von dem Gradientenmagnetfeld abhängig.

Verallgemeinernd lassen sich die Komponenten der komplexen Magnetisierung nach einer aus mehreren Elementarsequenzen bestehenden Sequenz als komplexe Reihen der Form

$$\left. \begin{aligned} \underline{M}_{xy} &= \underline{F}_0 + \sum_{i=1}^{N} \underline{F}_i + \underline{F}_{-i} \\ \underline{M}^*{}_{xy} &= \underline{F}^*{}_0 + \sum_{i=1}^{N} \underline{F}^*{}_{-i} + \underline{F}^*{}_i \\ M_z &= \underline{Z}_0 + \sum_{i=1}^{N} \underline{Z}_i + \underline{Z}_{-i} \end{aligned} \right\} \qquad (4.7)$$

schreiben. Die Summanden $\underline{F}_{\pm i}$ und $\underline{Z}_{\pm i}$ sind im eindimensionalen Fall gegeben durch

$$\underline{F}_i = \underline{a}_i \cdot \exp(-2\pi\mathrm{j} \cdot i\Delta k_x x), \ \ \forall i \tag{4.8}$$

und

$$\underline{Z}_i = \begin{cases} \underline{b}_i \cdot \exp(-2\pi\mathrm{j} \cdot i\Delta k_x x), & (\forall i \neq 0) \\ \underline{b}_i + M_0 \cdot [1 - \exp(-\tau/T_1)], & (i = 0) \end{cases} \tag{4.9}$$

mit $i$ ganzzahlig.

Im Kontext des K-t-Formalismus' werden die Summanden $\underline{F}_i$ und $\underline{Z}_i$ als Transversal- bzw. Longitudinal*konfigurationen* bezeichnet; die Koeffizienten $\underline{a}_i$ und $\underline{b}_i$ heißen die *Population* der Konfiguration mit der *Ordnung* $i$. Die Populationen hängen von den Relaxationseigenschaften des Objekts und von den Flip- und Phasenwinkeln der HF-Pulse der Elementarsequenzen ab, während Gradientenmagnetfelder ausschließlich die Ordnung der Transversalkonfigurationen beeinflussen.

Im zwei- und dreidimensionalen Fall wird die Ordnung für jede der Raumrichtungen getrennt definiert. Die Konfigurationen erhalten dazu für jede zu berücksichtigende Richtung einen zusätzlichen Index. Im dreidimensionalen Fall entstehen so beispielsweise die Konfigurationen

$$\underline{F}_{i,m,n} = \underline{a}_{i,m,n} \cdot \exp(-2\pi\mathrm{j} \cdot i\Delta k_x x - 2\pi\mathrm{j} \cdot m\Delta k_y y - 2\pi\mathrm{j} \cdot n\Delta k_z z), \ \ \forall(i,m,n) \tag{4.10}$$

und

$$\underline{Z}_{i,m,n} = \begin{cases} \underline{b}_{i,m,n} \cdot \exp(-2\pi\mathrm{j} \cdot i\Delta k_x x - 2\pi\mathrm{j} \cdot m\Delta k_y y - 2\pi\mathrm{j} \cdot n\Delta k_z z), & \forall(i,m,n) \neq (0,0,0) \\ \underline{b}_{i,m,n} + M_0 \cdot [1 - \exp(-\tau/T_1)], & (i,m,n) = (0,0,0). \end{cases}$$
$$\tag{4.11}$$

Die Indizes $m$ und $n$ werden analog zu $i$ für jede Raumrichtung getrennt ermittelt. Im folgenden wird auf die Diskussion höherer Dimensionen verzichtet.

## 4.2.2   Eigenschaften komplexer Konfigurationen

Wird zunächst die Wirkung von HF-Pulsen betrachtet, so gilt für die komplexe Magnetisierung $\vec{\underline{M}}(t)$ nach dem HF-Puls in Abhängigkeit von der komplexen Magnetisierung $\vec{\underline{M}}(t_0)$ vor dem Puls

$$\vec{\underline{M}}(t) = \underline{T}'_{\mathrm{HF}} \cdot \vec{\underline{M}}(t_0). \tag{4.12}$$

Aus Gleichung (4.7) und (4.12) ergibt sich

$$
\begin{pmatrix} \underline{E}^+{}_0 \\ \underline{E}^{*+}{}_0 \\ \underline{Z}^+{}_0 \end{pmatrix} + \sum_{i=1}^{N} \begin{pmatrix} \underline{E}^+{}_i \\ \underline{E}^{*+}{}_{-i} \\ \underline{Z}^+{}_i \end{pmatrix} + \sum_{i=1}^{N} \begin{pmatrix} \underline{E}^+{}_{-i} \\ \underline{E}^{*+}{}_i \\ \underline{Z}^+{}_{-i} \end{pmatrix} = \boldsymbol{T}'_{\mathrm{HF}} \cdot \begin{pmatrix} \underline{E}^-{}_0 \\ \underline{E}^{*-}{}_0 \\ \underline{Z}^-{}_0 \end{pmatrix} + \sum_{i=1}^{N} \boldsymbol{T}'_{\mathrm{HF}} \cdot \begin{pmatrix} \underline{E}^-{}_i \\ \underline{E}^{*-}{}_{-i} \\ \underline{Z}^-{}_i \end{pmatrix} +
$$

$$
\sum_{i=1}^{N} \boldsymbol{T}'_{\mathrm{HF}} \cdot \begin{pmatrix} \underline{E}^-{}_{-i} \\ \underline{E}^{*-}{}_i \\ \underline{Z}^-{}_{-i} \end{pmatrix}. \tag{4.13}
$$

Gleichung (4.12) gilt somit in formaler Übertragung auch für Vektoren von Konfigurationen der Ordnung $i$ bzw. deren Populationen, also

$$
\begin{pmatrix} \underline{E}^+{}_i \\ \underline{E}^{*+}{}_{-i} \\ \underline{Z}^+{}_i \end{pmatrix} = \boldsymbol{T}'_{\mathrm{HF}} \cdot \begin{pmatrix} \underline{E}^-{}_i \\ \underline{E}^{*-}{}_{-i} \\ \underline{Z}^-{}_i \end{pmatrix} \quad \text{und} \quad \begin{pmatrix} \underline{a}^+{}_i \\ \underline{a}^{*+}{}_{-i} \\ \underline{b}^+{}_i \end{pmatrix} = \boldsymbol{T}'_{\mathrm{HF}} \cdot \begin{pmatrix} \underline{a}^-{}_i \\ \underline{a}^{*-}{}_{-i} \\ \underline{b}^-{}_i \end{pmatrix} \tag{4.14}
$$

für beliebige ganzzahlige $i$. HF-Pulse verursachen, abhängig von ihrem Flip- und Phasenwinkel, einen gewichteten Austausch der Populationen $\underline{a}_i$ und $\underline{b}_i$ zwischen Konfigurationen[1] mit betraglich gleicher Ordnung $|i|$, verändern aber die Ordnung nicht.

Relaxationseffekte beeinflussen die Populationen von sowohl longitudinalen als auch transversalen Konfigurationen und verändern die Ordnung nicht. Analog zu Gleichung (4.14) gilt

$$
\begin{pmatrix} \underline{E}^+{}_i \\ \underline{E}^{*+}{}_{-i} \\ \underline{Z}^+{}_i \end{pmatrix} = \boldsymbol{T}_{\mathrm{Relax}} \cdot \begin{pmatrix} \underline{E}^-{}_i \\ \underline{E}^{*-}{}_{-i} \\ \underline{Z}^-{}_i \end{pmatrix} \quad \text{und} \quad \begin{pmatrix} \underline{a}^+{}_i \\ \underline{a}^{*+}{}_{-i} \\ \underline{b}^+{}_i \end{pmatrix} = \boldsymbol{T}_{\mathrm{Relax}} \cdot \begin{pmatrix} \underline{a}^-{}_i \\ \underline{a}^{*-}{}_{-i} \\ \underline{b}^-{}_i \end{pmatrix} \tag{4.15}
$$

für $i \neq 0$. Die Konfigurationen nullter Ordnung bilden wegen des additiven Vektors $\vec{T}_{\mathrm{Relax},\,z}$ aus Gleichung (4.5), der sich nur auf die longitudinale Konfiguration $\underline{Z}_0$ auswirkt, eine Ausnahme; für sie gilt

---

[1] Hennig läßt in [41] und [42] die Phasen von HF-Pulsen unberücksichtigt bzw. regt eine Erweiterung der Anzahl von Konfigurationstypen an, um beliebige HF-Phasen unterstützen zu können. Dies wird mit der hier verwendeten komplexwertigen HF-Transformationsmatrix $\boldsymbol{T}'_{\mathrm{HF}}$ unnötig.

$$
\begin{pmatrix} \underline{E}^+{}_0 \\ \underline{E}^{*+}{}_0 \\ \underline{Z}^+{}_0 \end{pmatrix} = \boldsymbol{T}_{\text{Relax}} \cdot \begin{pmatrix} \underline{E}^-{}_0 \\ \underline{E}^{*-}{}_0 \\ \underline{Z}^-{}_0 \end{pmatrix} + \vec{T}_{\text{Relax},\,z} \,,
\tag{4.16}
$$

wobei auch hier die Konfigurationen $\underline{E}$ und $\underline{Z}$ durch ihre Populationen $\underline{a}$ und $\underline{b}$ ersetzt werden können.

Gradientenmagnetfelder führen nach Gleichung (4.3) zu einer Phasenänderung der transversalen Konfigurationen nach Gleichung (4.8), lassen aber longitudinale Konfigurationen unbeeinflußt. Die Phasenänderung ist äquivalent zu einer Änderung der Ordnung einer Konfiguration. Wird ein Gradientenmagnetfeld mit dem Zeitintegral $q \cdot \Delta k_x$ eingesetzt, so ist die diesen Vorgang beschreibende Rotationsmatrix ($q$ ganzzahlig)

$$
\boldsymbol{T}_{\text{grad}}^q = \begin{pmatrix} \exp(-2\pi\mathrm{j} \cdot q \cdot \Delta k_x \cdot x) & 0 & 0 \\ 0 & \exp(2\pi\mathrm{j} \cdot q \cdot \Delta k_x \cdot x) & 0 \\ 0 & 0 & 1 \end{pmatrix};
\tag{4.17}
$$

es kommt zu den Übergängen $\underline{E}_{\pm i} \to \underline{E}_{\pm\,i+q}$ für beliebige ganzzahlige $i \geq 0$. Gradientenmagnetfelder verschieben die Ordnung aller transversalen Konfigurationen asymmetrisch in eine Richtung. Transversale Konfigurationen mit betraglich gleicher Ordnung $\underline{E}_i$ und $\underline{E}_{-i}$ besitzen daher i. a. verschiedene Populationen. Longitudinale Konfigurationen mit betraglich gleicher Ordnung $\underline{Z}_i$ und $\underline{Z}_{-i}$ ($i \neq 0$) tragen hingegen immer vorzeichenverschiedene, aber ansonsten gleiche Populationen. Konfigurationen haben somit folgende Eigenschaften:

$$
\begin{aligned}
\underline{E}_0 &= \underline{E}_0^* \\
\underline{E}_i &\neq \underline{E}_{-i}, \ (\forall i \neq 0) \\
\underline{Z}_i &= -\underline{Z}_{-i}, \ (\forall i \neq 0).
\end{aligned}
$$

Hieraus kann geschlossen werden, daß stets $M_z = \underline{Z}_0$ gilt und somit $\underline{Z}_0$ und $\underline{b}_0$ reell sein müssen. Die sich bei der Longitudinalmagnetisierung gegenseitig aufhebenden Konfigurationen $\underline{Z}_{\pm i}$ sind von praktischer Bedeutung, da sie die als *Phasengedächtnis der Longitudinalmagnetisierung* genannte Eigenschaft von Spin-Ensembles beschreiben.

Transversalmagnetisierung mit einer Position $k_x$ (oder auch eine Transversalkonfiguration mit der Ordnung $i$) kann mittels eines HF-Pulses als Longitudinalmagnetisierung gespeichert werden. Durch das Phasengedächtnis kann aus dieser Longitudinalmagnetisierung durch einen weiteren HF-Puls wieder Transversalmagnetisierung erzeugt werden, die die gleiche Position $k_x$ innehat wie die zuvor gespeicherte transversale Ausgangsmagnetisierung. Ein aus einer sol-chen Transversalmagnetisierung erzeugtes Echo ist in der Literatur unter dem Namen *Stimu-liertes Echo* geläufig und wird z. B. in STEAM-Sequenzen genutzt [83].

### 4.2.3  Zeitlich veränderliche Gradientenmagnetfelder

Bisher wurde angenommen, daß die eingesetzten Gradientenmagnetfelder während einer Ele-mentarsequenz zeitlich konstant sind. Damit erschließen sich die Konfigurationsordnungen $i$ auf einfache Art und Weise. Tatsächlich ändern die Konfigurationen ihre Position $k_x(t)$ jedoch kontinuierlich, im Falle von rechteckförmigen Gradienten linear mit der Zeit. Transver-sale und longitudinale Konfigurationen halten jeweils untereinander den minimalen Abstand $\Delta k_x$ ein. Die Positionen $k_x(t)$ erreichen zu den Zeiten der HF-Pulse ganzzahlige Vielfache von $\Delta k_x$.

Bei Gradientenmagnetfeldern mit beliebigen Zeitfunktionen ergibt sich die K-Raum-Position während einer Elementarsequenz zu

$$k_x(t) = k_x(t_0) + \gamma \cdot \int_{t_0}^{t} G_x(\tau) \, \mathrm{d}\tau, \tag{4.18}$$

wobei $t_0$ der Zeitpunkt direkt nach dem harten HF-Puls der Elementarsequenz ist. Zur Festle-gung der Konfigurationsordnung ist auch hier eine Größe $\Delta k_x$ erforderlich. Zu ihrer Bestim-mung werden zunächst die Gradienten-Zeitintegrale $\Delta k_{x,l}$ aller Elementarsequenzen $l$ ermit-telt. Nachfolgend läßt sich eine Bezugsgröße $k_{\mathrm{Bezug}}$ so finden, daß alle $\Delta k_{x,l}$ ganzzahlige Vielfache der Bezugsgröße sind. Die so bestimmte Bezugsgröße eignet sich als von $l$ unab-hängige Einheit $\Delta k_x$ zur Ermittlung der Konfigurationsordnung $i$.

# 4.3  K-t-Diagramme

Die zeitliche Entwicklung aller Konfigurationen kann anschaulich in *K-t-Diagrammen* visuali-siert werden. Dazu wird die Position $k_x(t)$ aller Konfigurationen über der Zeit aufgetragen.

### 4.3.1  Qualitative K-t-Diagramme

Für *qualitative K-t-Diagramme* wird angenommen, daß Flip- und Phasenwinkel der HF-Pulse einer MR-Sequenz beliebig sind. Sie zeigen alle potentiellen Trajektorien von Konfigurationen in der K-t-Ebene, visualisieren aber nicht Betrag oder Phase der Konfigurationspopulationen. Das qualitative K-t-Diagramm zu einer einfachen eindimensionalen Sequenz nach Bild 4.2 ist in Bild 4.3 für den Fall gezeigt, daß sich vor Beginn der Sequenz das Spinsystem im thermi-schen Gleichgewicht befindet, also nur eine komplexe Magnetisierung in *z*-Richtung mit $\vec{\underline{M}} = (0, 0, M_0)^T$ vorliegt. Die Größe $\Delta k_x$ stellt in diesem Diagramm den minimalen Abstand zweier transversaler bzw. zweier longitudinaler Konfigurationen dar.

Am zweiten HF-Puls ist gut zu erkennen, daß jede vor einem HF-Puls vorhandene Konfiguration mit der Ordnung $i$ in bis zu vier neue Konfigurationen $\underline{E}_{\pm i}$ und $\underline{Z}_{\pm i}$ aufgesplittet wird. Weniger neue Konfigurationen entstehen für $i = 0$, wo die Konfigurationen $\underline{E}_{\pm 0}$ sowie $\underline{Z}_{\pm 0}$ jeweils identisch sind. Überlagern sich vor einem HF-Puls einzelne Konfigurationen, so überlagern sich nach ihm die gesplitteten Konfigurationen; wegen *Konfigurationsinterferenzen* entstehen in diesem Fall weniger als vier neue Konfigurationen pro Eingangskonfiguration.

Bei nicht rechteckförmiger Gradientenzeitfunktion verlaufen die Trajektorien im K-t-Diagramm nicht mehr linear, sondern nach Gleichung (4.18) abhängig vom Integral über $G_x(t)$.

### 4.3.2   Quantitative K-t-Diagramme

Qualitative K-t-Diagramme zeigen unabhängig von Flip- und Phasenwinkeln der verwendeten HF-Pulse und unabhängig von Relaxationseffekten alle potentiell möglichen Konfigurationen. Für die Untersuchung einer Sequenz ist es jedoch oft auch wichtig, Aussagen über den Grad des Einflusses einzelner Konfigurationen auf die entstehenden Echos treffen zu können.

Bei bekannten Parametern der HF-Pulse und bekannten Relaxationseigenschaften können mit den Gleichungen (4.14) bis (4.17) die Populationen aller Konfigurationen und die Bewegung der Konfigurationen durch den K-t-Raum berechnet werden. Werden in ein K-t-Diagramm, z. B. durch Farbkodierung, die Größen der Populationen (Betrag, Phase, Real- oder Imaginärteil), integriert, entsteht ein *quantitatives K-t-Diagramm*. Ein Beispiel für ein quantitatives K-t-Diagramm ist in Bild 4.4 für die Sequenz nach Bild 4.2 gezeigt. Zu sehen ist neben den aus Bild 4.3 bereits bekannten Trajektorien der transversalen und longitudinalen Konfigurationen der Betrag ihrer jeweiligen Populationen in farbkodierter Form.

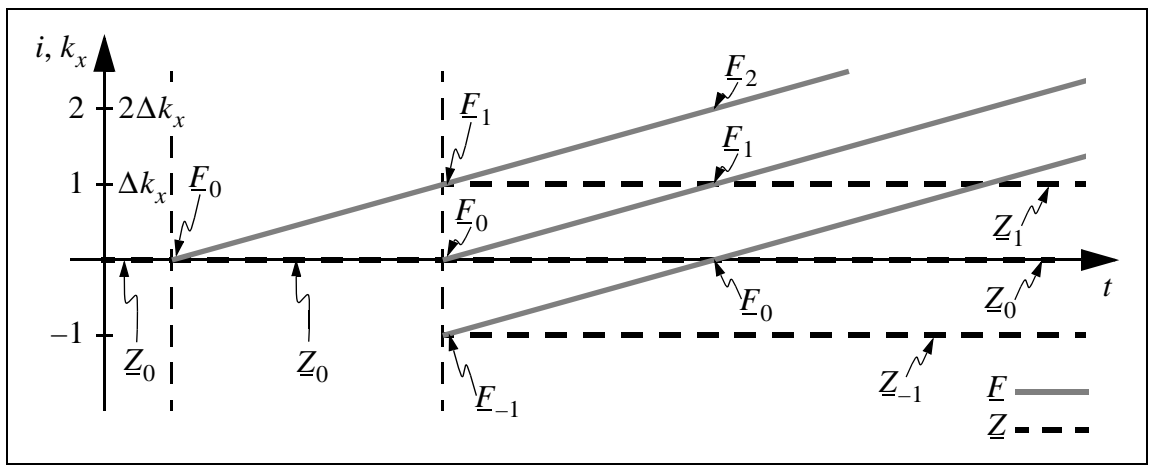

Bild 4.3: K-t-Diagramm zu der Sequenz aus Bild 4.2

Longitudinale Konfigurationen mit Ordnungen ungleich Null und transversale Konfigurationen zerfallen mit den Relaxationskonstanten $T_1$ und $T_2$; dies ist sehr gut an der kontinuierlichen Farbänderung erkennbar. Longitudinale Konfigurationen nullter Ordnung streben das thermische Gleichgewicht mit der Ruhemagnetisierung $M_0$ an, auch dies kann anhand des Diagramms gut beobachtet werden.

Sprunghafte Farbänderungen resultieren aus dem Splitting von Konfigurationen durch HF-Pulse. Sehr gut kann das Splitting beobachtet werden, das die vor dem zweiten HF-Puls vorhandene transversale Konfiguration zu gleichen Teilen in neue transversale und longitudinale Konfiguration zerlegt. Gut zu sehen ist außerdem, daß die *Longitudinal*magnetisierung nullter Ordnung, die durch $T_1$-Relaxation während des Zeitintervalls zwischen den beiden HF-Pulsen entstanden ist, durch den zweiten HF-Puls in eine neue *Transversal*konfiguration nullter Ordnung umgewandelt wird.

### 4.3.3   K-t-Formalismus und K-Raum-Konzept

Die Gesamtmagnetisierung eines Objekts ergibt sich durch Integration der Gleichungen (4.7) über die Ausdehnung des Objekts. Für große Werte von $\Delta k_x$ tragen die transversalen Konfigurationen $\underline{F}_i$ nach Gleichung (4.8) nicht mehr zur meßbaren Transversalmagnetisierung $\underline{M}_{xy}$ aus Gleichung (4.7) bei; relevant für die meßbare Magnetisierung ist in diesem Fall nur die Transversalkonfiguration nullter Ordnung. MR-Signale sind demnach dann meßbar, wenn sich die Trajektorie einer Transversalkonfiguration und die Zeitachse schneiden. Aus einem qualitativen K-t-Diagramm lassen sich auf diese Weise sehr rasch alle potentiellen Zeitpunkte ablesen, an denen Maxima der Echosignale erwartet werden können.

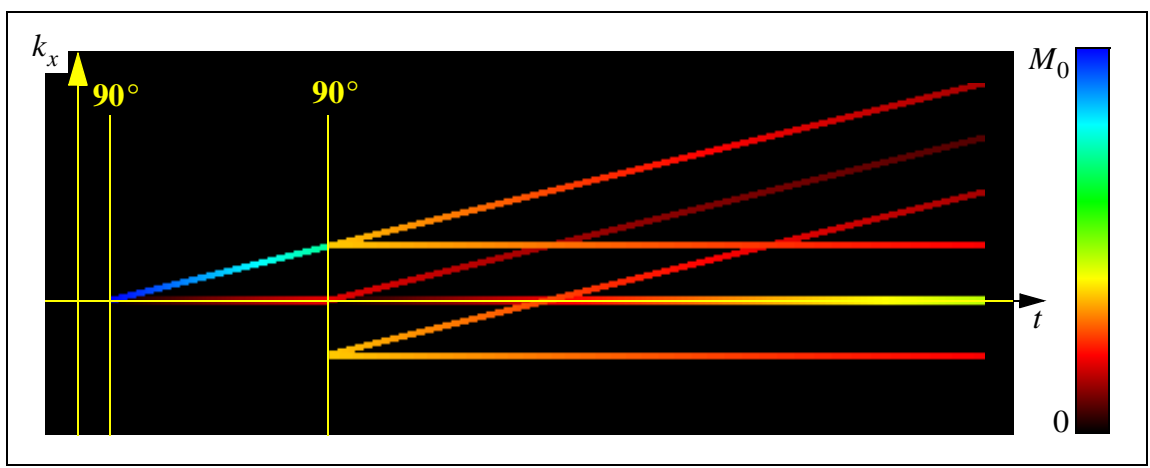

Bild 4.4: Quantitatives K-t-Diagramm zu der Sequenz aus Bild 4.2 für die Parameter $\alpha_1 = \alpha_2 = 90°$, $\phi_1 = \phi_2 = 0°$. Die Länge des dargestellten Bereichs entspricht 400 ms bei verwendeten Relaxationszeiten von $T_1 = 500$ ms und $T_2 = 300$ ms.

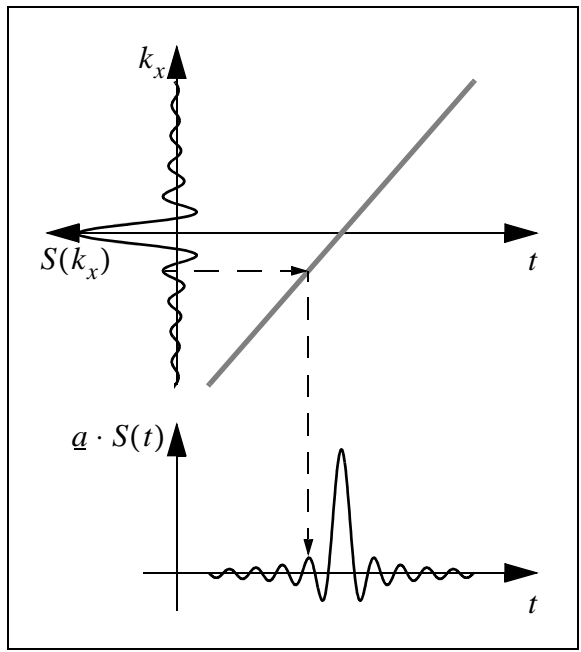 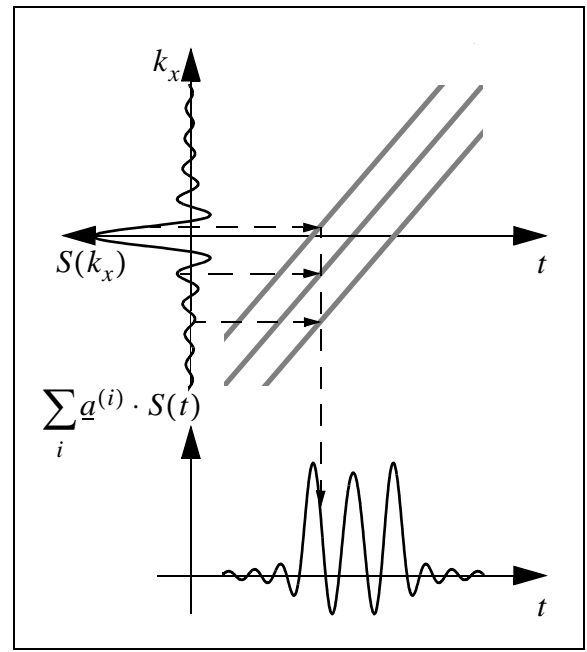

Bild 4.5: Existiert nur eine transversale Kon-
figuration mit Nulldurchgang bzw. ist der
Abstand $\Delta k_x$ zu benachbarten Konfiguratio-
nen groß, ergibt sich das Echosignal aus der
mit der Population $\underline{a}$ gewichteten Ortsfre-
quenzfunktion des Objekts.

Bild 4.6: Bei mehreren benachbarten trans-
versalen Konfigurationen mit Nulldurch-
gang ($\Delta k_x$ ist klein) ergibt sich das
Echosignal aus der mit der Summe der
Populationen gewichteten Ortsfrequenz-
funktion des Objekts.

K-t-Diagramme haben einen sehr engen Bezug zu dem in der Analyse von MR-Sequenzen
weitverbreiteten, aber meist nur für den zweidimensionalen Fall eingesetzten K-Raum-Kon-
zept [60] [98], vgl. Abschnitt 3.6.

Die Kombination des K-t-Formalismus mit dem K-Raum-Konzept läßt eine genauere Analyse
der Echogenerierung zu. Die transversalen Konfigurationen können als punktförmige Sonden
aufgefaßt werden, die die unendlich ausgedehnte Ortsfrequenzfunktion des Objekts entspre-
chend ihrer Position $k_x$ abtasten und entsprechend ihrer Population gewichten. Das gesamte
transversale MR-Signal ergibt sich aus der Summe über alle transversalen Konfigurationen,
jeweils gewichtet mit der Ortsfrequenzfunktion des Objekts an ihrer Position.

Ist die Distanz zwischen transversalen Konfigurationen groß, hat nur die Konfiguration in der
Nähe des K-Raum-Ursprungs einen Einfluß auf das Echo, da der Betrag der Ortsfrequenzfunk-
tion des Objekts an der Position der benachbarten Konfigurationen nahe oder gleich Null ist,
siehe Bild 4.5. Bei mehr als einer Konfiguration in der Nähe des K-Raum-Ursprungs haben
alle Konfigurationen einen Einfluß auf das Echo, siehe Bild 4.6. Solche aus mehreren Konfi-
gurationen gebildeten Echos treten in der Praxis meist unbeabsichtigt auf und sind schwierig zu handhaben, da die in ihnen enthaltene Information nicht in ihre Bestandteile separiert werden kann.

Ein Spezialfall von Mehrfachechos sind Hauptmagnetfeldinhomogenitäten, die man auch als örtlich nichtlineare, zeitlich konstante Gradientenmagnetfelder auffassen kann. In diesem Fall wird die Größe $\Delta \vec{k}$ aus Gleichung (4.4) ortsabhängig, und man erhält im qualitativen K-t-Diagramm eine Kurvenschar, in Bild 4.7 gezeigt am Beispiel einer Spin-Echo-Sequenz, in Bild 4.8 gezeigt an einer Gradientenechosequenz.

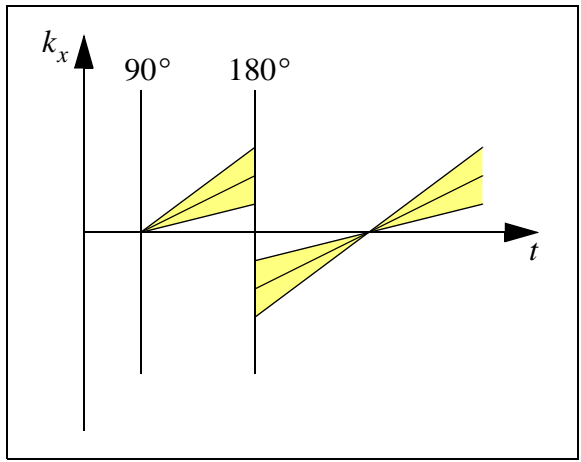
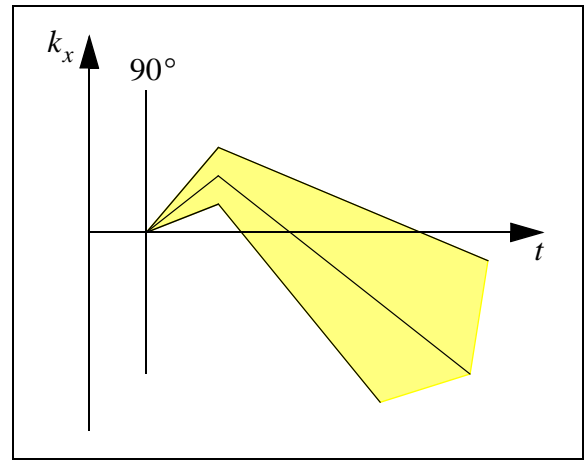

Bild 4.7: K-t-Diagramm einer eindimensionalen Spin-Echo-Sequenz unter dem Einfluß von Inhomogenitäten des Hauptmagnetfelds

Bild 4.8: K-t-Diagramm einer eindimensionalen Gradienten-Echo-Sequenz unter dem Einfluß von Inhomogenitäten des Hauptmagnetfelds

Es ist zu erkennen, daß sich bei einer „Spin-Echo-Sequenz" nach dem zweiten HF-Puls alle Konfigurationen in einem Punkt schneiden; das Maximum des entstehenden Echos kann somit genau lokalisiert werden. Im Gegensatz dazu kommt es bei der „Gradienten-Echo-Sequenz" zu einer Verbreiterung des Echomaximums, die zu einem verschlechterten Signal-Rausch-Abstand und zu Verzerrungen bei der Bildrekonstruktion führt.

Eine weitere, gleichwertige Interpretation der Kurvenschar ist die einer sich vergrößernden oder verkleinernden Konfiguration. Eine Konfiguration ist beim Vorhandensein von Inhomogenitäten i. a. im K-Raum nicht mehr punktförmig aufzufassen, sondern nimmt – im zweidimensionalen Fall – eine Fläche ein. Der Beitrag einer solchen Konfiguration zu einem Echo zu einer Zeit $t$ ergibt sich als gewichtetes Integral über die Ortsfrequenzfunktion innerhalb der Fläche der Konfiguration.

# 4.4    Anwendungen des K-t-Formalismus'

Der K-t-Formalismus stellt im Rahmen dieser Arbeit die Grundlagen für die korrekte Wahl der Simulationsparameter bereit, die die Abbildung von in der Natur kontinuierlichen Untersuchungsobjekten auf diskretisierte, für Digitalrechner geeignete Datenstrukturen spezifizieren. Die systemtheoretisch korrekte Diskretisierung kontinuierlicher Objekte wird in Abschnitt 5.2 behandelt.

Daneben macht die mathematische Formulierung des K-t-Formalismus' für homogene statische, Gradienten- und HF-Felder eine effiziente Simulation von MR-Bildgebungssequenzen möglich. Hierzu werden mit den Gleichungen (4.14) bis (4.17) Populationen und Positionen der Konfigurationen in ihrer zeitlichen Entwicklung verfolgt. Unabhängig von der Geometrie eines zu simulierenden Objekts, aber unter Berücksichtigung seiner Relaxationseigenschaften, können bereits aus den Konfigurationspopulationen Zeitpunkt des Auftretens sowie Amplitude und Phase der mit einer Sequenz erzeugten Echos ermittelt werden. Diese Informationen können als i. a. nicht-äquidistante Folge von komplex gewichteten diracförmigen Impulsen aufgefaßt werden.

Um den Wert des Echos zu einem bestimmten Zeitpunkt zu ermitteln, werden alle transversalen Konfigurationen im komplexen Raum mit dem Wert der Ortsfunktion des Objekts an der Stelle gewichtet, an der sich die Konfigurationen aktuell befinden. Diese gewichteten Werte werden danach superponiert. Im eindimensionalen Fall ergibt sich der Wert des Echos $\underline{s}(t)$ zu einer Zeit $t$ aus

$$\underline{s}(t) \;=\; \sum_{i=-\infty}^{\infty} \underline{a}_i \cdot \underline{S}[k_x(t,i)]\,; \qquad\qquad (4.19)$$

hierbei ist $\underline{S}[k_x(t,i)]$ der Wert der Ortsfrequenzfunktion des Objekts an der K-Raum-Position der Konfiguration mit der Population $\underline{a}_i$, vgl. Bild 4.6.

Der theoretische Zugang zu einer Reihe von ultraschnellen MR-Bildgebungsverfahren mit Multipuls-HF-Anregungen ist erst mit dem erweiterten Phasengraph-Algorithmus und dem K-t-Formalismus möglich geworden. Einige dieser Sequenzen beruhen auf dem Prinzip der exponentiellen Vergrößerung der Konfigurations- und damit der Echoanzahl [22] [39]. Hauptschwierigkeit bei der Bildgebung ist die stark unterschiedliche Population der echobildenden Konfigurationen. Dieser Umstand ist in Bild 4.9 an der stark unterschiedlichen Farbgebung der die $t$-Achse schneidenden Trajektorien nach dem letzten HF-Puls erkennbar.

Für die Analyse und Optimierung derartiger Multipuls-Sequenzen eignet sich der K-t-Formalismus besonders [10] [11] [55], da er die inhärent komplexen Vorgänge veranschaulichen kann. Die Betrachtung von ortsabhängigen Magnetisierungsvektoren ist hier für das Verständnis prinzipiell ungeeignet [41] [42].

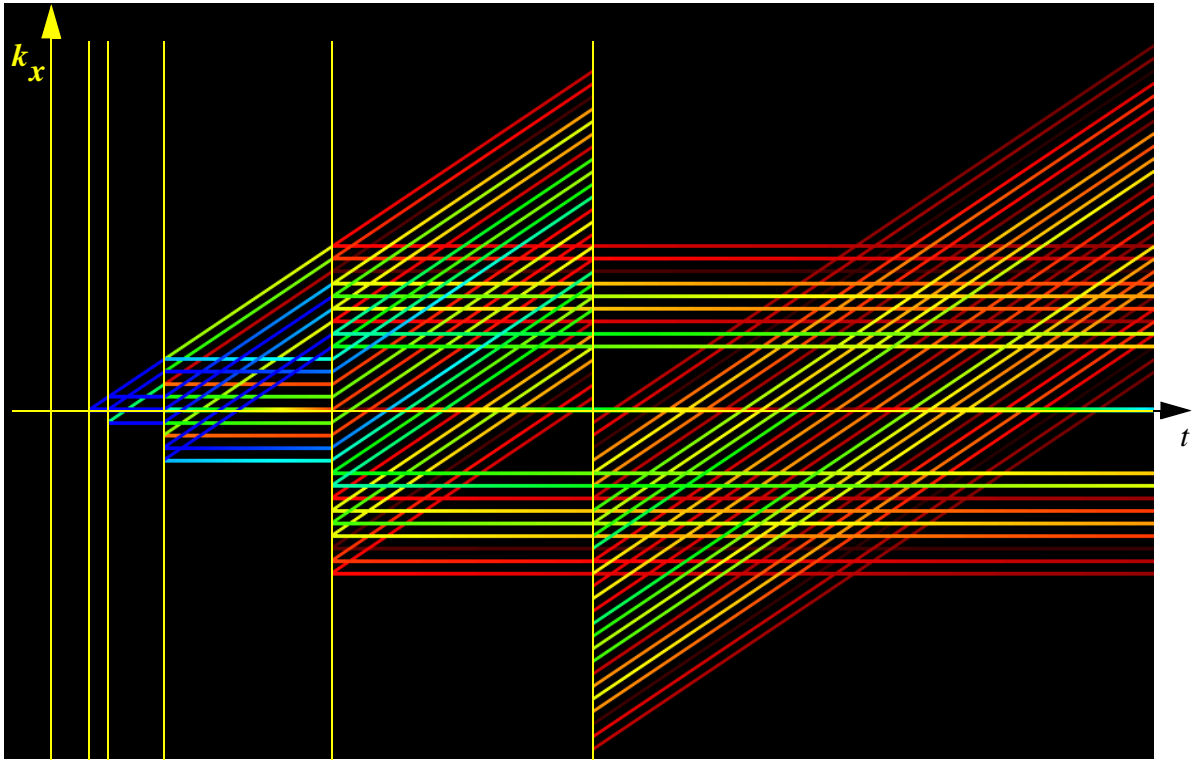

Bild 4.9: Quantitatives K-t-Diagramm einer 27 Echos generierenden PREVIEW-Sequenz nach [22]. Die Akquisition der 27 Echos erfolgt nach dem letzten HF-Puls.

# 5 Konzepte für einen Magnetresonanz-Simulator

Das vorliegende Kapitel beschäftigt sich mit den Konzepten zur Realisierung eines MR-Simulators, die für die Erfüllung der Aufgabenstellung in Frage kommen. Die Eigenschaften der möglichen Simulationsansätze werden erörtert, und die Ansätze werden hinsichtlich ihrer Eignung für diese Arbeit bewertet.

Einer der Ansätze, die spinbasierte Simulation, wird in vielen anderen Arbeiten eingesetzt. Nur in einer bekannten Veröffentlichung [5] wird jedoch die Problematik der systemtheoretisch korrekten Ortsdiskretisierung für einen einfachen Fall behandelt. In einem weiteren Abschnitt dieses Kapitels wird eine im Rahmen dieser Arbeit entwickelte Vorschrift vorgestellt, die präzise Zahlenwerte für den Höchstabstand zwischen punktförmigen Spins in Abhängigkeit von der eingesetzten Bildgebungssequenz liefert.

Anschließend wird gezeigt, wie ein auf quasi-freien Spins basierender Simulator durch eine Parallelisierung in seinem Laufzeitverhalten drastisch verbessert und hinsichtlich der Komplexität der Simulationsaufgabe und der Hardware skalierbar gestaltet werden kann.

Abgeschlossen wird das Kapitel mit der Skizzierung der in dieser Arbeit eingesetzten Methoden, die zur Validierung und zur Sicherung der Aussagekraft eines MR-Simulators eingesetzt wurden. Der Schwerpunkt liegt hier auf den numerischen Besonderheiten in einem parallelisierten Umfeld.

## 5.1 Ansätze zur Simulation von MR-Experimenten

Die Bandbreite möglicher Ansätze wurde bereits in Kapitel 2 anhand der vorgestellten Arbeiten anderer Arbeitsgruppen aufgezeigt. Im folgenden wird ein Überblick über Eigenschaften und potentielle Leistungsfähigkeit der vier Klassen von Simulationswerkzeugen gegeben.

### 5.1.1 Spinbasierte Simulationen

Das zu simulierende Objekt wird bei der spinbasierten Simulation im Ortsraum durch ortsdiskrete Spins modelliert. Die Spindynamik kann sowohl klassisch mit der Blochschen Gleichung als auch quantentheoretisch modelliert werden. Eine quantentheoretische Modellierung, mit der über den Blochschen Ansatz hinausgehende Spin-Spin-Kopplungen berücksichtigt werden können, ist zur Erfüllung der Aufgabenstellung nicht erforderlich und wird im folgenden nicht weiter betrachtet. Es genügt für die Aufgabe, von quasi-freien Spins auszugehen, deren schwache Kopplung sich mit der Blochschen Gleichung nachbilden läßt.

Wird mit der Blochschen Gleichung modelliert, kommt sie in der Schreibweise für reellwertige Magnetisierungsvektoren zum Einsatz. Das Blochsche DGL-System muß für jeden Spin unter Berücksichtigung der an seinem Ort jeweils vorliegenden Randbedingungen gelöst werden. In dieser Arbeit werden, wie bereits in Kapitel 3 erläutert wurde, analytische Lösungen verwendet, die am Ort eines Spins numerisch ausgewertet werden.

Eine spinbasierte Simulation ist potentiell außerordentlich flexibel, da das am Ort eines Spins herrschende äußere Magnetfeld aus beliebig vielen Quellen stammen kann. Werden die Magnetfelder aller Quellen zunächst linear superponiert, ist es möglich, z. B. örtliche Inhomogenitäten von statischen, Gradienten- und hochfrequenten Magnetfeldern gleichzeitig zu simulieren. Auch mehr als eine HF-Sendespule oder Spulenarrays für den Empfang des Signals können auf einfache Weise berücksichtigt werden.

Besonderes Augenmerk bei spinbasierten Simulationsansätzen muß auf die Positionierung der Spins zueinander gelegt werden, genauer auf die maximalen Abstände zwischen Spins. Hintergrund ist, daß hier eine Diskretisierung des in der Natur ortskontinuierlichen Objekts vorgenommen wird, für die das Abtasttheorem der Systemtheorie berücksichtigt werden muß.

In der bekannten Literatur wird dieser Aspekt nicht oder nur unbefriedigend behandelt. Die Bandbreite reicht hier von heuristischen Annahmen bezüglich der Anzahl der Spins pro Bildpunkt [96], über die sequenzabhängige Ausblendung störender Signalbestandteile [74], bis hin zur exakten Berücksichtigung des Abtasttheorems für eine Zwei-Puls-Sequenz im eindimensionalen Fall [5]. Die Herleitung des maximalen Spinabstands im dreidimensionalen Fall bei beliebigen Sequenzen findet sich in Abschnitt 5.2.

**Ablauf einer spinbasierten Simulation**

Für eine spinbasierte Simulation mit analytischen Lösungen der Blochschen Gleichung wird zunächst der Magnetisierungsvektor jedes Spins auf den Wert seiner Magnetisierung im thermischen Gleichgewicht $\vec{M}_0$ initialisiert. Anschließend werden Betrag und Richtung der Magnetisierungsvektoren unter Verwendung von Gleichung (4.1) sequentiell mit den für die Elementarsequenzen erforderlichen Matrixoperatoren verändert.

In den Rotationsoperator für Gradienten $T_{\text{grad}}$ wird durch lineare Superposition die Wirkung von Chemical Shift und von Inhomogenitäten des statischen Magnetfelds in die spinabhängige Kreisfrequenz $\omega(\vec{x})$ einbezogen (Gleichungen (3.7) und (3.10)). Bei einer Elementarsequenz mit Datenakquisition werden Rotation und Relaxation der Magnetisierung in eine Anzahl von

Schritten aufgeteilt, die der Anzahl der Abtastpunkte entspricht. Nach jedem Schritt wird die aktuelle Magnetisierung des Spins zur späteren Verwendung gespeichert.

Zur Berechnung eines Echosignals werden die während der Datenakquisition gespeicherten Magnetisierungsvektoren aller Spins für jeden Abtastpunkt additiv superponiert. Bei Inhomogenitäten der Empfangssensitivität erfolgt vor der Superposition eine für jeden Spin individuelle Gewichtung des Magnetisierungsvektors mit der Sensitivität $\vec{S}_{\mathrm{HF}}$ nach Gleichung (3.25).

**Überlegungen zum Laufzeitverhalten**

Die Laufzeit eines spinbasierten Simulationswerkzeugs wird im wesentlichen von der Anzahl der Spins und der Anzahl der Vektorrotationen bestimmt. Die Anzahl der auftretenden Vektorrotationsoperationen hängt ausschließlich von der Komplexität der zu simulierenden Sequenz, d.h. der Anzahl der für ihre Modellierung erforderlichen Elementarsequenzen, ab.

Die Anzahl der Spins $N_{\mathrm{Spin}}$, in die ein Objekt aufgeteilt wird, ist abhängig von der Größe des Objekts und dem vom Abtasttheorem vorgegebenen Höchstabstand zwischen den Spins. Sie steigt beim Übergang von einem ein- zu einem zweidimensionalen und von einem zwei- zu einem dreidimensionalen Objekt stark an. Unabhängig von der Sequenz ist innerhalb des vom Objekt eingenommenen Raums stets mindestens ein Spin pro Volumenelement (*Voxel*) nötig, damit sich im diskreten Bildraum der Eindruck eines kontinuierlichen Objekts ergibt. Abhängig von der konkreten Sequenz können wesentlich mehr Spins pro Voxel erforderlich werden, siehe Abschnitt 5.2.

Ein besonderer Vorzug einer spinbasierten Implementierung mit quasi-freien Spins ist die inhärent gute Skalierbarkeit. Durch Aufgliederung der Gesamtmenge der Spins in Untermengen läßt sich ein Simulationsauftrag gut parallelisieren und damit die Gesamtausführungszeit drastisch reduzieren. Eine für diese Arbeit entworfenen Parallelisierungsstrategie wird in Abschnitt 5.3 vorgestellt.

## 5.1.2 Simulationen mit dem K-t-Formalismus

Die Simulation nach dem K-t-Formalismus (vgl. Kapitel 4) setzt weitgehend ideale Feldverläufe voraus. Statische und Gradientenmagnetfelder dürfen keine transversalen Komponenten besitzen, Gradienten- und hochfrequente Magnetfelder dürfen keine Inhomogenität aufweisen.

**Ablauf einer K-t-basierten Simulation**

Bei einer Simulation mit dem K-t-Formalismus werden die von einer Sequenz erzeugten Konfigurationen und deren zeitliches Verhalten berechnet. Von den Eigenschaften des Objekts fließen in einem ersten Schritt lediglich die Relaxationskonstanten in die Berechnung ein. Die so berechnete zeitliche Entwicklung der Konfigurationen läßt sich für alle Teile des Objekts mit derselben Kombination von $T_1$ und $T_2$ verwenden. In einem zweiten Schritt werden zu den interessierenden Zeitpunkten die aktuellen Konfigurationen mit der Ortsfrequenzfunktion multipliziert und anschließend aufsummiert, vgl. Abschnitte 4.3.3 und 4.4. Diese komplexwertige Summe ist der Wert des Echosignals zum betrachteten Zeitpunkt.

Inhomogenitäten der $z$-Komponente des statischen Magnetfelds können auf zwei Weisen berücksichtigt werden:

1. Hinzunahme einer weiteren Dimension für die Magnetfeldinhomogenität zu den bis zu drei Dimensionen des Ortsfrequenzraums [79], oder

2. Auffassen des inhomogenen statischen Magnetfelds als zeitlich konstantes Gradientenmagnetfeld innerhalb jeweils eines Voxels.

Für diese Arbeit wurde der zweite Fall näher untersucht, vgl. Abschnitt 4.3.3 sowie die Bilder 4.7 und 4.8. Hier muß für jedes interessierende Voxel das statische Magnetfeld in eine Reihe zerlegt werden; der lineare Term dieser Reihe stellt ein zusätzliches, zeitlich konstantes Gradientenmagnetfeld dar. Ist das Voxel klein genug, können die höheren Terme der Reihenentwicklung vernachlässigt werden. Für jedes so gebildete Voxel muß der K-t-Formalismus nachfolgend getrennt ausgeführt werden. Die voxelweise ermittelten Simulationsergebnisse werden abschließend durch lineare Superposition zu einem Gesamtergebnis zusammengefügt.

**Überlegungen zum Laufzeitverhalten**

Der Aufwand bei der Simulation mit dem K-t-Formalismus setzt sich aus dem Ermitteln der vorhandenen Konfigurationen mit ihren Populationen und Ordnungen und der Überlagerung der von ihnen abgetasteten Ortsfrequenzfunktion des zu simulierenden Objekts nach Gleichung (4.19) zusammen.

Das Splitten von Konfigurationen und das Bestimmen ihrer Position im K-Raum erfolgt durch Vektorrotationen der komplexen Magnetisierungen und nachfolgender Änderung ihrer Ordnungen. Zur Ermittlung der aktuellen komplexwertigen Echoamplitude müssen alle transversalen Konfigurationen anschließend mit der Ortsfrequenzfunktion an ihrer Position multipliziert werden.

Bei einer Elementarsequenz ohne Datenakquisition kann die Wirkung des Gradientenmagnetfelds nach dem HF-Puls durch einmaliges Verschieben aller transversalen Konfigurationen im K-Raum berechnet werden. Bei Elementarsequenzen mit Datenakquisition erfolgt dies schrittweise für jeden Abtastwert. Um Inhomogenitäten des statischen Magnetfelds berücksichtigen zu können, müssen diese Operationen i. a. getrennt für jedes Voxel ausgeführt werden, das vom Objekt ganz oder teilweise eingenommen wird.

Die Anzahl der Konfigurationen ist im Verlaufe der Simulation einer Sequenz i. a. nicht konstant. Sie kann, abhängig von der simulierten Sequenz, linear bis exponentiell mit der Anzahl der HF-Pulse steigen. Bei jedem HF-Puls kann es zu einer Aufspaltung einer Konfiguration in bis zu vier neue Konfigurationen kommen. Zeichnet man die Entstehungsgeschichte der Konfigurationen auf, so bilden die Konfigurationen einen bis zu quaternären Baum.

Für Simulationsaufgaben ohne Inhomogenitäten des statischen Magnetfelds bietet ein K-t-basierter Ansatz, verglichen mit einem nicht parallelisierten, spinbasierten Ansatz, i. a. einen drastischen Laufzeitvorteil. Grund ist, daß der erste Schritt einer K-t-Simulation nahezu ohne ortsabhängige Parameter auskommt. Lediglich die örtlich variablen Relaxationszeiten haben einen Einfluß, sind aber konstant innerhalb von Regionen im Ortsraum, die meist größer als ein Voxel sind.

Wird eine skalierbare, parallelisierte Struktur des Simulationsprogramms angestrebt, so muß es aufgrund der baumartigen Abhängigkeiten zwischen den Konfigurationen Datenaustausch zwischen Rechenknoten geben. Zusätzlich müssen Mechanismen zur Verfügung gestellt werden, die die Elternknoten des Konfigurationsbaums von den Rechenknoten entfernen, sobald sie nicht mehr benötigt werden, und die vorhandenen Rechenaufträge müssen neu auf die freigewordenen Rechenknoten aufgeteilt werden. Es ist daher zu erwarten, daß die Parallelisierung eines K-t-basierten Simulators im Vergleich zu einem spinbasierten Werkzeug eine wesentlich größere Belastung der Kommunikationspfade zwischen den Rechenknoten erzeugt.

### 5.1.3  Zusammenfassung

In diesem Abschnitt wird die Eignung der beiden Bloch-basierten Ansätzen aus den Abschnitten 5.1.1 und 5.1.2 hinsichtlich des Anforderungsprofils genauer untersucht. Der in Abschnitt 1.3 definierte Katalog von Anforderungen an das Modell eines Simulators umfaßt, hier in Kurzform wiedergegeben, die Berücksichtigung von

- ortsabhängigen Magnetisierungsvektoren $\vec{M}$ ;

- Spindichte bzw. Ruhemagnetisierung $\vec{M}_0$ ;

- schwacher Spinkopplung (quasi-freie Spins), modelliert mit den Relaxationszeiten $T_1$ und $T_2$ ;

- Chemical Shift;

- Grenzflächen zwischen Materialien mit unterschiedlicher magnetischer Suszeptibilität;

- größtmöglicher Flexibilität bei der Sequenzgestaltung in Bezug auf die Abfolge von Sequenzelementen,;

- HF- und Gradientenmagnetfeldern mit speziellen Zeitabhängigkeiten;

- Abweichungen der Magnetfelder in Betrag und Richtung vom Idealfall, d.h. Feldlinien der Magnetfeldtypen, die nicht senkrecht aufeinander stehen bzw. parallel zueinander liegen, sowie Inhomogenitäten des Betrags der magnetischen Flußdichte.

Die K-t-basierte Simulation deckt eine ganze Reihe der genannten Anforderungen ab. Sofern Magnetfeldinhomogenitäten nicht berücksichtigt werden brauchen, bietet er i. a. einen Laufzeitvorteil vor spinbasierten Ansätzen. Leider bestehen in den bekannten Formulierungen des K-t-Formalismus' Lücken bei der Modellierbarkeit. So können Inhomogenitäten des hochfrequenten Magnetfelds und Komponenten des Gradientenmagnetfelds transversal zu $\vec{B}_0$ nicht in die Berechnung eines Echos miteinbezogen werden.

Müssen Inhomogenitäten des Hauptmagnetfelds, Suszeptibilitätsgrenzflächen oder Chemical Shift nachgebildet werden, verschwindet der Laufzeitvorteil vor spinbasierten Ansätzen bereits unter einfachen Bedingungen. Der K-t-Formalismus ist daneben wegen der baumartigen Abhängigkeit der Konfigurationen voneinander nicht so gut parallelisierbar wie ein spinbasierter Ansatz mit quasi-freien Spins, da ein Datenaustausch zwischen Eltern- und Kindkonfigurationen erforderlich ist.

Hingegen deckt der spinbasierte Ansatz alle in der Liste aufgeführten Anforderungen ab. Für komplizierte Experimentbedingungen wie Inhomogenitäten verschiedener Arten weist er keinen Laufzeitnachteil gegenüber einem K-t-basierten Simulationswerkzeug auf. Die Modellierung mit quasi-freien Spins nach der Blochschen Gleichung liefert einen inhärent guten Ansatzpunkt für eine effiziente Parallelisierung der nötigen Arithmetik.

Als Grundlage für das im Rahmen dieser Arbeit zu erstellende Simulationswerkzeug wurde aus diesen Gründen der spinbasierte Ansatz ausgewählt und in ein Simulationswerkzeug umgesetzt.

## 5.2   Korrekte Ortsdiskretisierung bei spinbasierter Simulation

Die im Sinne des Abtasttheorems korrekte Positionierung der Spins zueinander spielt bei spinbasierten Ansätzen eine bedeutende Rolle. Wegen der unbefriedigenden Behandlung des Themas in der Literatur wurde ihm in der vorliegenden Arbeit besonderes Gewicht beigemessen.

Dieser Abschnitt leitet mit Hilfe des K-t-Formalismus' eine Bedingung für den Höchstabstand zwischen Spins her, bis zu dem das Simulationsergebnis von der Diskretisierung des Objekts unbeeinflußt bleibt. Die gefundene Beziehung ist für beliebige Sequenzen in alle drei Raumrichtungen einsetzbar und stellt eine allgemeingültige Voraussetzung für spinbasierte Simulationen dar.

Bei einem realen Experiment ist eine Diskretisierung nur im Zeitbereich erforderlich. Dort dient sie zur Umsetzung von Daten des digitalen Steuerrechners in analoge Signale und zur Weiterverarbeitung von analogen Empfangsdaten in einem zeitdiskreten System. Die im folgenden beschriebene räumliche Diskretisierung des zu untersuchenden Objekts hat keine Ent-sprechung in der Realität, sondern ist eine Besonderheit der numerischen, spinbasierten Simulation.

## 5.2.1   Ortsfrequenzfunktion eines abgetasteten Objekts

In der Natur sind die zu untersuchenden Objekte, zumindest makroskopisch gesehen, kontinuierlich. Mikroskopisch sind die zur Bildgebung genutzten [1]H-Nukleonen in sehr großer Dichte vorhanden, d.h. größenordnungsmäßig mehr als $10^{20}\,\mathrm{cm}^{-3}$ Teilchen. Bei der numerischen Simulation von Spins muß der Rechen- und Speicherbedarf stark reduziert werden. Dazu werden die Objekteigenschaften mit Hilfe von punktförmigen Spins räumlich abgetastet, und jedem abtastenden Spin werden die Eigenschaften des Objekts an seiner Position zugewiesen. Durch die Abtastung wird die Ortsfrequenzfunktion $S(k_x)$ des simulierten Objekts periodisch mit einem Ortsfrequenzabstand $\Delta K_x = 2\pi/\Delta x$ wiederholt, der umgekehrt proportional zum Abtastabstand $\Delta x$ ist, siehe Bild 5.1. Es entsteht das periodische Spektrum $\tilde{S}(k_x)$.

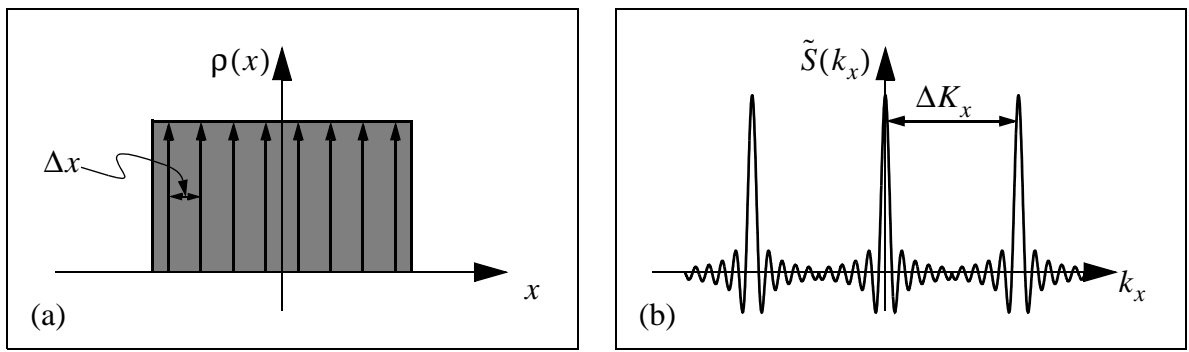

Bild 5.1: Die räumliche Abtastung eines Objekts mit dem Abtastabstand $\Delta x$ (a) führt zu einer mit dem Abstand $\Delta K_x = 2\pi/\Delta x$ periodisch wiederholten Ortsfrequenzfunktion (b).

Bei einer spinbasierten Simulation muß dafür Sorge getragen werden, daß sich während des Experiments transversale Konfigurationen nicht im Bereich der periodischen Wiederholungen der Ortsfrequenzfunktion aufhalten. Ist dies nicht gewährleistet, kommt es im Simulationsergebnis zu Aliasing-Artefakten. Deckt beispielsweise eine echobildende Transversalkonfiguration nach Bild 5.1 das Ortsfrequenzintervall $[-3/2 \cdot \Delta K_x, 3/2 \cdot \Delta K_x]$ ab, so erhält man als Echosignal die dreifach wiederholte Ortsfrequenzfunktion des Objekts, siehe Bild 5.2.

## 5.2.2   Einfluß der Sequenz

Bei der Festlegung der Ortsdiskretisierung muß offenbar die tatsächlich eingesetzte Bildgebungssequenz berücksichtigt werden, um diesen Fehler zu verhindern. Der Abtastabstand $\Delta x$ ist ausreichend klein, wenn die periodische Wiederholung der Ortsfrequenzfunktion des Objekts außerhalb des K-Raum-Bereichs liegt, in dem sich longitudinale oder transversale Konfigurationen aufhalten.

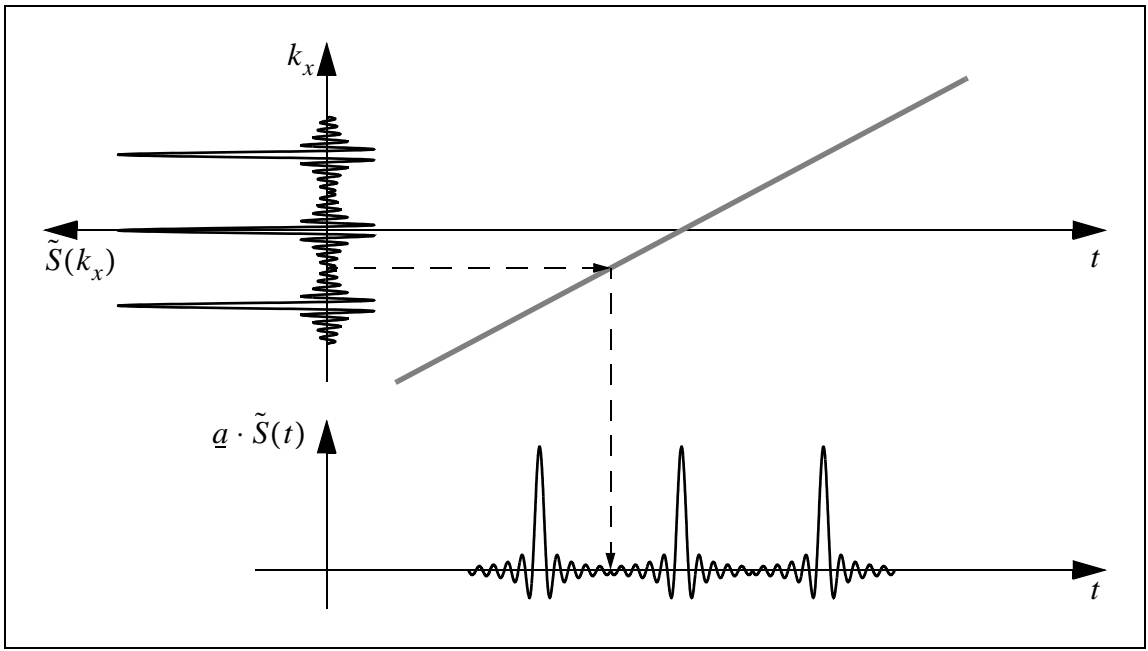

Bild 5.2: Entstehung eines fehlerhaften Echosignals,
das aus einer unzureichender Ortsabtastung resultiert.

Der von den Konfigurationen einer Sequenz genutzte K-Raum-Bereich kann größer sein als derjenige, in dem die Ortsfrequenzfunktion nach den Maßgaben von Field-of-View und Auflösung aufgezeichnet wird, siehe Anhang B.1. Im Fall der Preview-Sequenz nach Bild 4.9 ist beispielsweise lediglich ein sehr schmaler $k_x$-Bereich um die Zeitachse für die Echobildung nötig. Ein fehlerfreies Simulationsergebnis ergibt sich jedoch nur, wenn die periodische Wiederholung der Ortsfrequenzfunktion des Objekts zu jeder Zeit der Preview-Sequenz außerhalb des von irgendeiner Konfiguration benutzen Bereichs liegt.

### 5.2.3   Mindestabstand von Spins

Definiert man das Maximum über alle Konfigurationen der während einer Sequenz eingenommenen K-Raum-Positionen zu

$$K_{x,\,max} = \max_{\text{alle Konfigurationen}} \left\{ |k_x(t)| \right\},\qquad(5.1)$$

so ist die Bedingung für die erforderliche Ortsabtastung

$$\Delta x < \frac{\pi}{K_{x,\,max}} = \Delta x_{max}.\qquad(5.2)$$

Für die Raumrichtungen $y$ und $z$ gelten sämtliche Überlegungen analog. Die Höchstabstände $\Delta x_{max}$, $\Delta y_{max}$ und $\Delta z_{max}$ der Spins in den drei Raumrichtungen sind voneinander unabhängig.

In Regionen mit Inhomogenitäten des statischen Magnetfelds können sich statische und Gradientenmagnetfelder in ihrer Wirkung auf $k_x(t)$ einer Konfiguration verstärken. Die Größe $k_x(t)$ kann daher über $t$ stärker wachsen als ohne Inhomogenität. In diesem Fall ist die neue, weiter von der $t$-Achse entfernte Position der Konfiguration für die Ermittlung von $\Delta x$ maßgebend. Chemical Shift hat eine der Inhomogenität des statischen Magnetfelds vergleichbare Wirkung. Es muß somit bei der Ermittlung von $\Delta x$ auf analoge Weise berücksichtigt werden.

Olsson geht in [74] die Problematik fehlerhafter Echos bei einer geringen Anzahl von Spins pro Voxel mit Hilfe einer dem K-t-Formalismus verwandten Überlegung an. Magnetisierungsvektoranteile, die nicht der beabsichtigten Hauptwirkung von HF-Pulsen folgen, werden hier zu Null gesetzt, um ihren fehlerhaften Einfluß auf das Simulationsergebnis zu eliminieren. Das Verfahren ist allerdings sequenzabhängig und auf HF-Pulse mit einem Flipwinkel von 180° beschränkt. Zudem ist der Rechenaufwand sehr hoch, weil Überprüfungs- und Modifikationsprozeduren für jeden Spin angewendet werden müssen.

### 5.2.4   Reduktion des Simulationsaufwands bei Steady-State-Sequenzen

Bei vielen Sequenzen reicht es aus, die Höchstabstände der Spins zueinander aus den maximal erreichbaren K-Raum-Positionen zu ermitteln , ohne die tatsächliche Population der Konfigurationen zu berücksichtigen. Die so ermittelten Spinabstände sind i.a. zu klein (und führen damit zu einem überhöhten Simulationsaufwand), wenn die Populationen der äußeren Konfigurationen nahe Null sind. Dies ist von besonderer Bedeutung bei sogenannten Steady-State-Sequenzen.

Bei *Steady-State*-Sequenzen wird im Gegensatz zu den bisher vorgestellten Experimenten nach der Akquisition von Meßdaten nicht gewartet, bis sich im Spin-Ensemble erneut das thermische Gleichgewicht eingestellt hat. Vielmehr wird der Datenaufnahmezyklus umgehend wiederholt. Es stellt sich im Verlauf eines solchen Experiments nach einer Anzahl von Wiederholungen bzw. Zyklen ein *dynamisches* Gleichgewicht ein.

Das dynamische Gleichgewicht zeichnet sich dadurch aus, daß sich das Spin-Ensemble zu Beginn eines Zyklus' in demselben Zustand wie zu Beginn des vorangegangenen Zyklus befindet. Typisch für derartige Experimente ist, daß einige Dutzend Wiederholungen des Grundzyklus' des Experiments durchgeführt werden, die *Warmlaufzyklen*, bevor mit der Datenakquisition begonnen wird. Mit Steady-State-Sequenzen ist eine kontinuierliche Datenaufnahme möglich. Ein typisches Timing-Diagramm zeigt Bild 5.3.

Die nach dem K-t-Formalismus maximal erreichte K-Raum-Position steigt mit der Dauer des Experiments linear und theoretisch unbeschränkt an. Damit sinkt bei einer Simulation der nötige Höchstabstand zwischen den Spins mit der Dauer des Experiments, die Spinanzahl steigt, und die Simulationsdauer verlängert sich stark.

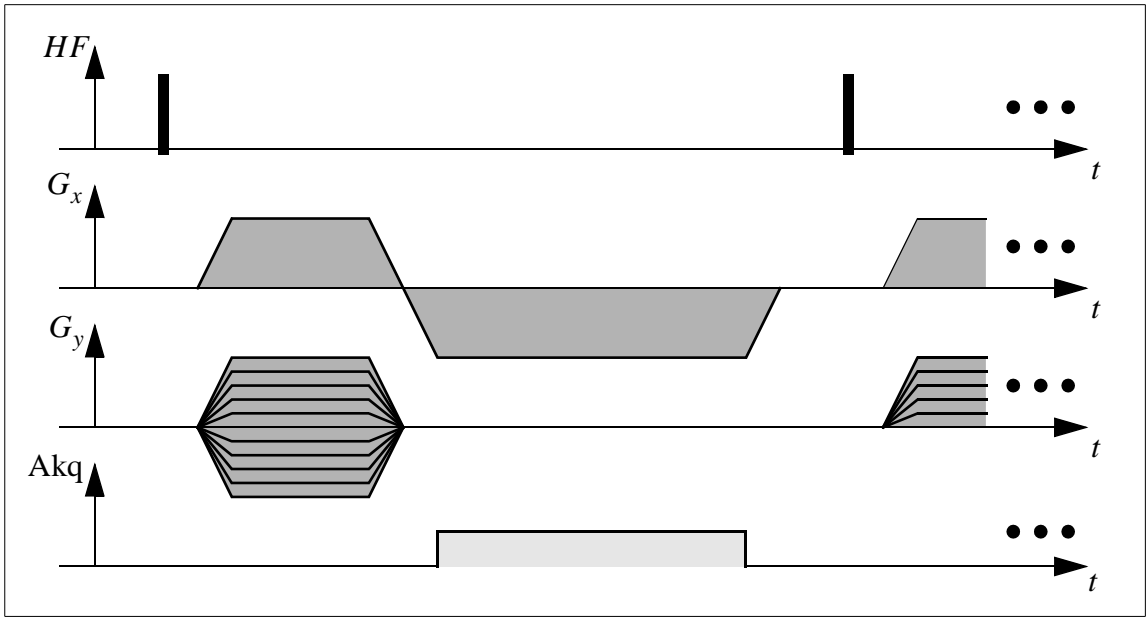

Bild 5.3: Sequenzdiagramm einer Gradienten-Echo-Sequenz für den Steady-State-Fall. Ein Zyklus reicht von einem HF-Puls bis zum nächsten. Der Flipwinkel ist gewöhnlich klein und liegt im Bereich von einigen zehn Grad.

Bild 5.4 zeigt, daß schon während der Warmlaufzyklen der Sequenz die Populationen von transversalen Konfigurationen höherer Ordnungen nur klein sind. Konfigurationen mit sehr kleiner Population führen, wenn der Ortsraum durch Verletzung von Gleichung (5.2) unterabgetastet ist, nur zu schwachen Aliasing-Effekten, die im Hinblick auf kleine Simulationszeiten tolerierbar sein können. Im folgenden wird gezeigt, wie abgeschätzt werden kann, welche Konfigurationen für die Ermittlung des Spinhöchstabstands berücksichtigt werden müssen.

Hierzu wird exemplarisch angenommen, daß Gleichung (5.2) dadurch verletzt ist, daß es Konfigurationen gibt, die sich während der Datenakquisition im Bereich der periodischen Wiederholungen des periodischen Ortsfrequenzspektrums $\tilde{S}(k_x)$ aufhalten, vgl. Bild 5.5. Im gezeigten Fall stören die durch die Konfigurationen mit den Populationen $\underline{a}_1$ bis $\underline{a}_3$ hervorgerufenen Signale das gewünschte Signal $\underline{a}_0 \cdot \tilde{S}[k_x(t)]$ maximal.

Im Bildbereich entsteht aus der Konfiguration mit der Population $\underline{a}_0$ das Nutzbild; die Konfigurationen mit den Populationen $\underline{a}_1$ bis $\underline{a}_3$ sind für Bilder geringer Amplitude verantwortlich, die sich zu einem Störbild überlagern. Das Verhältnis zwischen dem Maximum $M_{\text{Nutz}}$ des Nutzbilds und den Maxima $M_{\text{Stör}, i}$ der Komponenten des Störbildes läßt sich wegen der Linearität der Fourier-Transformation aus den Populationen ermitteln, und es gilt

$$\frac{M_{\text{Stör}, i}}{M_{\text{Nutz}}} = \frac{|\underline{a}_i|}{|\underline{a}_0|} . \qquad (5.3)$$

Das Verhältnis der Maxima von Nutz- und Störbild läßt sich dann nach oben abschätzen durch

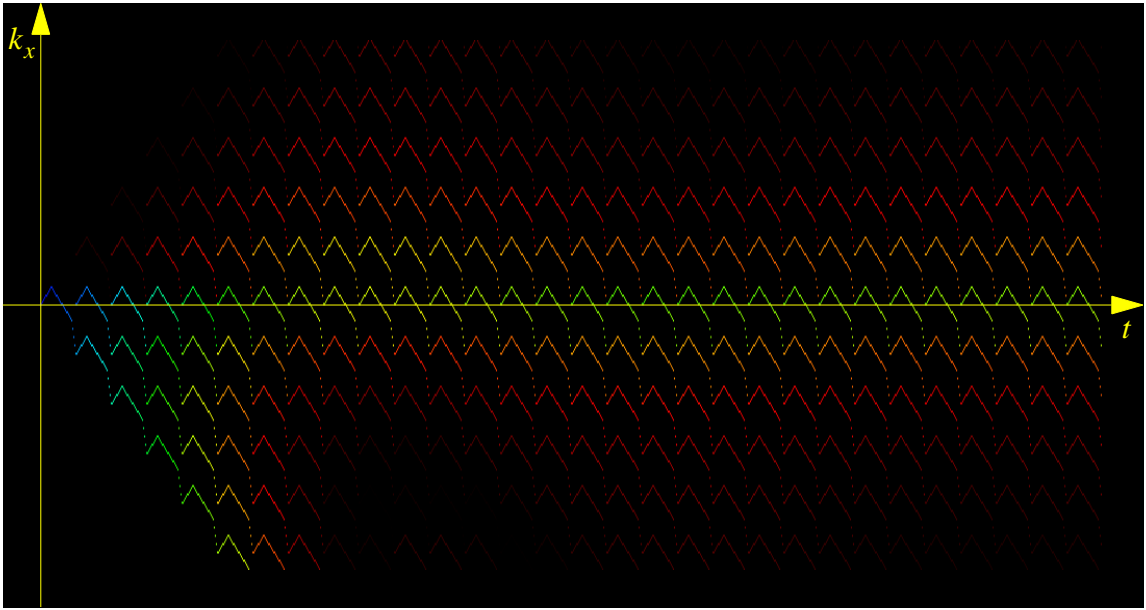

Bild 5.4: K-t-Diagramm der transversalen Konfigurationen $a_n$ über die ersten 30 Zyklen einer Steady-State-Gradienten-Echo-Sequenz.

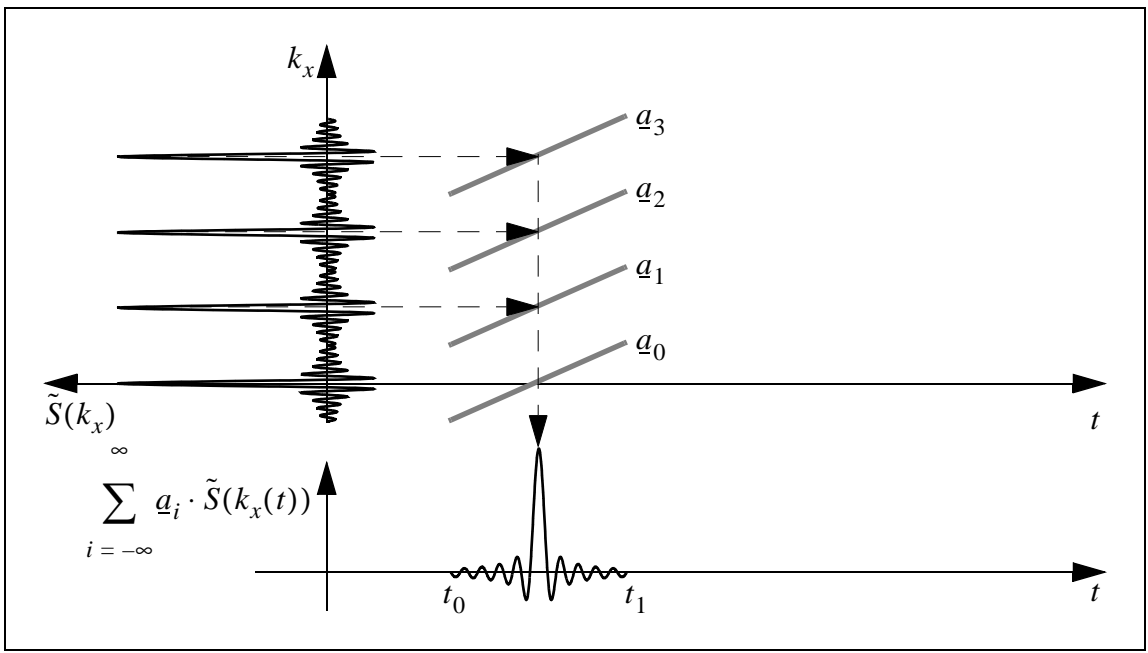

Bild 5.5: Überlagerung des von vier Konfigurationen verursachten Signals bei unzureichender Ortsabtastung.

$$\frac{M_{\text{Stör}}}{M_{\text{Nutz}}} \leq \frac{|\underline{a}_1| + |\underline{a}_2| + |\underline{a}_3|}{|\underline{a}_0|} \, . \tag{5.4}$$

Die Bilddarstellung erfolgt i.a. in Form von Grauwertbildern mit einer Quantisierung $\Delta\rho$, bei der der gewünschte Grauwertbereich in eine endliche Zahl von Graustufen, z.B. 256, untergliedert wird. Das Störbild kann in einer solchen quantisierten Darstellung nicht mehr ausgemacht werden, wenn sein Maximum dem Grauwert mit dem Index 0, also einem schwarzen Pixel, zugeordnet wird. Unter der Voraussetzung, daß der hellste Grauwert im Bild dem Maximum des Nutzbilds entspricht, liegt das Maximum des Störbild unterhalb der Schwelle zum Grauwert mit dem Index 1, wenn gilt

$$\frac{M_{\text{Stör}}}{M_{\text{Nutz}}} \leq \frac{\Delta\rho}{2} \,. \tag{5.5}$$

Mit Hilfe der Beziehungen (5.4) und (5.5) lassen sich transversale Konfigurationen bei der Ermittlung des Spinabstands ausklammern und so die Anzahl der Spins für eine Simulation und damit den Simulationsaufwand bei Steady-State-Sequenzen deutlich reduzieren.

# 5.3   Strategien zur Laufzeitreduktion durch Parallelisierung

Trotz der sich rasch weiterentwickelnden Computer-Technik stößt die numerische Simulation vieler technischer und naturwissenschaftlicher Probleme immer wieder an die Grenze dessen, was mit modernen Prozessoren in annehmbarer Zeit abgearbeitet werden kann. Die Kopplung mehrerer Zentraleinheiten zu einem parallelverarbeitenden System bietet hier prinzipielle Abhilfe und erlaubt es, komplexere Experimente bei geringerem Zeitaufwand zu bewältigen.

Die parallelisierte Abarbeitung eines Problems erfolgt allerdings i. a. nicht automatisch. Vielmehr muß ein Computerprogramm explizit so gestaltet sein, daß es die parallelverarbeitenden Möglichkeiten eines Rechnersystems nutzt. Für die Gestaltung eines parallelisierten Simulationswerkzeugs gibt es meist mehrere Herangehensweisen, von denen problemabhängig eine optimale Strategie auszuwählen ist.

Eine für diese Arbeit geeignete Parallelisierungsstrategie muß sicherstellen, daß die verfügbare Rechenleistung möglichst effizient genutzt werden kann, also die vorhandenen Prozessoren ohne Unterbrechung beschäftigt werden. Dies setzt voraus, daß Wartezeiten vermieden oder zumindest so kurz wie möglich gehalten werden. Die langsamste Komponente der Infrastruktur ist das die Rechenknoten verbindende Netzwerk. Oberstes Ziel muß es sein, das Netzwerk möglichst selten und nur für kleine Datenmengen zu nutzen.

Für diese Arbeit wurden zwei Ansätze für eine Parallelisierung untersucht; sie werden im vorliegenden Abschnitt kurz beschrieben und hinsichtlich ihrer Eignung bewertet.

## 5.3.1   Functional Decomposition

Bei einem Functional Decomposition Approach führen die Rechenknoten einer parallelen Rechenanlage verschiedene Operationen auf eine zentrale Menge von Daten aus. Die Operato-ren können dabei einerseits verschiedenartig, andererseits aber auch von gleicher Art, aber mit unterschiedlichen Zahlenwerten sein. Fixe Größe bei einem Functional Decomposition Approach ist die hardwaremäßig vorgegebene Anzahl von Prozessoren. Sie ist zunächst unabhängig von der durch die Sequenz vorgegebene Anzahl von Operatoren. Die Hardware-Ebene muß daher mit Rechenknoten abstrahiert, und eine Abbildungsvorschrift zwischen Rechenknoten und Prozessoren muß angegeben werden.

**Abbildung von Rechenknoten auf Prozessoren**

Im folgenden wird angenommen, daß ein Rechenknoten alle nötigen Operationen für eine Elementarsequenz durchführt. Die Rechenknoten sind dann in ihrem Aufbau gleich, unterscheiden sich aber in ihren Zahlenwerten für HF-Pulse und Gradientenmagnetfelder sowie in ihrer Länge. Übersteigt die Anzahl der Rechenknoten die der Prozessoren, so muß ein Prozessor die Aufgaben mehrerer Knoten übernehmen.

Dazu können einerseits mehrere aufeinanderfolgende Elementarsequenzen bzw. Rechenknoten zu einer größeren Einheit zusammengefaßt und als solche auf einem Prozessor angeordnet werden. Andererseits können unter Verwendung der Multitasking-Fähigkeiten des Betriebssystems mehrere Knoten einen Prozessor bei gleichmäßiger Aufteilung der Rechenleistung unabhängig voneinander nutzen.

**Pipeline von Rechenknoten**

Die Abbildung von Elementarsequenzen auf Rechenknoten setzt grundsätzlich voraus, daß alle Knoten jederzeit lesenden und schreibenden Zugriff auf jeden einzelnen Spin haben. Dies erfordert auf den ersten Blick einen stetigen Datenaustausch mit einer zentralen Instanz, die alle Spins ständig vorhält.

Allerdings müssen die Elementarsequenzen in einer vorgegebenen Reihenfolge abgearbeitet werden. Werden die Rechenknoten in einer Pipeline hintereinander angeordnet, bedarf es keiner zentralen Instanz für die Spin-Speicherung. Die zu bearbeitenden Spins werden stattdessen nacheinander in die Pipeline eingefügt und anschließend von Knoten zu Knoten weitergereicht. Unabhängig davon müssen die von jedem Spin erzeugten Echosignale von einer zentralen Instanz gesammelt und superponiert werden.

**Leistungsbegrenzende Faktoren**

Datenaustausch über das Netzwerk tritt beim Functional Decomposition Approach zum einen beim Weiterreichen eines Spins von einem Knoten zu seinem Nachfolger, zum anderen beim Abliefern eines Simulationsergebnisses bei der zentralen Superpositionseinheit auf. Bei gefüllter Pipeline haben sämtliche Knoten Kommunikationsbedarf; die resultierende Netzwerkbelastung ist damit hoch.

Der Zeitabstand zwischen zwei in die Pipeline eintretenden Spins ist im Idealfall so groß wie der Zeitaufwand für eine einzelne Operation. Die aus dem Zeitabstand resultierende Durchsatzrate abgearbeiteter Spins ist damit hoch. Sie wird jedoch von zwei Faktoren begrenzt:

- Die hohe Belastung eines von allen Prozessoren gleichzeitig genutzten Netzwerks verhindert, daß alle Knoten die Spins synchron an ihre Nachfolger weitergeben. Es kann zu Wartezeiten kommen, bei denen Knoten an langsameren Netzwerksegmenten tendenziell benachteiligt werden. Die Effizienz der Nutzung der Prozessoren sinkt mit steigenden Wartezeiten.

- Der langsamste Knoten in der Pipeline bestimmt die Durchsatzrate der Spins, da ihn alle Spins passieren müssen. Schnellere Prozessoren müssen auf langsamere Prozessoren warten; auch dies senkt die Effizienz der Nutzung.

Die Gesamtlaufzeit einer Simulation ist im wesentlichen die Zeit zwischen Injektion des ersten Spins in die Pipeline und Entnahme des letzten Spins aus der Pipeline. Während des Füllens der Pipeline zu Beginn der Simulation werden die Rechenknoten erst nach und nach aktiv, sobald sie der erste Spin erreicht. Umgekehrt werden die Knoten beim Leeren der Pipeline am Ende einer Simulation nach und nach inaktiv, sobald sie den letzten Spin bearbeitet haben. Die Nutzungseffizienz sinkt durch diese ungleichmäßige Ausnutzung zu Beginn und am Ende einer Simulation weiter ab.

## 5.3.2   Domain Decomposition

Bei einem Domain Decomposition Approach führen alle Rechenknoten die gleichen Operationen durch. Die vorhandene Datenmenge wird nach applikationsabhängigen Kriterien auf die verfügbaren Rechenknoten aufgeteilt.

Im vorliegenden Fall ist die Datenmenge die Menge der Spins eines Objekts. Den Spins ist gemeinsam, daß sie alle Elementarsequenzen durchlaufen müssen; die auf jeden Spin anzuwendende Operation ist somit die Sequenz. Für einen Domain Decomposition Approach ist diese Definition die im vorliegenden Fall einzig denkbare. Im Gegensatz zum Functional Decomposition Approach kann bereits ein einzelner Rechenknoten eine vollständige Simulation bearbeiten, da er alle nötigen Informationen besitzt. Ein Rechenknoten kann somit direkt auf einen Prozessor abgebildet werden.

Um die Menge von Spins vollständig die Sequenz durchlaufen zu lassen, werden die Spins den verfügbaren Rechenknoten einzeln oder in Blöcken zugewiesen. Die Rechenknoten liefern anschließend das Simulationsergebnis für einen Spin oder einen Block von Spins bei einer zentralen Instanz ab. Die Größe der Spinblöcke muß so klein gewählt werden, daß jedem Rechenknoten ein Spinblock zugewiesen werden kann. Für eine Nivellierung unterschiedlicher

Rechenleistungen der beteiligten Knoten ist es sinnvoll, jedem Knoten potentiell mehr als einen Block zuweisen zu können.

**Leistungsbegrenzende Faktoren**

Die Rechenknoten benötigen keine Informationen voneinander, da jeder Knoten autark eine Simulation bearbeiten kann. Es kann daher nicht zu Wartezeiten dadurch kommen, daß schnelle Knoten auf Ergebnisse langsamerer warten. Im Gegensatz zum Functional Decomposition Approach gibt es keine Netzwerkbelastung durch das Weiterreichen von Spin-Informationen von Knoten zu Knoten. Netzwerklast entsteht im wesentlichen durch zwei Vorgänge:

- Jedem rechenbereiten Knoten muß ein neuer Spinblock zugewiesen werden. Alle Spinblöcke (also jeder Spin) müssen einmal über das Netzwerk transportiert werden.

- Für jeden Spinblock entsteht ein Ergebnisdatensatz, der über das Netzwerk an eine zentrale Instanz transportiert werden muß. Die Größe des Datensatzes hängt nicht von der Größe der Spinmenge ab, sondern von der Anzahl der Elementarsequenzen mit Datenakquisition und der Anzahl von Echoabtastwerten pro Elementarsequenz.

Da die Rechenknoten voneinander entkoppelt sind, kommt es bei der Übermittlung von Simulationsergebnissen nur zufällig zu gleichzeitigem Netzwerkzugriff. Auch bei gleicher Rechenleistung aller Rechenknoten sind Kollisionen selten, da die Spinblöcke den Knoten nacheinander und nicht gleichzeitig zugewiesen werden, und somit auch die Ergebnisdatensätze mit einem Zeitversatz entstehen.

Zusammengefaßt können Wartezeiten, die zu einer weniger effizienten Nutzung eines Prozessors führen, entstehen, wenn

1. ein Ergebnisdatensatz wegen gleichzeitiger Netzaktivität eines anderen Knotens nicht versandt werden kann; dieses tritt in der Praxis nicht häufig auf, oder

2. wenn einem rechenbereiten Knoten nicht sofort nach dessen Ergebnisversand ein neuer Spinblock zugestellt wird, weil gleichzeitig ein anderer Knoten versorgt wird. Auch diese Konstellation ist selten.

## 5.3.3 Zusammenfassung

Im Vergleich zum Functional Decomposition Approach kann erwartet werden, daß eine Parallelisierungsstrategie nach dem beschriebenen Domain Decomposition Approach zu einem wesentlich geringeren Wartezeitanteil an der Gesamtlaufzeit führt. Wegen der besseren Nutzungseffizienz der eingesetzten Prozessoren und der daraus resultierenden kürzeren Laufzeit einer Simulation kommt in dieser Arbeit der Domain Decomposition Approach zum Einsatz.

# 5.4    Validierung des Simulators

Von großer Bedeutung bei dem in dieser Arbeit vorgestellten Simulationswerkzeug ist das Sicherstellen der Genauigkeit des Rechenergebnisses. Sie ergibt sich aus der Korrektheit der Simulationsergebnisse einerseits im Vergleich zu realen Experimenten und andererseits bezüglich der theoretischen Grundlagen.

Bei der Implementierung eines Simulationswerkzeugs müssen die in der Software-Technik gängigen Testmethoden eingesetzt werden, um die Richtigkeit der berechneten Ergebnisse sicherstellen zu können. Dazu zählen White- und Black-Box-Tests der Module des Programms genauso wie ein Integrationstest, mit dem das korrekte Zusammenspiel aller Komponenten überprüft wird. Sehr wichtig sind *Regressionstests*, mit denen bei Programmänderungen und Programmerweiterungen kontrolliert werden kann, ob bisherige Ergebnisse konsistent neu erzeugt werden können. Regressionstests bauen auf Sätzen von Referenzdaten, sogenannten *Szenarien*, auf, mit denen die Simulationsergebnisse neuer Programmversionen automatisiert verglichen werden können.

Die Regressionstests stellen bei der vorliegenden Arbeit auf zwei Weisen eine besondere Herausforderung dar. Zum einen müssen initiale Szenarien sorgfältig validiert werden, ohne daß es möglich ist, sie ihrerseits mit Referenzdaten vergleichen zu können. Zum anderen handelt es sich bei Simulationsergebnissen um das Resultat einer Abfolge von numerischen Operationen im parallelisierten Umfeld; dies erschwert den Vergleich eines Testdatensatzes mit einem Szenario.

## 5.4.1    Gewinnung initialer Referenzdaten

Initiale Referenzdaten sind Sätze von Simulationsergebnissen, die mit einer ersten Version eines Simulationsprogramms erzeugt und einer intensiven Validitätsüberprüfung unterzogen werden. Die Validitätsüberprüfungen, die in dieser Arbeit durchgeführt werden müssen, basieren auf folgenden Säulen:

- Analyse und Tracking einzelner Rechenschritte,
- Plausibilität des Ergebnisses, und
- Vergleich mit Ergebnissen anderer Simulationswerkzeuge.

Für die Analyse und das Tracking von Rechenschritten wird das analytisch zu erwartende Rechenergebnis mit dem vom Simulationswerkzeug tatsächlich ermittelten Ergebnis schrittweise verglichen. Bei der Plausibilitätsüberprüfung wird getestet, ob das Simulationsergebnis einer Reihe von Kriterien entspricht. Eine wichtige Rolle spielt die systemtheoretische Betrachtung des Ergebnisses. Beispielsweise weisen Überschwinger an Objektkanten (Gibbs-sches Phänomen) eine spezielle Symmetrie auf, die davon abhängt, ob die Kante auf dem Pixelraster der Abbildung liegt oder nicht.

Der Vergleich mit Simulationsergebnissen anderer Werkzeuge stellt eine schnelle Testtechnik dar. Der Vergleich ist besonders aussagekräftig, wenn die miteinander verglichenen Werkzeuge auf unterschiedliche Weisen zu einem Ergebnis gekommen sind. Für den hier betrachteten spinbasierten Ansatz ist es sinnvoll, Werkzeuge einer anderen Klasse oder von anderen Autoren zum Vergleich heranzuziehen. Aus Vorarbeiten [54] [107] standen derartige Vergleichwerkzeuge zur Verfügung.

## 5.4.2   Simulationsergebnisse bei parallelverarbeitender Realisierung

Bei einer parallelisierten, mit Gleitkommazahlen operierenden Anwendung auf einem heterogenen Netzwerk von Prozessoren können bereits zwei aufeinanderfolgende Simulationsläufe mit unverändertem Programmcode zu verschiedenen Ergebnissen führen. Grund hierfür ist, daß die Aufteilung der Prozesse auf die verfügbaren Prozessoren Schwankungen unterliegt, also beispielsweise die einzelnen Prozesse nicht immer auf den jeweils selben Prozessoren gestartet werden. Schwankungen der Hauptprozessorbelastung durch System- und sekundäre Benutzerprozesse oder Fluktuationen der Belastung eines Netzwerkpfads durch eine Datenübertragung können zudem dazu führen, daß einzelne Datenblöcke bei verschiedenen Programmläufen auf verschiedenen Recheneinheiten verarbeitet werden.

Unterschiede in der numerischen Genauigkeit der Gleitkommarechenwerke verschiedener Prozessoren und numerische Schwankungen in den Ergebnissen, die z. B. von der Reihenfolge von Additionsoperationen abhängen, führen zu Fluktuationen im Gesamtergebnis. Diese Variationen sind i. a. klein, können aber bei Zahlenwerten in der Nähe der Maschinengenauigkeit $\varepsilon$ zu großen relativen Fehlern führen.

Für den Vergleich von Szenarien mit aktuellen Testdatensätzen im Rahmen von Regressionstests wurden in dieser Arbeit Vergleichsmethoden entwickelt, die gegenüber derartigen Gleitkommafehlern tolerant sind, aber dennoch Abweichungen zwischen Referenz- und Testdaten klar aufzeigen können. Dazu wurden Maßzahlen definiert, die einerseits einen Überblick über die Übereinstimmung bieten, andererseits Hinweise auf einzelne fehlerhafte Zahlenwerte geben.

Für einen Überblick eignet sich der *Störabstand* sehr gut. Er wird hier definiert als die Energie der Differenz zwischen Referenz- und Testdatensatz bezogen auf die Energie des Referenzsignals, angegeben in Dezibel. Die Ergebnisdatensätze enthalten Werte von Magnetisierungsvektoren; der Störabstand wird damit definiert zu

$$\Delta E_{\text{stör}} \;=\; 10 \text{ dB} \cdot \log \frac{\sum_i \left| \vec{M}_{\text{ref}}(i) - \vec{M}_{\text{test}}(i) \right|^2}{\sum_i \left| \vec{M}_{\text{ref}}(i) \right|^2} \; . \tag{5.1}$$

Gezielte Hinweise auf einzelne fehlerhafte Abtastwerte ergeben sich aus dem relativen Fehler $f$ jeder Komponente jedes Abtastwerts. Wegen der großen Menge von so ermittelten Vergleich­merkmalen dient als erster Anhaltspunkt für eine gute oder schlechte Qualität eines Testdaten­satzes die Anzahl der Abtastwerte, bei denen der relative Fehler eine konstante Schwelle über­schritten hat. Liegen zwei zu vergleichende Komponenten im Bereich der Maschinengenauig­keit $\varepsilon$, wird für sie der relative Fehler nicht ermittelt.

# 6 PARSPIN — Ein skalierbarer Magnetresonanz-Simulator

Eine Festlegung auf ein spinbasiertes Simulationswerkzeug erschließt nicht inhärent die Umsetzung des Konzepts in ein lauffähiges Programm. Randbedingungen der softwaretechnischen Realisierung und Überlegungen zur Benutzbarkeit beeinflussen, wie die Implementierung tatsächlich erfolgt. Im vorliegenden Kapitel finden sich Implementierungsdetails zu dem in dieser Arbeit entstandenen Simulationswerkzeug PARSPIN (Parallelized Spin-based MR-Simulator).

Zunächst wird beleuchtet, wie Objekt, Sequenz und System sinnvoll modelliert werden können. Besonderes Augenmerk wird auf ein möglichst flexibles Modell gerichtet, daß die für ein Hilfsmittel der Forschung erforderliche Variabilität bietet. Anschließend wird das Benutzungsumfeld erläutert, d.h. die verfügbare Infrastruktur, die Zielgruppe der Benutzer und ihr Simulationsbedarf. Hieraus werden Schlußfolgerungen bezüglich Portabilität und der Gestaltung der Benutzerschnittstelle gezogen. Die Beschreibung der für eine parallelisierte Implementierung nötigen Software-Komponenten, ihrer Aufgaben und Funktionsweise rundet das Kapitel ab.

## 6.1 Modellbildung

Alle für einen MR-Simulator nötigen mathematischen und konzeptionellen Werkzeuge sind in den vorangegangenen Kapiteln vorgestellt wurden. Dieser Abschnitt faßt die wesentlichen Aspekte des in PARSPIN benutzten Modells zusammen und beschreibt einige weiterführende Details der gewählten Modellierung von Objekt, Sequenz und System.

### 6.1.1 Objektmodellierung

Die bisherigen Ausführungen zum Objekt bezogen sich vornehmlich auf einzelne Spins und ihre Eigenschaften. Diese mikroskopische Sicht muß ergänzt werden durch eine makroskopische Struktur, die es dem Benutzer mit einfachen Mitteln erlaubt, zusammenhängende Objektgebiete mit bestimmten Eigenschaften zu spezifizieren.

Die für diese Arbeit grundlegenden Spins können als punktförmige Objekte aufgefaßt werden. Systemtheoretisch stellen sie einen idealen Abtaster dar, mit dem das Objekt diskretisiert werden kann, und auf den die Regeln des Abtasttheorems über den Abstand zwischen Abtastpunk-ten unmittelbar angewendet werden können, siehe Abschnitt 5.2. Jeder Spin besitzt eine Reihe von Eigenschaften; diese sind seine Ortskoordinaten, ein Magnetisierungsvektor, eine Ruhemagnetisierung $M_0$ und die Relaxationskonstanten $T_1$ und $T_2$. Zur Berücksichtigung von Chemical Shift oder Verzerrungen des statischen Felds an Suszeptibilitätsgrenzflächen des Objekts dient eine ortsabhängige Präzessionskreisfrequenz

$$\Delta\omega(\vec{x}) \;=\; \Delta\omega_{CS}(\vec{x}) - \gamma \cdot \Delta B_{susz}(\vec{x}) \,. \tag{6.1}$$

In der Natur lassen sich in Teilgebieten eines Objekts, z. B. innerhalb eines Voxels, häufig zwei oder mehr Materialien mit unterschiedlichen Magnetresonanzeigenschaften nachweisen. Dies tritt meist dann auf, wenn eine Grenzfläche zwischen zwei strukturellen Komponenten des Objekts innerhalb des Voxels liegt, in der Medizin beispielsweise an Organgrenzen. Aber auch ohne makroskopisch vorhandene Strukturierung gibt es häufig Areale in einem Objekt, in denen sich *multiexponentielle* Relaxation finden läßt. Die Relaxationseigenschaften und ihre Wirkung auf die Magnetisierung solcher Objektteilgebiete lassen sich als gewichtete Summe von Exponentialfunktionen mit unterschiedlichen Zeitkonstanten modellieren.

Eine im Rahmen dieser Arbeit entwickelte Methode, das zu simulierende Objekt flexibel definieren zu können, ist seine Aufteilung in Quader, deren Größe frei wählbar ist. Ein Quader wird mit Hilfe von Spins unter Berücksichtigung des Abtasttheorems diskretisiert. Ruhemagnetisierung, Relaxationskonstanten und Präzessionskreisfrequenz können funktional vom Ort abhängen, so daß die Eigenschaften der Spins eines Quaders voneinander verschieden sein können. Multiexponentielle Relaxation oder verschiedene Kreisfrequenzen können durch Überlagerung zweier oder mehrerer Quader modelliert werden. Bild 6.1 zeigt beispielhaft für den zweidimensionalen Fall, wie ein Objekt umgesetzt werden kann, in welchem es in einem Teilgebiet zu bi-exponentieller Relaxation kommt.

Der Ansatz, das Objekt in Quader zu gliedern, bietet einerseits maximale Variabilität bei der Konstruktion eines virtuellen Objekts bei hohen Anforderungen an die Komplexität und Strukturiertheit. Andererseits läßt sich durch die Definition der Eigenschaften nur eines einzelnen Quaders schnell und einfach die Grundlage für eine Simulation schaffen, bei der die Objektstruktur eine nur untergeordnete Rolle spielt.

## 6.1.2   Sequenzmodellierung

In Abschnitt 4.1 wurde gezeigt, daß neben Ereignislisten auch die im Rahmen dieser Arbeit entwickelten Elementarsequenzen ein universelles Mittel sind, eine Sequenz abzubilden. Elementarsequenzen können in Form von sequentiellen Matrixmultiplikationen nach Gleichung (4.1) unmittelbar in Programm-Quelltext umgesetzt werden. Ihnen wurde daher in dieser Arbeit der Vorzug vor Ereignislisten gegeben.

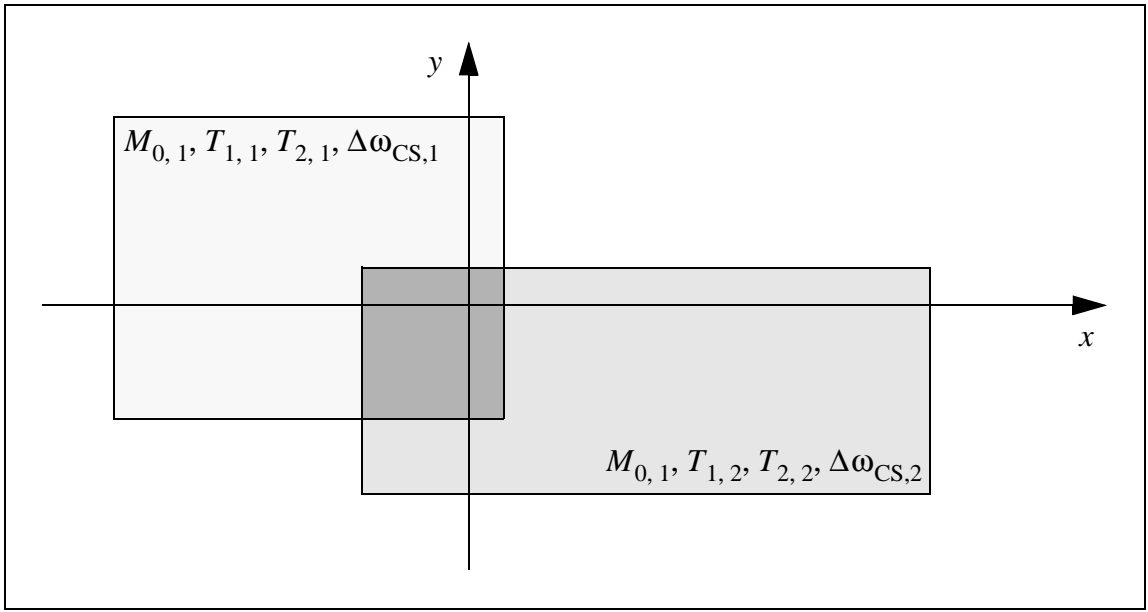

Bild 6.1: Zweidimensionales, aus zwei Rechtecken bestehendes Objekt. Im Überlappungsbereich kommt es zu bi-exponentiellen Relaxationseffekten.

Im realen Experiment ist das von den HF-Empfangsspulen empfangene MR-Signal zeit- und wertkontinuierlich. Für die Weiterverarbeitung in einem Digitalrechner wird es zeitlich abgetastet und mit einem Analog-Digitalwandler wertdiskretisiert. Umgekehrt liegt die Hüllkurve von amplituden- und phasenmodulierten HF-Pulsen in zeitdiskreter Form in einem Digitalrechner vor und wird mit einem Digital-Analog-Wandler in ein zeitkontinuierliches Signal umgesetzt. Für beide Vorgänge muß das Abtasttheorem eingehalten werden. Die hierfür nötigen mathematischen Zusammenhänge sind in Anhang B.2 zusammengefaßt.

### 6.1.3 Modellierung des Bildgebungssystems

#### Statisches Magnetfeld und seine Inhomogenitäten

Die magnetische Flußdichte des statischen Magnetfelds kann aus physikalischen Gründen bei Anordnungen mit endlichen Abmessungen örtlich nicht absolut homogen sein. Sie kann als Summe aus einer konstanten Nominalflußdichte $B_0$ und einer ortsabhängigen Abweichung $\Delta B_0(\vec{x})$ geschrieben werden. Die Präzessionskreisfrequenz der Magnetisierung eines Spins ist damit, zusammen mit Gleichung (6.1),

$$\begin{aligned}
\omega(\vec{x}) &= -\gamma \cdot [B_0 + \Delta B_0(\vec{x}) + \Delta B_{susz}(\vec{x})] + \Delta\omega_{CS}(\vec{x}) \\
&= \omega_0 + \Delta\omega_0(\vec{x}) + \Delta\omega_{susz}(\vec{x}) + \Delta\omega_{CS}(\vec{x}).
\end{aligned}$$

(6.1)

Die Realisierung vereinfacht sich, wenn der virtuelle Experimentaufbau im mit $\omega_{HF}$ rotierenden Koordinatensystem betrachtet wird, vgl. Abschnitt 3.4.4.1. Im rotierenden Koordinatensystem gilt

$$\omega'(\vec{x}) \;=\; \omega(\vec{x}) - \omega_{\mathrm{HF}}$$
$$\;=\; (\omega_0 - \omega_{\mathrm{HF}}) + \Delta\omega_0(\vec{x}) + \Delta\omega_{\mathrm{susz}}(\vec{x}) + \Delta\omega_{\mathrm{CS}}(\vec{x}). \tag{6.2}$$

Meist sind $\omega_0$ und $\omega_{\mathrm{HF}}$ annähernd gleich. Ihre Differenz liegt damit nahe bei Null, und Gleichung (6.2) wird von Magnetfeldinhomogenitäten, Chemical Shift und Suszeptibilität dominiert.

Exemplarisch soll eine realitätsnahe Beziehung für die Inhomogenität $\Delta B_0(\vec{x})$ vorgestellt werden, die im folgenden Kapitel für einige der Beispiele eingesetzt wird. Ausgegangen wird von einem klinischen Ganzkörpertomographiesystem. Das Spulensystem für das statische Magnetfeld ist so gestaltet, daß das Feld innerhalb einer Kugel mit dem Radius $R$ homogen[1] bezüglich Richtung und Betrag der magnetischen Flußdichte ist.

In erster Näherung nimmt der Betrag der Inhomogenität mit dem Abstand $r$ von der Kugelmitte exponentiell zu. Eine genauere Analyse findet sich in [86]. Hier wird die magnetische Feldstärke in Koeffizienten von *sphärischen Harmonischen* entwickelt [25]. Für die Abweichung der magnetischen Flußdichte von der Nominalflußdichte gilt dann

$$\Delta B_0(\vec{x}) \;=\; -C \cdot \left(\frac{r}{R}\right)^{12} \cdot P_{12}(\cos\theta)\,, \tag{6.3}$$

mit dem Abstand $r = |\vec{x}|$ vom Zentrum der Homogenitätskugel, dem Deklinationswinkel bezüglich der $z$-Achse $\theta$ und dem Legendre-Polynom zwölfter Ordnung $P_{12}(x)$ (siehe [17]). Die Größe $C$ ist eine frei wählbare, positive Konstante mit der Dimension 1 Tesla. Bild 6.2 zeigt den Verlauf von $\Delta B_0$ auf der Ebene $z = 0$ in Abhängigkeit von $x$ und $y$.

**Inhomogenitäten der Empfangssensitivität**

Die Simulation einer inhomogenen Empfangssensitivität der für den Empfangsfall vorgesehenen HF-Spulen folgt eng der mathematischen Formulierung in Gleichung (3.25). An die Stelle des Volumenintegrals tritt allerdings eine Summe über das Skalarprodukt aus Sensitivität und Magnetisierungsvektor am Ort jedes Spins. Jeder Magnetisierungsvektor wird somit vor einer linearen Superposition aller Magnetisierungen mit der Spulensensitivität gewichtet.

## 6.2   Gestaltung der Benutzerschnittstelle

Die Zielgruppe für ParSpin setzt sich aus technisch orientierten Forscherinnen und Forschern im Bereich der Magnetresonanz-Tomographie und aus Studentinnen und Studenten techni-

---

[1] Bei heutigen Ganzkörpertomographen sind innerhalb der Homogenitätskugel örtliche Schwankungen der magnetischen Flußdichte im Bereich von 3 ppm der Nominalflußdichte üblich.

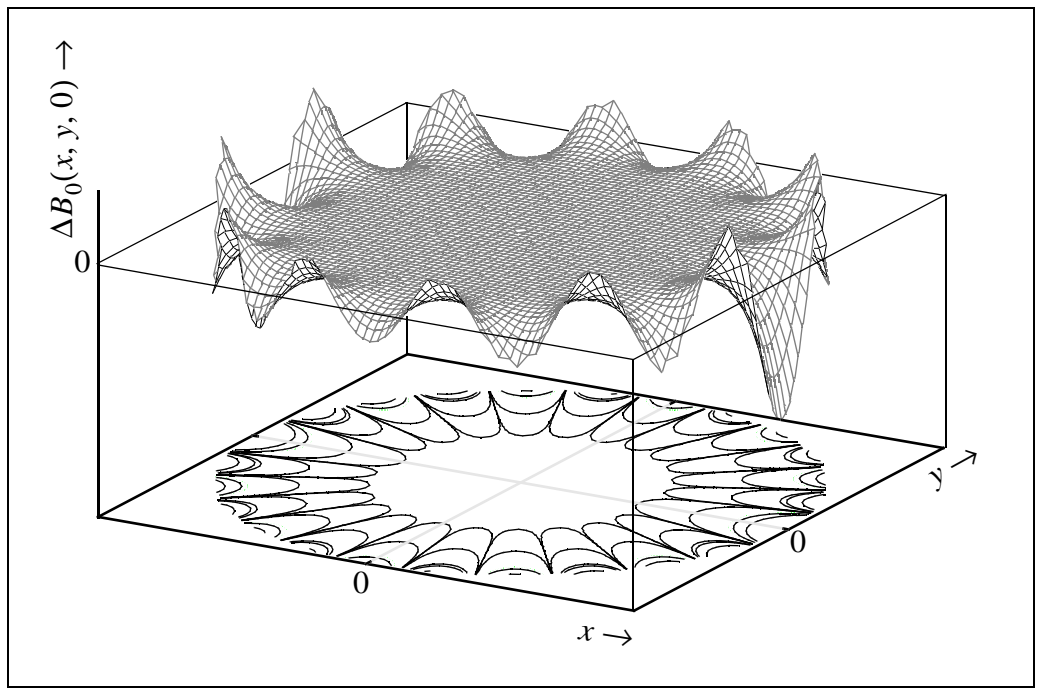

Bild 6.2: Legendre-Inhomogenität nach Gleichung (6.3), bewertet auf der $x$-$y$-Ebene innerhalb eines Homogenitätskreises mit dem Durchmesser 0.5 m.

scher Fachrichtungen zusammen. Anwender in der Forschung stellen dabei die Hauptnutzergruppe dar. Beim Entwurf der Schnittstelle zwischen Benutzer und dem Simulationswerkzeug mußten die folgenden Anforderungen erfüllt werden:

- Erweiterungen des Funktionsumfangs dürfen nur geringe und programmtechnisch unaufwendige Änderungen des Benutzerinterfaces erfordern.

- Die Eingangsdaten eines Simulationslaufs, also die Zahlenwerte für Objekt-, Sequenz- und Systemmodell, müssen gemeinsam mit den Simulationsergebnissen archivierbar und bei Bedarf wiederverwendbar sein.

- Das Interface darf auch bei sehr komplexen Experimentbedingungen, z. B. Sequenzen mit komplizierter K-Raum-Trajektorie, nicht die Simulierbarkeit des Experiments und die Auswertbarkeit des Ergebnisses behindern.

- Mehrere Simulationsläufe sollen sich batchgesteuert nacheinander oder parallel starten lassen, z. B. über Nacht, über ein Wochenende oder für Untersuchungen der Bildgebungsqualität in Abhängigkeit von einem variablen Parameter.

Insbesondere die Archivier- und Wiederverwendbarkeit sprachen für einen pragmatischen Ansatz, bei dem die Experimentbeschreibung in textbasierter Form in der Gestalt von Beschreibungsdateien erfolgt.

Aus Aufwandsgründen wurde das hier beschriebene Simulationswerkzeug als textkonsolenbasiertes Programm realisiert, dessen Betrieb UNIX-typisch mit Kommandozeilenoptionen gesteuert wird, und das die Experimentbedingungen aus lesbaren, veränderbaren Beschreibungsdateien entnimmt. Die vom Benutzer bereitgestellten Beschreibungsdateien müssen hierbei einer spezifischen Syntax genügen, die leicht erlern- und implementierbar ist.

Das Simulationsergebnis wird nach Ende eines Simulationslaufs ebenfalls in Dateien bereitgestellt, die ein sehr einfaches und damit universell weiterverwendbares Format besitzen. Eine Weiterverarbeitung der berechneten Echosignale erfolgt nicht im Simulationswerkzeug selbst, da es keine Kenntnis der beabsichtigten Zuordnung zwischen den Echosignalen im Zeitbereich und dem K-Raum hat.

Ergänzt werden muß das Simulationswerkzeug um Back-End-Software, die aus dem simulierten Echosignal das gewünschte Bild rekonstruiert. Im Rahmen dieser Arbeit wurden auf einige Sequenztypen (Spin-Echo, Gradienten-Echo, Gradienten-Echo-Planar-Imaging) zugeschnittene Auswertungsprogramme konzipiert und realisiert, die mit Hilfe der inversen Fouriertransformation ein Bild erzeugen. Diese Programme stehen dem Anwender im Quelltext zur Modifikation zur Verfügung, erlauben eine Anpassung an die K-Raum-Trajektorie der untersuchten Sequenz und ermöglichen sogar den vollständigen Ersatz der Fouriertransformation durch andere Rekonstruktionsverfahren.

Für sehr komplexe Experimente, z. B. im Bereich der stochastischen Bildgebung [16] oder mit komplizierten Objektstrukturen, reicht die einfache Syntax der Beschreibungsdateien zuweilen nicht aus. Für diese Fälle ist zusätzlich Front-End-Software erforderlich, die die gewünschte Sequenz oder das gewünschte Objekt auf die Primitive der PARSPIN-Beschreibungsdateien umsetzt. Die Wahl der Umgebung, die diese Beschreibungen erzeugt, ist beliebig.

## 6.3   Komponenten des Simulators

Rechenknoten sind eine wichtige Komponente des Simulators nach dem Domain Decomposition Approach (vgl. Abschnitt 5.3). Neben dieser die arithmetischen Operationen einer Simulation durchführenden Komponente muß es zwei weitere Elemente geben, die das korrekte Funktionieren der Rechenknoten ermöglichen.

Erstens ist eine allgemeine Kontrollinstanz erforderlich, die die Gesamtübersicht und -kontrolle über einen Simulationslauf besitzt. Sie erzeugt Blöcke von Spins, weist sie den einzelnen Rechenknoten zu und führt über sie Buch, um zu gewährleisten, daß die Beiträge aller Spinblöcke in das Endergebnis einfließen. Zweitens muß eine zentrale Instanz bereitstehen, die erzeugten Teilsimulationsergebnisse von den Rechenknoten anzunehmen, zu superponieren und sie damit zu einem Gesamtergebnis zusammenzufassen. Weitere Aufgaben existieren, die auf eine der drei Komponenten aufzuteilen sind. So müssen die Beschreibungsdateien des Benutzers gelesen und ausgewertet, Rückmeldungen über den Fortschritt der Simulation erzeugt und das Simulationsergebnis gespeichert werden.

Die drei Komponenten Kontrollinstanz, Rechenknoten und Superpositionsinstanz wurden bei der Implementierung von PARSPIN unmittelbar in die Programmstruktur übernommen. Sie finden sich in den PARSPIN-Komponenten *Master*, *Compute-Slave* und *File-Slave* wieder; diese Bezeichnungen haben ihren Ursprung in der technischen Realisierung mit Hilfe der Bibliothek PVM [31]. Die Aufteilung der Aufgaben auf die Komponenten und der Entwurf des Ablaufs einer Simulation erfolgte nach softwaretechnischen Notwendigkeiten, die hier nicht weiter abgeleitet werden sollen. Die Vorgehensweise wird im folgenden kurz beschrieben.

Wie in Abschnitt 5.3.2 erläutert, verfügen alle Rechenknoten über die vollständige Information bezüglich Sequenz und System. Die Rechenknoten unterscheiden sich insoweit, als ihnen verschiedene Spinblöcke zugewiesen werden. Beim Interpretieren der Sequenzinformation ist es aus Laufzeitgründen sinnvoll, eine Reihe von Spin-unabhängigen Berechnungen durchzuführen, bevor ein Spinblock bearbeitet wird, z. B. die trigonometrischen Funktionen für HF-Pulse in Gleichung (3.19). Da die nötigen Operationen für jeden Knoten bzw. Compute-Slave gleich sind, werden sie bereits vorher im Master durchgeführt.

Den weitaus größten Anteil an der Laufzeit des Simulators haben die Compute-Slaves. Aus diesem Grund muß besonderes Augenmerk auf eine effiziente Implementierung dieser Komponente des Simulators gelegt werden. Folgende Aspekte sind in diesem Zusammenhang von großer Bedeutung:

- innerhalb der drei inneren Schleifen über alle Elementarsequenzen, alle Spins und alle zeitlichen Abtastpunkte werden Funktionsaufrufe vermieden, um moderne Prozessoren mit Pipeline-Architekturen gut nutzen zu können;

- innerhalb der drei inneren Schleifen werden Entscheidungen vermieden, da sie zu unterschiedlichen Ausführungspfaden führen würden. Ein in PARSPIN eingesetzter Ansatz hierfür sind spezialisierte Code-Fragmente, sogenannte *Codelets*, die einem Konzept aus [27] und [28] entlehnt sind;

- die Reihenfolge der drei inneren Schleifen ist so gewählt, daß bei einem neuen Durchlauf durch eine innere Schleife der Prozeßkontext (z. B. Inhalte der Prozessorregister) möglichst wenig Änderungen erfährt und die Anzahl der notwendigen Neuberechnungen gering bleibt.

Der Datenfluß innerhalb von PARSPIN, beginnend bei der benutzerdefinierten Experimentbeschreibung und endend bei der Ausgabe des Gesamtergebnisses, ist in Bild 6.3 veranschaulicht. Zusätzliche Informationen über Aspekte des praktischen Einsatzes finden sich in der PARSPIN-Benutzerdokumentation [26].

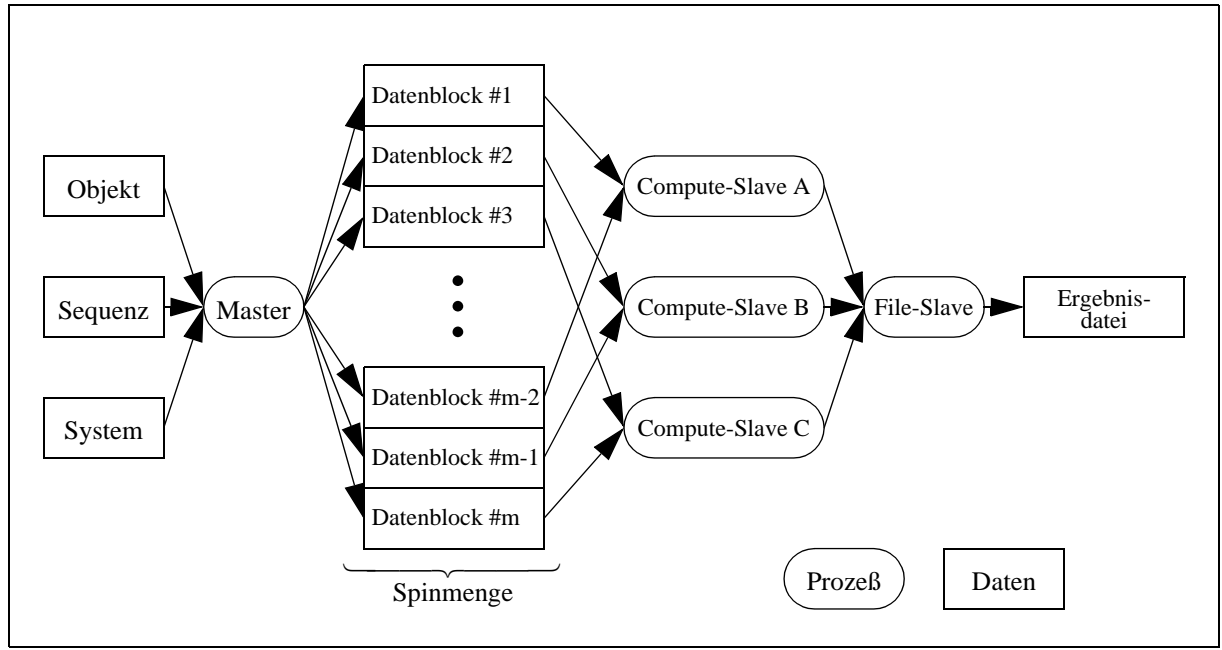

Bild 6.3: Übersicht über den Datenfluß in PARSPIN. Sequenz, System und Objekt enthalten die Experimentbeschreibung; sie wird vom Master ausgewertet. Auf ihrer Grundlage erzeugt er die das Objekt modellierende Spinmenge, teilt sie in Datenblöcke auf und verschickt sie nacheinander an die Compute-Slaves. Diese berechnen das zum Datenblock gehörende Teilergebnis und liefern es beim File-Slave ab. Der File-Slave sammelt alle Teilergebnisse und stellt das Gesamtergebnis für den Benutzer bereit.

# 7 Simulationsergebnisse

Dieses Kapitel demonstriert einige der Einsatzmöglichkeiten und Anwendungen des MR-Simulators PARSPIN. Durch eine Vielzahl von Simulationsbeispielen wird die mögliche Einsatzbandbreite des Simulators verdeutlicht, auch wenn bei weitem nicht alle Anwendungsmöglichkeiten gezeigt werden können. Das Kapitel wird durch das nachfolgende ergänzt. Dort werden zusätzlich Simulations- mit Meßergebnissen verglichen.

Hier werden zunächst Ergebnisse von Bildgebungsexperimenten präsentiert, die aus der von PARSPIN erzeugten Echoinformation abgeleitet werden. PARSPIN ermöglicht über die Bereitstellung von Echoinformation hinaus, zu jedem beliebigen Zeitraum einer Sequenz auf die Magnetisierungsvektoren der Spins zugreifen zu können. Simulationsbeispiele, die diese Fähigkeit ausnutzen, werden im Anschluß an die gezeigten Bildgebungsergebnisse präsentiert.

Ein wichtiger Punkt bei der Entwicklung von PARSPIN war die Optimierung seiner Rechengeschwindigkeit durch eine skalierbare, parallelisierte Architektur und durch eine effiziente Formulierung der arithmetischen Anweisungen. In einem weiteren Abschnitt dieses Kapitels werden die Laufzeiteigenschaften des Simulators anhand von Laufzeitmessungen analysiert. Die mit PARSPIN erzielbare Simulationsgeschwindigkeit wird anhand eines Vergleichs seiner Laufzeiten mit denen anderer Simulatoren bewertet.

## 7.1 Erzeugung und Auswertung von MR-Signalen

Die weitaus häufigste Anwendung von PARSPIN ist die Simulation von Bildgebungssequenzen. Hierzu berechnet PARSPIN anhand des vom Benutzer definierten Experiments Echoinformation, d.h. den Verlauf der Magnetisierung bzw. der Empfangsspulenspannung über der Zeit. Aus den berechneten Echosignalen wird anschließend mit einem geeigneten, i. a. sequenzabhängigen Verfahren ein Bild rekonstruiert. In vielen Fällen ist eine inverse Fourier-Transformation gut zur Bildrekonstruktion geeignet.

### 7.1.1 Das Shepp-Logan-Phantom

Zur Bewertung und zum Vergleich von Bildrekonstruktionsverfahren in der auf Röntgenstrahlung basierenden Computertomographie (CT) entwickelten Shepp und Logan in [89] ein virtuelles Phantom als Grundlage für Simulationsstudien. Es bildet einen aus zehn Ellipsen zusammengesetzten, axialen Schnitt durch den menschlichen Kopf nach, siehe Bild 7.1.

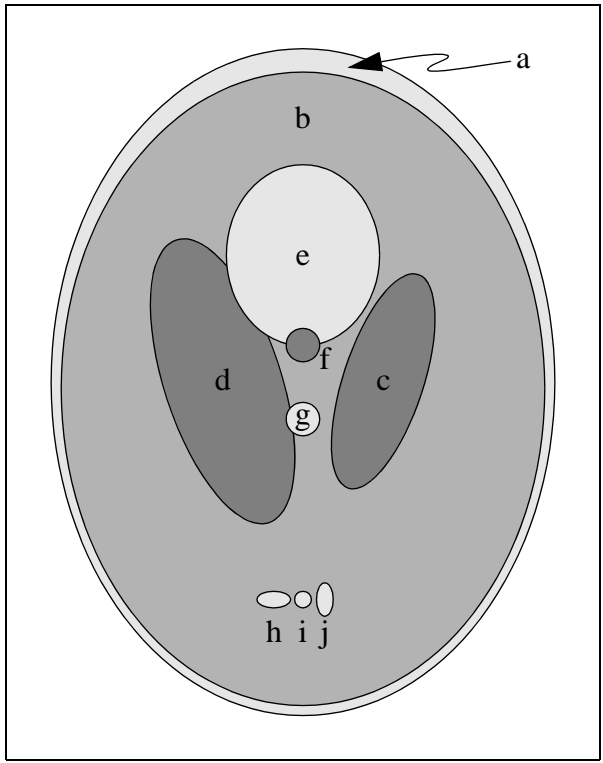

| Ellipse | $M_0$ | $T_1/\text{s}$ | $T_2/\text{s}$ |
|---------|-------|-------|-------|
| a | 1.00 | 1.0 | 0.2 |
| b | 0.51 | 1.0 | 0.2 |
| c | 0.40 | 1.0 | 0.2 |
| d | 0.40 | 1.0 | 0.2 |
| e | 0.60 | 1.0 | 0.2 |
| f | 0.80 | 1.0 | 0.2 |
| g | 0.80 | 1.0 | 0.2 |
| h | 0.70 | 1.0 | 0.2 |
| i | 0.70 | 1.0 | 0.2 |
| j | 0.70 | 1.0 | 0.2 |

Bild 7.1: Skizze des im folgenden
verwendeten Shepp-Logan-Phantoms

Tabelle 7.1:  MR-spezifische Parameter
des Shepp-Logan-Phantoms.

Beim Einsatz in der Magnetresonanz-Tomographie werden den vorgegebenen Arealen MR-spezifische Eigenschaften wie Ruhemagnetisierung/Spindichte und Relaxationszeiten zugeordnet. Die in den nachfolgenden Beispielen für zweidimensionale Bildgebungsverfahren benutzten, den einzelnen Ellipsen zugeordneten Parameter sind in Tabelle 7.1 zusammengefaßt.

### 7.1.2   Spin-Echo-Experimente

Spin-Echo-Experimente zählen zu den in der Praxis am häufigsten eingesetzten Bildgebungssequenzen. Ihre Struktur wurde bereits in Abschnitt 4.1 (Seite 33 ff.) beschrieben. Im vorliegenden Abschnitt werden Simulationsergebnisse gezeigt, die einen Eindruck von dem Einfluß von Inhomogenitäten des statischen Magnetfelds auf das Bild bei zweidimensionalen Spin-Echo-Sequenzen vermitteln. Außerdem wird das Ergebnis der Simulation einer dreidimensionalen Spin-Echo-Sequenz gezeigt.

**Zweidimensionale Sequenz**

Bei einem Spin-Echo-Experiment werden die entstehenden Echos zeilenweise in die K-Raum-Matrix einsortiert (Bild 7.2a). Mit einer inversen Fourier-Transformation wird anschließend das rekonstruierte Bild erzeugt (Bild 7.2b). Bei idealen Voraussetzungen, also homogenen statischen und hochfrequenten Magnetfeldern und Gradientenmagnetfeldern mit linearer Ortsab-

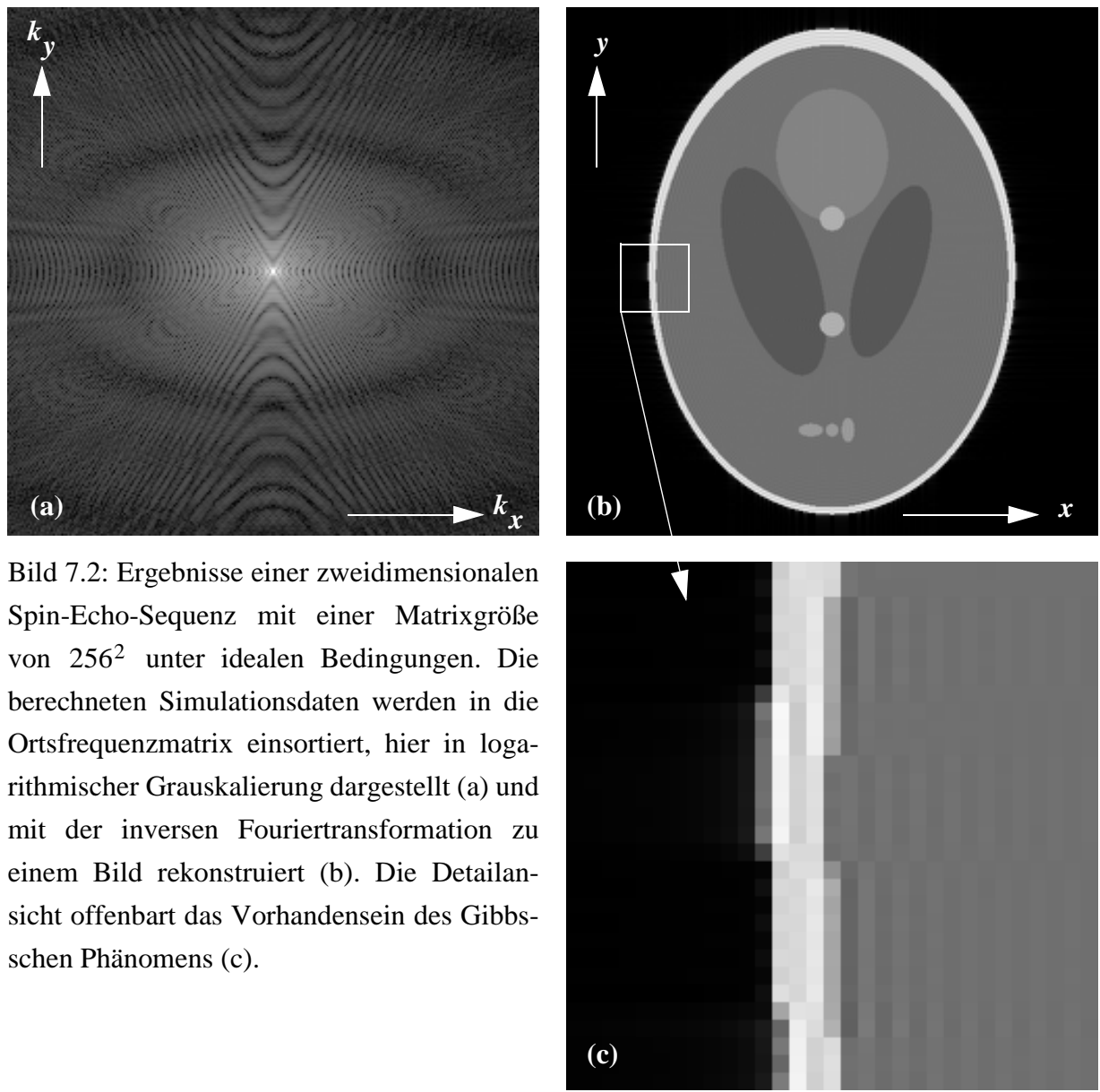

Bild 7.2: Ergebnisse einer zweidimensionalen Spin-Echo-Sequenz mit einer Matrixgröße von $256^2$ unter idealen Bedingungen. Die berechneten Simulationsdaten werden in die Ortsfrequenzmatrix einsortiert, hier in logarithmischer Grauskalierung dargestellt (a) und mit der inversen Fouriertransformation zu einem Bild rekonstruiert (b). Die Detailansicht offenbart das Vorhandensein des Gibbsschen Phänomens (c).

hängigkeit, entsteht ein artefaktarmes Bild. Erst die Detailansicht (Bild 7.2c) offenbart die erwarteten Überschwinger der Pixelintensität an Objektkanten (Gibbssches Phänomen), die durch zeitliche Fensterung des unendlich ausgedehnten Echos entstehen. Die verwendeten Experimentparameter sind in Tabelle 7.2 (Seite 79) zusammengefaßt.

Inhomogenitäten des statischen Magnetfelds führen im Vergleich zum ungestörten Experiment zu geometrischen Verzerrungen des Bildes in Bereichen mit großer Inhomogenität (Bild 7.3), also vor allem am Rand des kugelförmigen Homogenitätsbereichs, vgl. Bild 6.2 (Seite 71). Für das in Bild 7.3 gezeigte Ergebnis wurde die zuvor eingesetzte Sequenz nach Tabelle 7.2 und zusätzlich eine Inhomogenität nach Gleichung (6.3) mit $C = 20 \, \mu T$ verwendet.

Bei einer sehr langen Einwirkdauer der Magnetfeldinhomogenitäten, d.h. bei großen Echozeiten, entstehen im rekonstruierten Bild sogenannte *Hot-Spots*. Hierbei handelt es sich um sehr

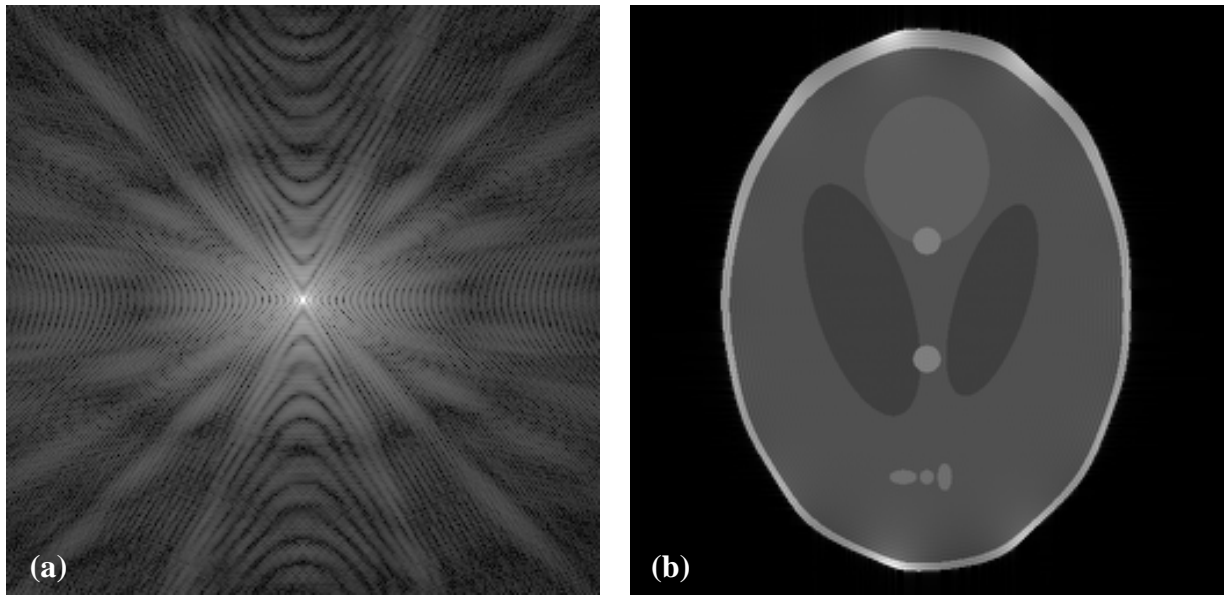

Bild 7.3: K-Raum (a) und rekonstruiertes Bild (b), erzeugt von einer Spin-Echo-Sequenz unter Einfluß von Hauptmagnetfeldinhomogenitäten.

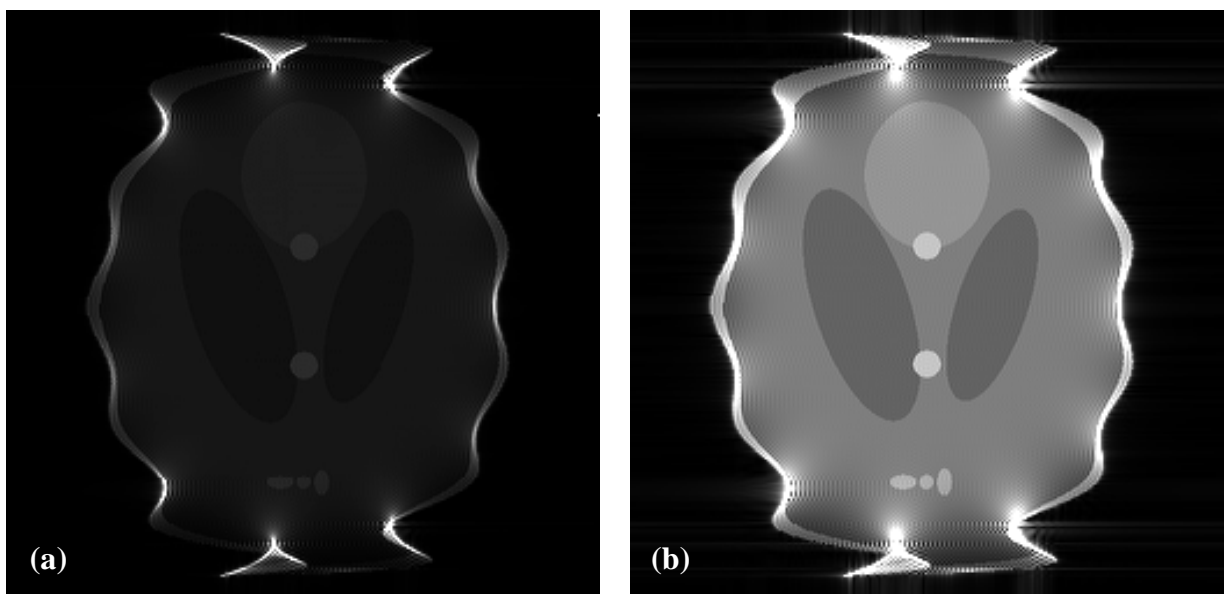

Bild 7.4: Bei sehr langer Einwirkdauer der Magnetfeldinhomogenitäten entstehen im Bild sogenannte Hot-Spots (a). Erst mit geeigneter Fensterung des Grauwertbereichs ist das Objekt gut zu erkennen (b), wenn auch geometrisch stark verzerrt.

helle Bildbereiche, die aufgrund ihrer Intensität das Nutzbild überstrahlen, siehe Bild 7.4a. Erst durch eine geeignete Fensterung des Grauwertebereichs wird das Nutzbild sichtbar, auch wenn es geometrisch stark verzerrt ist, siehe Bild 7.4b. Tabelle 7.3 zeigt die dazugehörigen Experimentparameter.

| Parameter | Wert |
|---|---|
| FOV | 0.5 m |
| Matrixgröße | $256^2$ |
| maximaler Auslesegradient | $0.239 \, \frac{\text{mT}}{\text{m}}$ |
| Echozeit $T_E$ | 50 ms |

Tabelle 7.2: Experimentparameter zu den Bildern 7.2 und 7.3

| Parameter | Wert |
|---|---|
| FOV | 0.5 m |
| Matrixgröße | $256^2$ |
| maximaler Auslesegradient | $0.024 \, \frac{\text{mT}}{\text{m}}$ |
| $T_E$ | 500 ms |

Tabelle 7.3: Experimentparameter zu Bild 7.4

**Dreidimensionale Sequenz**

Im Falle von dreidimensionalen Sequenzen erfolgt die Visualisierung schichtweise zweidimensional oder z. B. mit einer Ray-Tracing-Software [105] in pseudo-dreidimensionaler Darstellung. Das in Bild 7.5 gezeigte Beispiel ist das rekonstruierte Bild einer 3D-Spin-Echo-Sequenz mit einer Matrixgröße von $128^2$ in der $x$-$y$-Ebene und 64 in $z$-Richtung. Tabelle 7.4 gibt eine Übersicht über die verwendeten Experimentparameter.

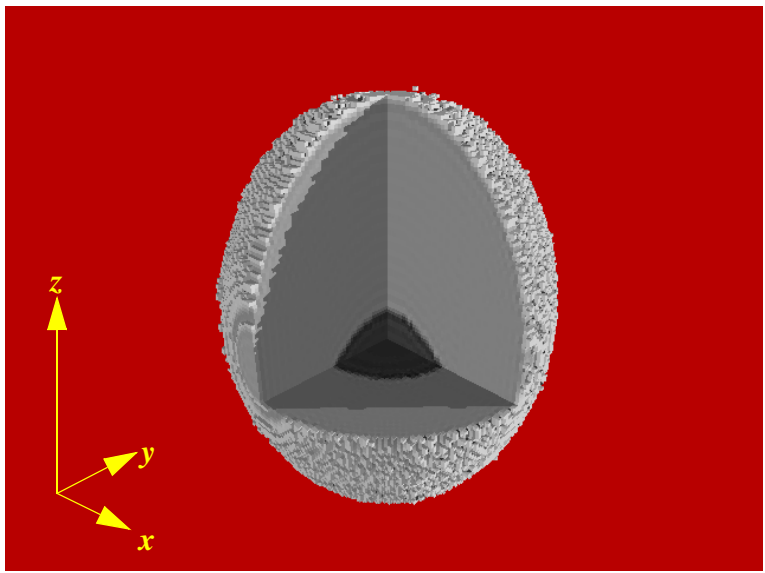

Bild 7.5: Mit einer 3D-Spin-Echo-Sequenz erzeugtes Bild. Nach der Rekonstruktion wurde zu Visualisierungszwecken ein Quadrant ausgeschnitten.

| Parameter | Wert |
|---|---|
| FOV | 0.5 m |
| Matrixgröße | $128^2 \times 64$ |
| maximaler Auslesegradient | $0.119 \, \frac{\text{mT}}{\text{m}}$ |
| $T_E$ | 50 ms |

Tabelle 7.4: Experimentparameter zu Bild 7.5

### 7.1.3   Turbo-Spin-Echo-Experimente

Bei Turbo-Spin-Echo-Sequenzen (TSE) wird das Spin-Ensemble, wie bei normalen Spin-Echo-Sequenzen, mit einem 90°-HF-Puls angeregt, aber es wird durch wiederholte Verwendung von 180°-HF-Pulsen mehr als ein Echo pro Anregung gewonnen. Die Anzahl der Echos pro Anregung ist ein Maß für die Meßzeitreduktion und wird *Turbo-Faktor* (TF) genannt.

Bei einem Turbo-Faktor von nur $TF = 2$ und einer sequentiellen Einsortierung der Echos in den K-Raum in der Reihenfolge ihres Entstehens kommt es zu einer Amplitudenmodulation der K-Raum-Zeilen mit einer zweizeiligen Periode. Ursache dafür ist, daß das zweite aufgenommene Echo wegen $T_2$-Relaxation eine geringere Amplitude als das erste hat. Im Bildbereich führt dies zu dem sogenannten *N/2-Ghost*, einem um die halbe Spaltenhöhe verschobenen Geisterartefakt mit geringerer Intensität als das Originalbild, siehe Bild 7.6. Tabelle 7.5 enthält die dazugehörigen Experimentparameter.

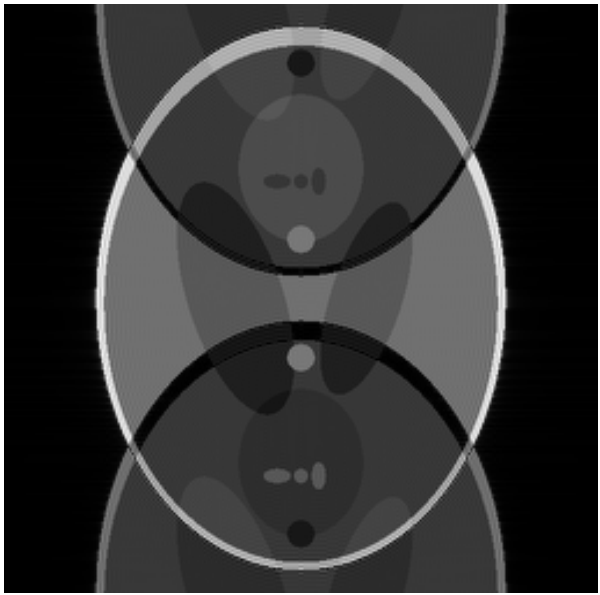

| Parameter | Wert |
|---|---|
| FOV | 0.5 m |
| Matrixgröße | $256^2$ |
| maximaler Auslesegradient | $0.060 \; \frac{\text{mT}}{\text{m}}$ |
| $T_E$ des ersten Echos | 310 ms |
| $T_E$ des zweiten Echos | 520 ms |

Bild 7.6: Artefaktbehaftetes Bild einer Turbo-Spin-Echo-Sequenz mit einem Turbo-Faktor von 2 und sequentieller K-Raum-Sortierung.

Tabelle 7.5:  Experimentparameter der TSE-Sequenz zu Bild 7.6

### 7.1.4   Gradienten-Echo-Planar-Imaging

Gradienten-Echo-Planar-Imaging-Sequenzen (Gradient-EPI) unterscheiden sich von TSE-Sequenzen durch eine andere Art der Magnetisierungsrefokussierung. Während bei TSE-Sequenzen HF-Pulse mit dem Flipwinkel 180° für die Echobildung sorgen, wird dies bei Gradient-EPI durch einen Vorzeichenwechsel des Auslesegradienten erreicht. Die K-Raum-Trajektorie während der Datenakquisition verläuft nun nicht mehr zeilenweise immer in eine Richtung, sondern mäanderförmig mit einem Richtungswechsel nach jeder akquirierten Zeile.

Gradienten-EPI-Sequenzen können i. a. schneller einen vollständigen Satz von K-Raum-Zeilen erzeugen, als dies TSE-Sequenzen vermögen. Sie sind jedoch anfälliger gegen die Auswirkungen von Hauptmagnetfeldinhomogenitäten, wie in Abschnitt 4.3.3 beim Vergleich der Bilder 4.7 und 4.8 (Seite 45) diskutiert wurde. Bild 7.7 zeigt das Simulationsergebnis einer derartigen Sequenz, Tabelle 7.6 gibt die wesentlichen Sequenzparameter wieder.

Der Vergleich der K-Raum-Matrix aus Bild 7.7a mit der des idealen Spin-Echo-Experiments aus Bild 7.2a zeigt deutliche „Verwaschungen" für größere Werte

| Parameter | Wert |
|---|---|
| FOV | 0.5 m |
| Matrixgröße | $128^2$ |
| maximaler Auslesegradient | $5.962 \, \frac{\mathrm{mT}}{\mathrm{m}}$ |
| Effektive Echozeit $T_E$ | 67.7 ms |

Tabelle 7.6: Experimentparameter zu Bild 7.7

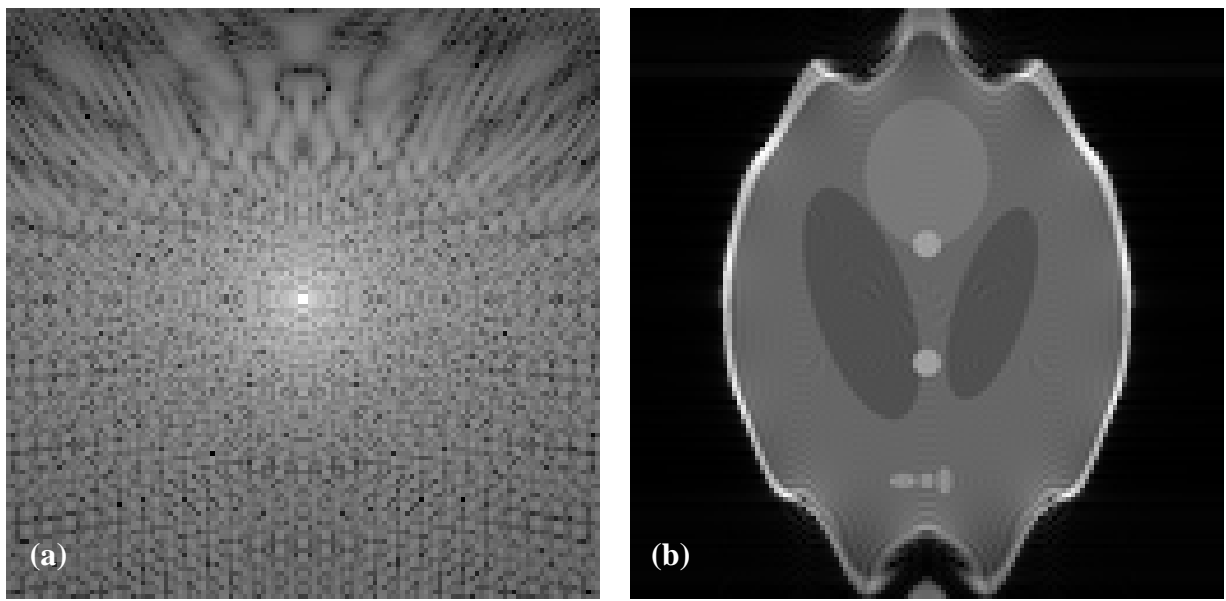

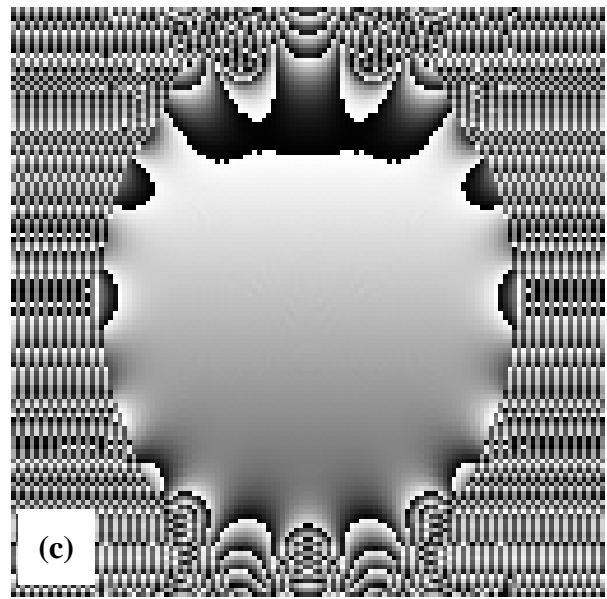

Bild 7.7: Simulationsergebnisse eines Gradienten-EPI-Experiments. Gezeigt sind K-Raum-Matrix in logarithmischer Grauwertskalierung (a) sowie Betrag (b) und Phase (c) des rekonstruierten Bildes.

von $k_y$; dies korrespondiert mit Bild 4.8. Resultierend aus den akkumulierten, im Gegensatz zu Spin-Echo-Experimenten nie rephasierten Phasenfehlern, zeigen sich im Ortsraum deutliche geometrische Verzerrungen. Auch im Fall der Spin-Echo-Sequenz mit großem $T_E$ (Bild 7.4) ist es zu starken geometrischen Verzerrungen gekommen; diese unterscheiden sich jedoch qualitativ von denjenigen im EPI-Fall. Daneben ist die effektive Echozeit beim EPI-Experiment zwar kürzer, der Einfluß der Magnetfeldinhomogenitäten auf das Bild jedoch wesentlich größer.

Das Phasenbild 7.7c zeigt große qualitative Ähnlichkeit mit der Auftragung der Inhomogenität des Magnetfelds in Bild 6.2. Dies liegt darin begründet, daß bei Gradienten-Echo-Sequenzen die Phase im Ortsraum, abgesehen von dem unvermeidlichen Phase-Wrapping, proportional zur Inhomogenität $\Delta B_0$ ist.

### 7.1.5 Ultraschnelle Multi-Puls-Sequenzen

Von großer Bedeutung in der medizinischen Anwendung sind Sequenzen, die in sehr kurzer Zeit ein Bild des interessierenden Objektausschnitts erzeugen können. Durch Weiterentwicklungen im Hardware-Bereich ist das Echo Planar Imaging auf aktuellen MR-Systemen möglich geworden und kommt i. a. für schnelle (im Zehntelsekundenbereich) und sehr schnelle (im Hundertstelsekundenbereich) Experimente zum Einsatz.

Ein alternativer Ansatz für ultraschnelle Bildgebung mit geringen Hardware-Anforderungen sind Multipulssequenzen. Sie regen mit einer Folge von hochfrequenten Pulsen mit kleinen Flipwinkeln das untersuchte Objekt an. Durch geeignete Steuerung der Gradientenmagnetfelder werden die von der Pulsfolge erzeugten transversalen Konfigurationen so im K-Raum angeordnet, daß mit einem zeitlich konstanten Gradientenmagnetfeld eine K-Raum-Matrix vollständig gefüllt werden kann. Ein Beispiel für eine derartige Sequenz ist die in Abschnitt 4.4 genannte PREVIEW-Sequenz.

Ein anderes, in der Rekonstruktion verglichen mit PREVIEW weniger problematisches Bildgebungsverfahren [32] ist OUFIS (Optimized Ultra-Fast Imaging Sequence) [106]. In der OUFIS-Sequenz erfolgt die Anregung mit 64 Anregungspulsen. Mit einem HF-Puls mit dem Flipwinkel 180° werden transversale Konfigurationen, analog zur Spin-Echo-Sequenz, am K-Raum-Ursprung gespiegelt, und schließlich die erzeugten Echos aufgezeichnet.

Allen Multipulssequenzen ist gemein, daß sich einige transversale Konfigurationen in Ausleserichtung $k_x$ sehr weit vom Koordinatenursprung entfernen. So werden die 64 von OUFIS generierten Konfigurationen jeweils in dem Abstand $2 \cdot k_{x,\,max}$ zueinander angeordnet, der nach Gleichung (B.2) (Seite 125) für die systemtheoretisch korrekte Abbildung des FOV erforderlich ist. Die äußere transversale Konfiguration erreicht dadurch im K-Raum eine maximale Position von $K_{x,\,max} = 64 \cdot 2 \cdot k_{x,\,max}$. Bedingt durch diesen großen Wert muß nach Glei-

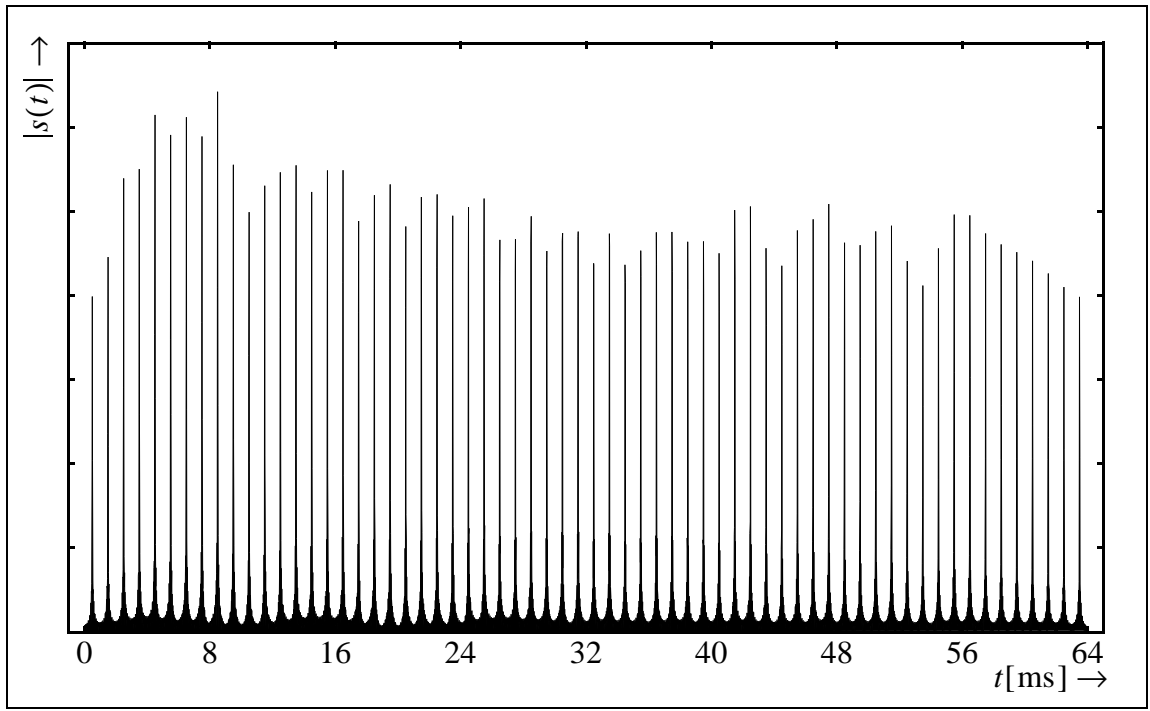

Bild 7.8: Folge der 64 ohne Phasenkodierung erzeugten Echos einer OUFIS-Sequenz nach [106] mit einer Matrixgröße von $64^2$ Punkten.

chung (5.2) der Abstand zwischen den Spins in $x$-Richtung sehr klein gewählt werden; dies führt zu einer hohen Anzahl von Spins. Hier ist die parallele Architektur von PARSPIN besonders nützlich und sorgt für eine verglichen mit anderen Arbeiten stark verringerte Gesamtlaufzeit.

Die von OUFIS erzeugten 64 Echos unterscheiden sich in ihrer maximalen Amplitude und ihrer Phase voneinander. Beim Einsortieren der Echos in die Zeilen der K-Raum-Matrix führt dies zu einer zeilenweisen Modulation, die, bleibt sie unkorrigiert, nach einer inversen Fourier-Transformation zu erheblichen Artefakten im Bild führt. Bild 7.8 zeigt die von PARSPIN simulierte Folge der 64 OUFIS-Echos bei einem nicht-phasenkodierten Experiment. Die Variation der Amplituden ist deutlich erkennbar. Die Grafik stimmt sehr gut mit experimentellen und theoretischen Ergebnissen aus [106] und [32] überein.

Die Rekonstruktion des Bildes erfordert eine Korrektur der phasenkodierten Daten. Dazu wird eine nicht-phasenkodierte Kalibrationsmessung eingesetzt [10] [11]. Ein von [18] abgeleitetes Verfahren berechnet aus der Kalibrationsmessung für jede Zeile eine Faltungsfunktion, mit der die phasenkodierten Daten korrigiert werden können. Die so korrigierte K-Raum-Matrix und das aus ihr rekonstruierte Bild sind in Abbildung 7.9 gezeigt.

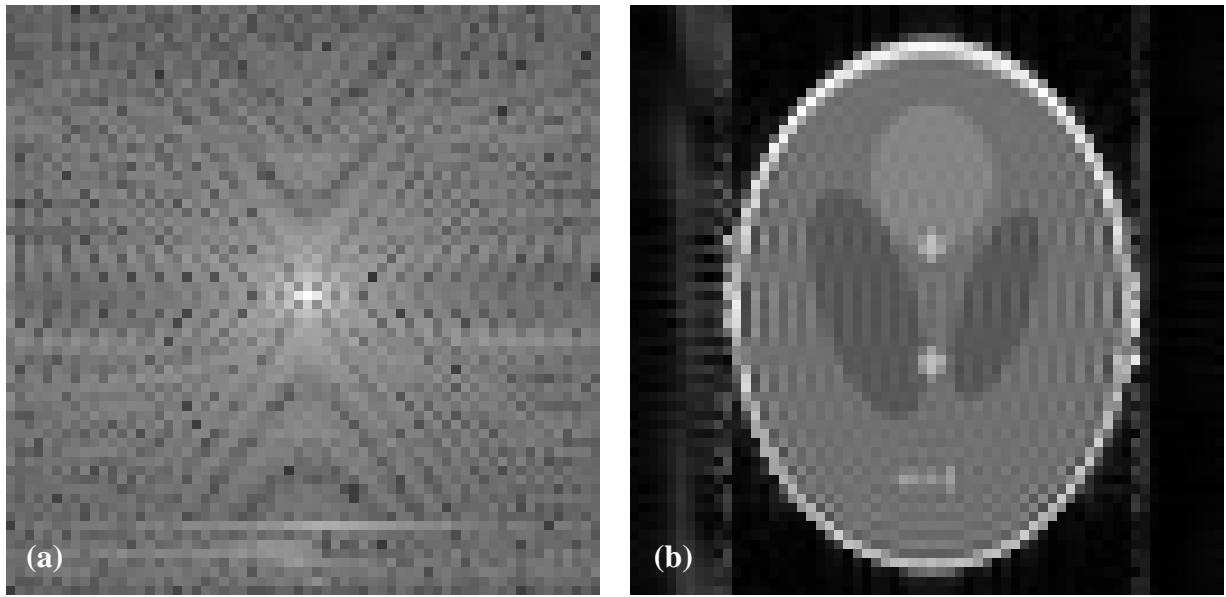

Bild 7.9: Betrag von K-Raum-Matrix (a) und Bildmatrix (b) eines phasenkodierten OUFIS-Experiments. Die K-Raum-Daten wurden vor Visualisierung und inverser Fourier-Transformation einer zeilenweisen Korrektur unterzogen.

Die Gesamtdauer des simulierten Experiments betrug hier 129 ms. Die Echos wurden im zeitlichen Abstand von 1 ms angeordnet; für ihre Aufzeichnung war somit eine Zeit von 64 ms notwendig. Die Flipwinkel der Anregungspulse lagen bei jeweils 11.25°, und ihre Phasenwinkel entsprachen denen einer Frank-Sequenz [8][63] der Länge 64. Weitere Experimentparameter sind in Tabelle 7.7 zusammengefaßt.

| Parameter | Wert |
|---|---|
| FOV | 0.5 m |
| Matrixgröße | $64^2$ |
| maximaler Auslesegradient | $3.0\ \dfrac{\mathrm{mT}}{\mathrm{m}}$ |
| Effektive Echozeit $T_E$ | 96 ms |

Tabelle 7.7:  Experimentparameter zu den Bildern 7.8 und 7.9

### 7.1.6   Inhomogenitäten des HF-Empfangsfelds

Zur Verbesserung der Bildqualität bei der Untersuchung oberflächennaher Strukturen werden häufig Empfangsspulen für das hochfrequente Magnetfeld eingesetzt, die von der Sendespule getrennt ausgeführt sind. Derartige Spulen haben i. a. die Größe des zu untersuchenden Areals und werden in seiner unmittelbaren Nähe angebracht. In einigen Anwendungen werden mehrere derartiger Spulen zu Spulen-Arrays gekoppelt [46].

Empfangsspulen für den oberflächennahen Bereich spielen nicht nur in der medizinischen Anwendung eine wichtige Rolle, sondern insbesondere in der DFG-Forschergruppe „NMR-Oberflächentomographie", an der der Lehrstuhl für Allgemeine Elektrotechnik und Datenverarbeitungssysteme der RWTH Aachen beteiligt ist. Durch die geringe Größe der Empfangs-spulen ist ihre Empfangssensitivität innerhalb des FOV inhärent inhomogen. Die Inhomogenität äußert sich im rekonstruierten Bild unmittelbar in einer ortsabhängigen Gewichtung der Pixelintensität.

Zur Demonstration wurde die Empfangssensitivität einer kreisrunden Spule mit dem Durchmesser $D = 150$ mm berechnet und bei der Simulation einer Spin-Echo-Sequenz verwendet. Bildgebungsebene war die $y$-$z$-Ebene, wobei das statische Magnetfeld parallel zur $z$-Achse gerichtet war. Matrixgröße war $128^2$, und die Echozeit war $T_E = 50$ ms. Das FOV war nahezu vollständig von dem simulierten Objekt ausgefüllt. Die Anordnung von Spule und Bildgebungsebene sowie das rekonstruierte Bild sind in Abbildung 7.10 gezeigt.

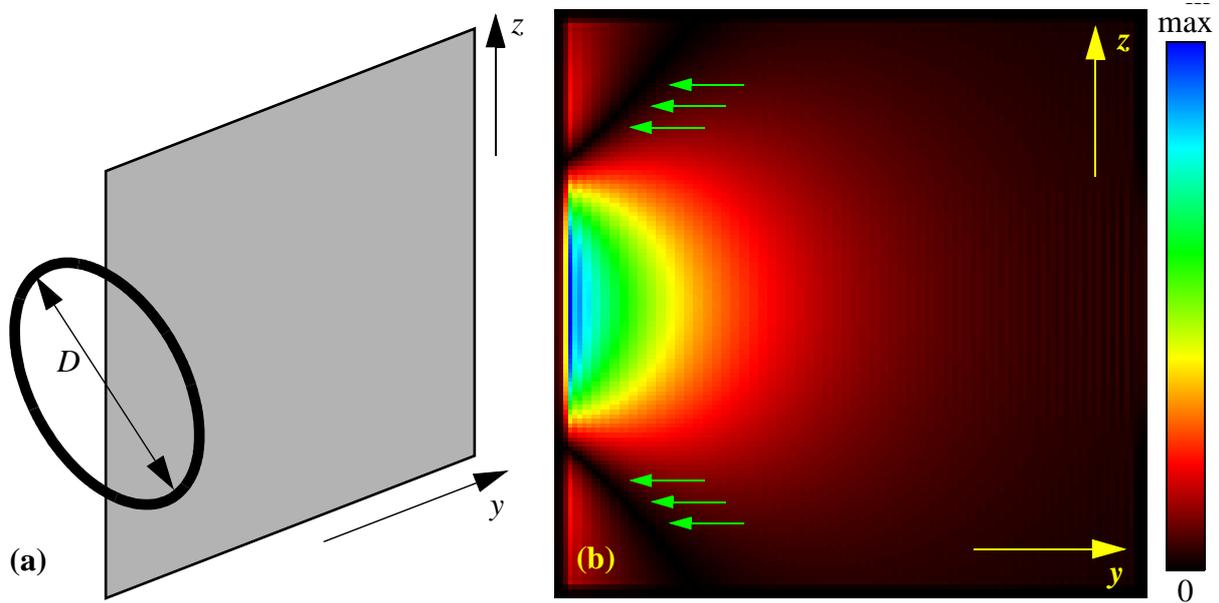

Bild 7.10: Mit einer Spin-Echo-Sequenz wird eine Schicht eines homogenen Objekts in der $y$-$z$-Ebene homogen angeregt. Eine kreisrunde Empfangsspule liegt im Raum parallel zur x-y-Ebene. (a). Das entstehende Bild zeigt das charakteristische Empfangsprofil der Spule (b).

Deutlich ist zu erkennen, daß die Intensität des Bildes in der Nähe der Spule am stärksten ist, von dort also am meisten Signalanteile aufgenommen werden. In $y$-Richtung fällt die Pixelintensität allmählich ab. Auffällig sind dunkle, im Bild mit grünen Pfeilen gekennzeichnete Bereiche, in denen kein Signal detektiert wird. In diesen Arealen sind die Vektoren der magnetischen Flußdichte der Spule parallel zur longitudinalen Magnetisierung gerichtet, die sich zeitlich nur langsam durch Spin-Gitter-Relaxation ändert. Nach Gleichung (3.25) tragen sie daher nicht zum Signal bei.

## 7.2    Erzeugung und Auswertung von Magnetisierungs- information

Völlig andere Visualisierungsmethoden bieten sich an, wenn PARSPIN anstelle von Echodaten Magnetisierungsdaten generiert. Hiermit hat der Benutzer Zugriff auf die Magnetisierungsvektoren aller Spins des Objekts zu beliebigen Zeiten. Wichtige Grundbegriffe der MR-Bildgebung lassen sich damit visualisieren, was insbesondere in der Ausbildung von Bedeutung ist.

### 7.2.1    Funktionsweise schichtselektiver HF-Pulse

Als erstes Beispiel dient ein amplitudenmodulierter HF-Puls mit einer rechteckgefensterten si-Funktion[1] als Hüllkurve. Ein solcher Puls wirkt zusammen mit einem Gradientenmagnetfeld schichtselektiv in Richtung des Gradienten. Die Steilheit des Gradienten bestimmt die Dicke der Schicht. Mit Hilfe von Animationen kann die Entwicklung des Schichtprofils genau verfolgt werden. Es zeigt sich, daß sich das gewünschte Schichtprofil nicht gleichmäßig aufbaut, sondern während des Pulses Schwankungen unterworfen ist. Bild 7.11 zeigt Einzelbilder aus der Animation des Betrags des Selektionsprofils zu den im Timing-Diagramm eingezeichneten Zeiten $t_1$ bis $t_6$.

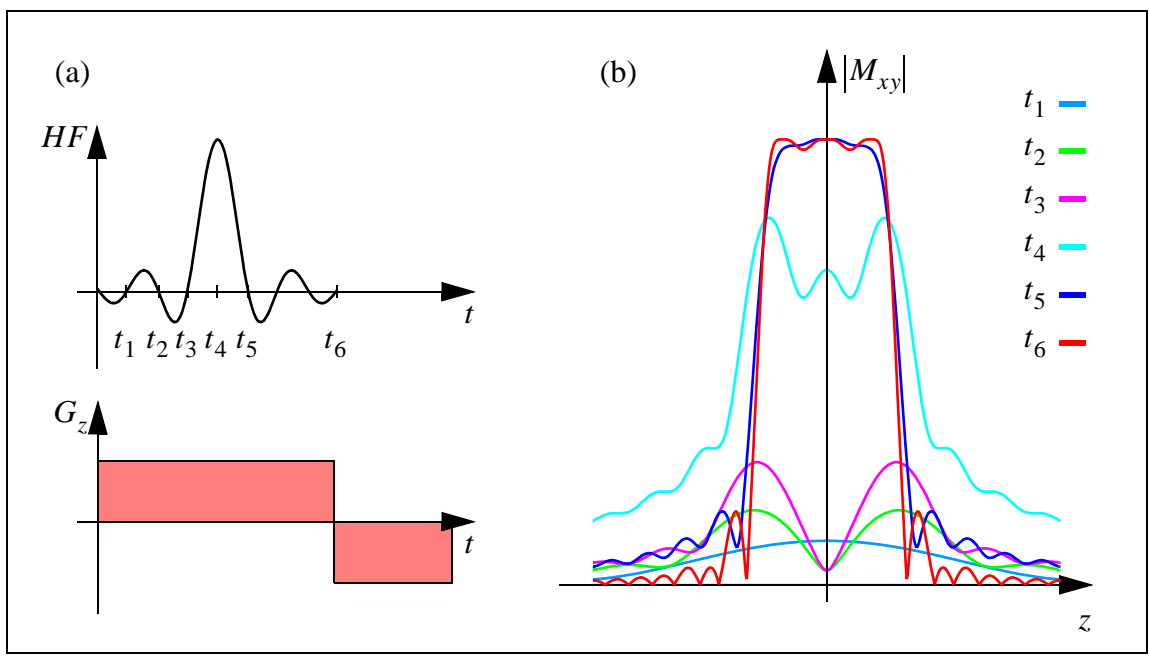

Bild 7.11: Sequenztiming (a) und Einzelbilder einer Animation der zeitlichen Entwicklung des Schichtprofils (b), das von einem amplitudenmodulierten HF-Puls erzeugt wird.

---

[1] $\mathrm{si}(x) = \sin(x)/x$

## 7.2.2   Entstehung des Eight-Ball-Echos

Ein wichtiges Experiment in der Magnetresonanz-Bildgebung und -Spektroskopie ist das STEAM-Experiment [83] (Stimulated Echo Acquisition Mode). Bei ihm können mit drei HF-Pulsen mit gleichen Flipwinkeln bis zu vier Echos erzeugt werden. Werden die Flipwinkel zu je 90° gewählt, tritt zwischen dem ersten und dem zweiten HF-Puls das bereits von Hahn [37] beschriebene *Eight-Ball-Echo* auf, bei dem die Magnetisierungsvektoren des Objekts die Ziffer Acht auf einer Kugel beschreiben.

Die Entwicklung des Eight-Ball-Echos und anderer von Multipulsanregungen stammenden Echos kann anhand der Blochschen Gleichung nur mit Mühe anschaulich nachvollzogen werden. Animationen erleichtern hier das Verständnis erheblich. Zur Demonstration wurde ein an die STEAM-Sequenz angelehntes Experiment verwendet, das zwei Echos erzeugt. Das Timing-Diagramm der Sequenz ist in Bild 7.12 zu sehen.

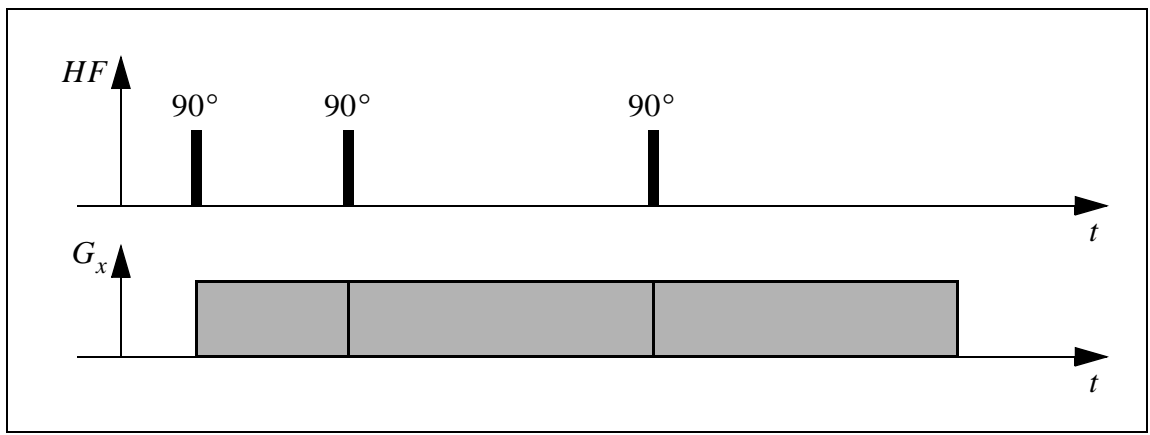

Bild 7.12: Timing-Diagramm einer Drei-Puls-Sequenz zur Demonstration der Entstehung des Eight-Ball-Echos.

Für die Erzeugung von Animationen der Spinbewegung wurden die von PARSPIN berechneten Magnetisierungsdaten mit einem Ray-Tracing-Programm [105] in Einzelbilder umgesetzt. Diese können anschließend mit einem Animationsprogramm betrachtet werden. Bild 7.13 zeigt einige Einzelbilder der Animation der Demonstrationssequenz als Bildfolge.

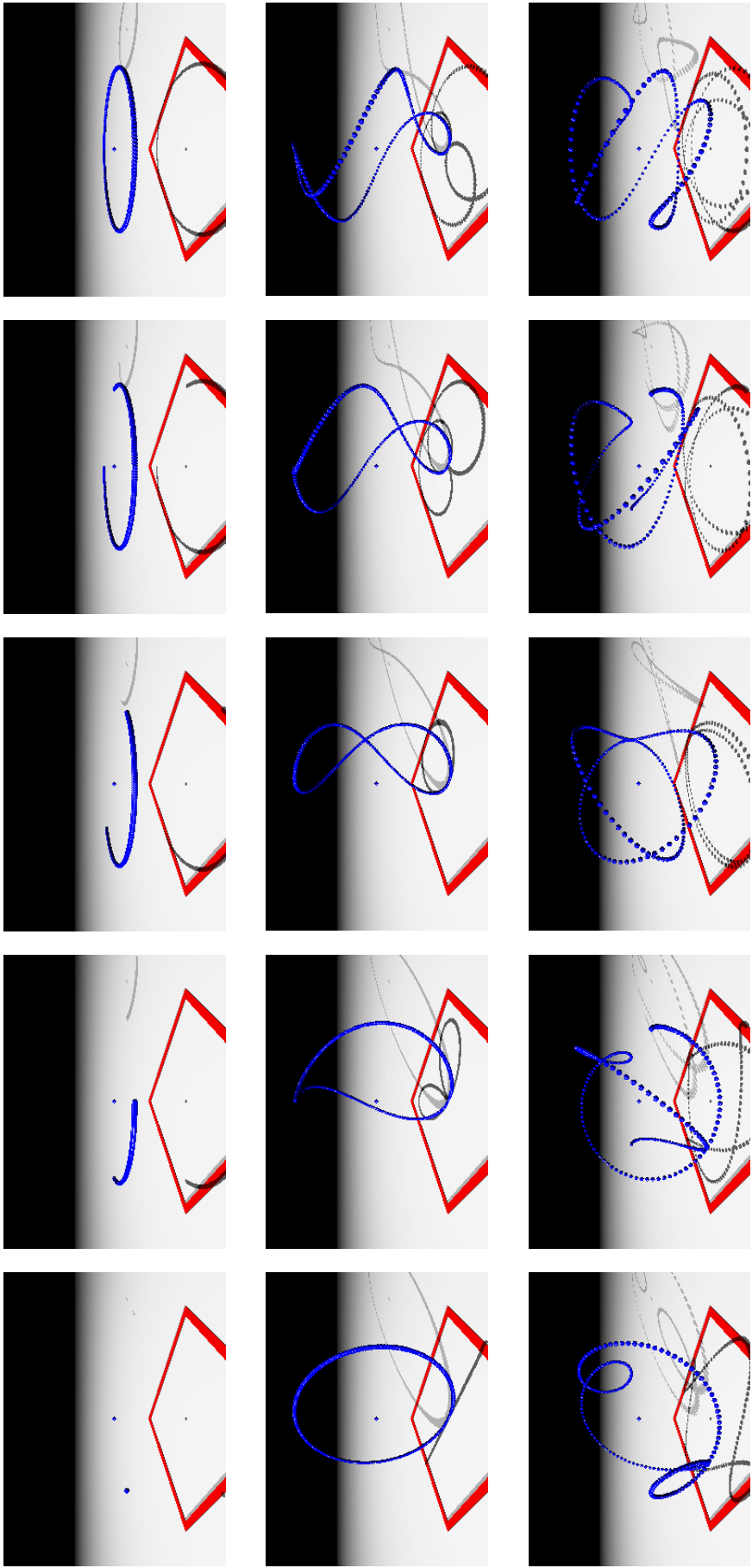

Bild 7.13: Einzelbilder aus der Animation der Magnetisierungsvektoren während der Demonstrationssequenz aus Bild 7.12. Vor jeder Bildreihe wurde ein HF-Puls eingestrahlt. Das mittlere Bild zeigt die Magnetisierungskonstellation des Eight-Ball-Echos.

# 7.3    Leistungsmessungen

Bei der Implementierung von PARSPIN war neben einem detaillierten physikalischen Modell auch die Optimierung des Laufzeitverhaltens von Bedeutung. Mit einem parallelverarbeitenden Ansatz wurde die verfügbare Rechenleistung gebündelt, um die Dauer einer Simulation gegenüber einer Einprozessorlösung zu verringern. Gegenstand dieses Abschnitts sind Analyse und Bewertung des erzielten Laufzeitvorteils sowie eine Einordnung der Leistungsfähigkeit in Bezug auf Arbeiten anderer Gruppen.

## 7.3.1    Leistungssteigerung durch Parallelisierung

Bei der parallelisierten Implementierung eines Algorithmus' ist es wichtig beurteilen zu können, wie gut die Qualität der Parallelisierung ist, wie groß also die mit mehreren Prozessoren erreichbare Laufzeitverringerung ist. Als Leistungskriterium zur Bewertung der Rechenleistung von PARSPIN wird der Spindurchsatz $\eta$ herangezogen. Er ist definiert durch die für ein Experiment benötigte Spinanzahl $N_{spins}$ bezogen auf die Zeit $T_{sim}$, die der Master vom Starten des Programms durch den Benutzer bis zum Programmende benötigt, und es gilt

$$\eta = \frac{N_{spins}}{T_{sim}}. \tag{7.1}$$

Der Spindurchsatz hängt parametrisch von der simulierten Sequenz und den Eigenschaften des simulierten MR-Systems ab. Da die Laufzeitumgebung moderner Computersysteme die Ausführung arithmetischer Operationen abhängig von den Zahlenwerten der Operanden beschleunigen kann, ist $\eta$ i. a. zusätzlich von der speziellen Wahl der Zahlenwerte einer Sequenz oder eines Systems abhängig. Unter idealen Bedingungen, also bei vollständiger Auslastung jedes Prozessors und identischer Leistungsfähigkeit, kann erwartet werden, daß $\eta$ linear mit der Anzahl der verfügbaren Prozessoren steigt.

Zur Verifikation wurde eine Simulation einer Turbo-Spin-Echo-Sequenz mit einem Shepp-Logan-Phantom und einer Legendre-Inhomogenität mehrfach mit verschiedenen Prozessoranzahlen ausgeführt. Die Simulation umfaßte 770 Elementarsequenzen und etwa 130 000 Spins. Verwendet wurde eine heterogene virtuelle Maschine, die aus bis zu 23 Sun-Ultra-Workstations mit unterschiedlicher Leistungsfähigkeit und Sekundärbelastung bestand. Die verfügbaren Rechner weisen ohne Sekundärlast einen Leistungsunterschied auf, der sich in Rechenzeitunterschieden von bis zu einem Faktor zwei äußert.

Die für jede Prozessoranzahl ermittelten Spindurchsätze wurden mit einer Regressionsgeraden approximiert und mit der theoretisch möglichen Steigerung des Spindurchsatzes verglichen. Spindurchsatzmeßwerte, Regressions- und Vergleichsgerade sind gemeinsam in Bild 7.14 gezeigt. Es ist zu erkennen, daß die gemessenen Werte für den Spindurchsatz sehr gut durch eine Gerade angenähert werden kann. Beim Übergang von 3 auf 23 Prozessoren wird eine Steigerung des Spindurchsatzes um den Faktor 6.3 erreicht. Bei gleicher Leistungsfähigkeit und Sekundärbelastung der zur virtuellen Maschine zusammengefaßten Workstations wäre theoretisch eine Steigerung um den Faktor 23 / 3 = 7.7 möglich. Dieser Idealfall ist in der Grafik durch eine Vergleichsgerade kenntlich gemacht

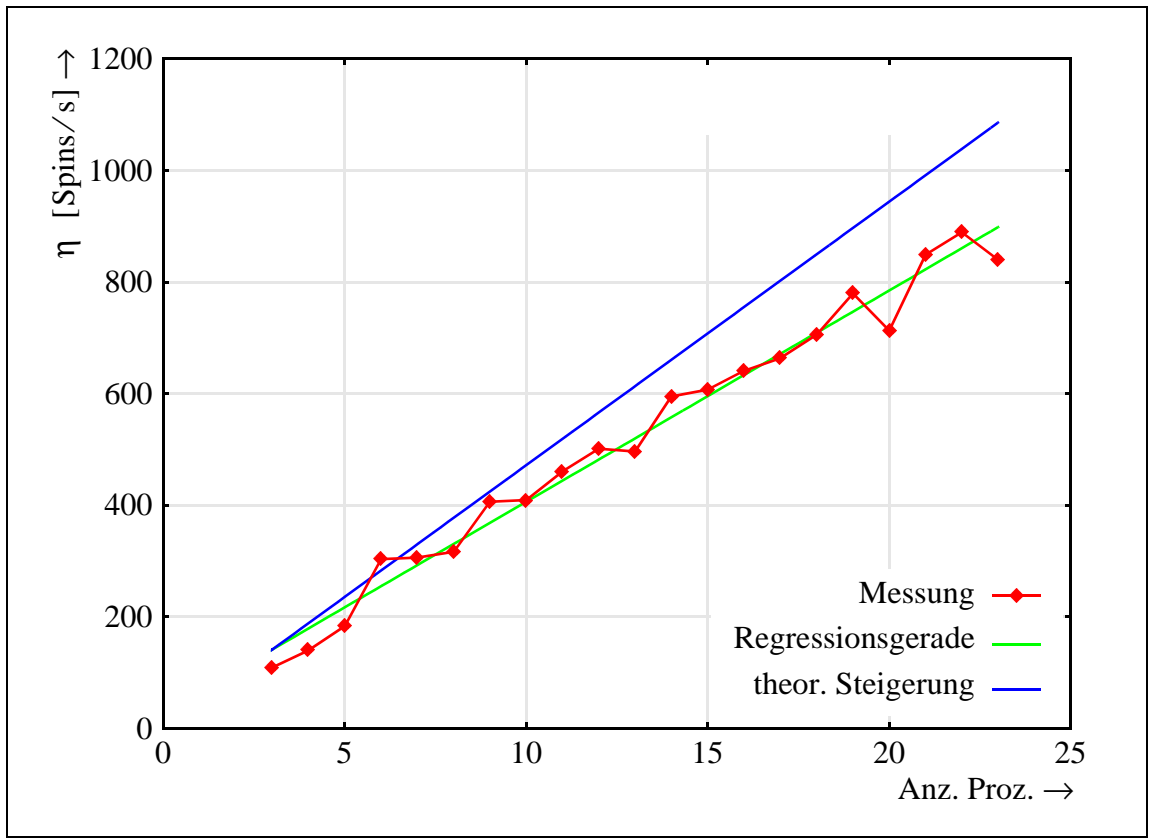

Bild 7.14: Vergleich von gemessener mit theoretisch erwarteter Steigerung des Spin­durchsatzes in Abhängigkeit von der Anzahl der Prozessoren.

Mit Hilfe der zum PVM-Paket [30] gehörenden Analysewerkzeuge konnten zwei Faktoren ausgemacht werden, die die mögliche Leistungssteigerung auf Werte unterhalb des theoreti­schen Optimums begrenzen. Der erste Faktor ist die unterschiedliche Leistungsfähigkeit der beteiligten Workstations, die im arithmetischen Mittel die Performance der gesamten virtuel­len Maschine senkt. Der zweite Faktor ist eine Gegenkopplung zwischen Master und File­Slave, die hin und wieder zu Wartezeiten bei den Compute-Slaves führt. Eine offenbar nur geringe Rolle spielt Overhead durch die Kommunikation der Programmkomponenten unter­einander.

Die – beabsichtigte – Gegenkopplung dient der Stabilität des Programms und sorgt dafür, daß die Eingangsspeicher der den File-Slave beherbergenden Workstation nicht an ihre Kapazitäts­grenze kommen. Dazu wird der Master über das erfolgreiche Bearbeiten eines Spinblocks erst dann unterrichtet, wenn der File-Slave ihn empfangen und sein Teilergebnis in das Gesamter-gebnis integriert hat. Ein Compute-Slave muß daher i. a. nach dem Ende der Berechnung eines Teilergebnisses bis zur Zuteilung eines neuen Spinblocks warten, bis der File-Slave das Teilergebnis ebenfalls bearbeitet hat. Insgesamt ist die knotenabhängige Steigerung dennoch mehr als befriedigend, vor allem im Vergleich mit dem ähnlichen Ansatz in [69], siehe Abschnitt 7.3.3.

## 7.3.2   Leistungssteigerung im Verlauf der Entwicklung

Die Entwicklung des Simulators PARSPIN erfolgte anfangs in der Programmiersprache Fortran 90, um durch Kompatibilität der Quelltexte aus anderen Projekten der Gruppe Medizinelektronik Nutzen ziehen zu können. Die entwicklungsunterstützenden Werkzeuge für Fortran 90 zur Fehlersuche und zur Analyse des Laufzeitverhaltens auf Instruktionsebene sind allerdings für Fortran 90 bei weitem nicht so ausgereift und vollständig wie für andere Sprachen, z. B. ANSI-C oder C++. Zudem lassen sich betriebssystemspezifische Funktionen unter Fortran 90 oft nur über den Umweg einer C-Unterroutine erreichen. PARSPIN wurde daher nach ANSI-C portiert und anschließend weiterentwickelt. Optimierungen des Laufzeitverhaltens erfolgten nach und nach im Verlauf der Weiterentwicklung.

| Experiment | Matrixgröße | Anzahl Spins | Anzahl Elementarsequenzen | Simulationsdauer | | | |
|---|---|---|---|---|---|---|---|
| | | | | 1.2.3 | 1.3.0 | 1.3.1 | 1.5.0 |
| 2D TSE | $128^2$ | 10 000 | 512 | 301 s | 45 s | 23 s | 18 s |
| 2D Radial | $128^2$ | 10 000 | 384 | 226 s | 41 s | 20 s | 13 s |
| 2D EPI | $128^2$ | 10 000 | 257 | 438 s | 38 s | 19 s | 11 s |
| 3D Spin-Echo | $128^2 \cdot 64$ | 176 000 | 24 576 | 184 h | 24 h | 4 h | 4 h |

Tabelle 7.8:  Verbesserung des Laufzeitverhaltens mit der PARSPIN-Version

Anhand von vier Experimenten wurde die Leistungssteigerung im Verlauf der Entwicklung regelmäßig überprüft. Typische Laufzeiten auf einer virtuellen Maschine mit fünf Workstations der Firma Sun Microsystems (Ultra-1 170) sind in Tabelle 7.8 gegenübergestellt. Die dort genannten PARSPIN-Versionsnummern sind folgendermaßen einzuordnen:

- 1.2.3: Programmiersprache Fortran 90 unter einem frühen, nicht optimierenden Compiler der Firma NAG;

- 1.3.0: Programmiersprache Fortran 90 unter einem optimierenden, nativen Compiler der Firma Sun Microsystems;

- 1.3.1: Programmiersprache ANSI-C unter dem C-Compiler von Sun Microsystems;

- 1.5.0: Programmiersprache ANSI-C unter dem C-Compiler von Sun Microsystems, zusätzliche Laufzeitoptimierung mit Codelets.

Mit jeder Änderung der Versionsnummer ging eine wesentliche Erweiterung der Funktionalität einher. Durch Optimierungen des Programms, hauptsächlich in den inneren Programmschleifen, und den Wechsel der Entwicklungsbasis konnte zusätzlich das Laufzeitverhalten deutlich und kontinuierlich verbessert werden. Simulationsdauern für Standardexperimente von wenigen zehn Sekunden sind bereits mit geringem Hardware-Aufwand erreichbar.

## 7.3.3    Vergleich mit anderen Arbeiten

In der Literatur zu anderen Simulationswerkzeugen spielt die Analyse des Laufzeitverhaltens eine nur untergeordnete Rolle. Meist wird für ein Simulationsbeispiel die Simulationszeit angegeben, die auf einem nicht immer präzise spezifizierten Rechnersystem benötigt wurde. Die Komplexität der simulierten Sequenz variiert hierbei, genauso wie Größe und Beschaffenheit des simulierten Objekts. Zudem entwickeln sich Computersysteme rasch fort, so daß der Vergleich mit einem älteren Aufsatz oft einen Vorteil zugunsten neuerer Ergebnisse ergibt, der lediglich aus der Entwicklung der Prozessortechnik resultiert.

Am ausführlichsten ist die Beschreibung des Laufzeitverhaltens in der Arbeit von Michiels et. al. [69]. Da bei dem dort beschriebenen Simulationswerkzeug ein ähnlicher, parallelisierter Ansatz verfolgt wurde wie bei PARSPIN, ist der detaillierte Vergleich beider Arbeiten am fruchtbarsten und nimmt hier den größten Raum ein. In einem weiteren Teil werden die Laufzeiten einiger Simulationen gegenübergestellt und mit PARSPIN verglichen.

### Vergleich mit einem spinbasierten, parallelisierten Werkzeug

Die in [69] dokumentierten Laufzeitmessungen wurden auf Computersystemen der Firma IBM vom Typ RISC System/6000 3AT unter dem UNIX-Derivat AIX durchgeführt. Ein gleichartiges System stand für die Vermessung von PARSPIN nicht zur Verfügung. Stattdessen wurden Laufzeitmessungen auf drei verschiedenen Systemen durchgeführt, einer Ultra-1 170 und einer Ultra-10 300 der Firma Sun Microsystems unter Sun Solaris 2.5 und 2.6 sowie einem handelsüblichen PC mit einem Intel-Prozessor 200 MMX unter Linux (Kernel 2.0.36).

Als Maß für die Leistungsfähigkeit dieser Systeme wurden die von der SPEC-Organisation unter [93] veröffentlichten Kennwerte für die Gleitkommarechenleistung von Standardsystemen herangezogen. Speziell diente der Kennwert `SPECfp_base95` als Grundlage für einen Vergleich. Er steht für alle vier betrachteten Systeme zur Verfügung. Bild 7.15 vergleicht die Verhältnisse der auf den vier Systemen zu erwartenden Rechenzeiten eines gedachten, gleitkommaintensiven Programms. Als Norm wurde ein Sun-Ultra-1-System gewählt; das Programm benötigt hier per definitionem 100 % der Norm-Rechenzeit.

In [69] wurde eine Gradient-Echo-Sequenz mit einer Bildmatrix von $256^2$ simuliert. Mit PAR-SPIN wurde zum Vergleich eine Spin-Echo-Sequenz mit ähnlicher Komplexität nachgebildet. Die Anzahl der Spins $N_{\text{spin}}$ in einem Objekt wurde zwischen $10^3$ und $10^5$ variiert. Bild 7.16 zeigt die Laufzeiten beider Werkzeuge bei Abarbeitung auf einem Prozessor für verschiedene $N_{\text{spin}}$ und – im Fall von PARSPIN – auf verschiedenen Computersystemen. Beide Werkzeuge skalieren im betrachteten Bereich von $N_{\text{spin}}$ linear mit der Anzahl der Spins. Deutlich ist zu erkennen, daß die Laufzeit des Werkzeugs aus [69] immer mindestens eine Größenordnung über der des langsamsten PARSPIN-Systems, des Pentium-Systems, liegt, obwohl der Vergleich der SPEC-Kennwerte einen deutlichen Vorteil des IBM-Systems gegenüber dem Pentium-System erwarten läßt. Auf den für PARSPIN genutzten Systemen spiegelt der Laufzeitvergleich die Rangfolge des SPEC-Benchmarks wieder.

Der große Laufzeitunterschied zwischen beiden Werkzeugen könnte darin begründet sein, daß das in [69] simulierte Experiment deutlich komplexer ist als angenommen. Es deutet jedoch nichts darauf hin, daß in [69] ein Effekt berücksichtigt, in der PARSPIN-Simulation aber vernachlässigt wurde, der einen Rechenzeitunterschied von einer Größenordnung erklärt. Eine weitere Interpretationsmöglichkeit ist, daß der über mehrere Anwendungen gemittelte SPEC-Kennwert die Leistungsfähigkeit des IBM-Systems hinsichtlich der vorliegenden Applikation zu hoch bewertet. Eine Laufzeitdifferenz von mehr als einer Größenordnung läßt sich jedoch auch unter Berücksichtigung beider Faktoren zusammen nicht erklären.

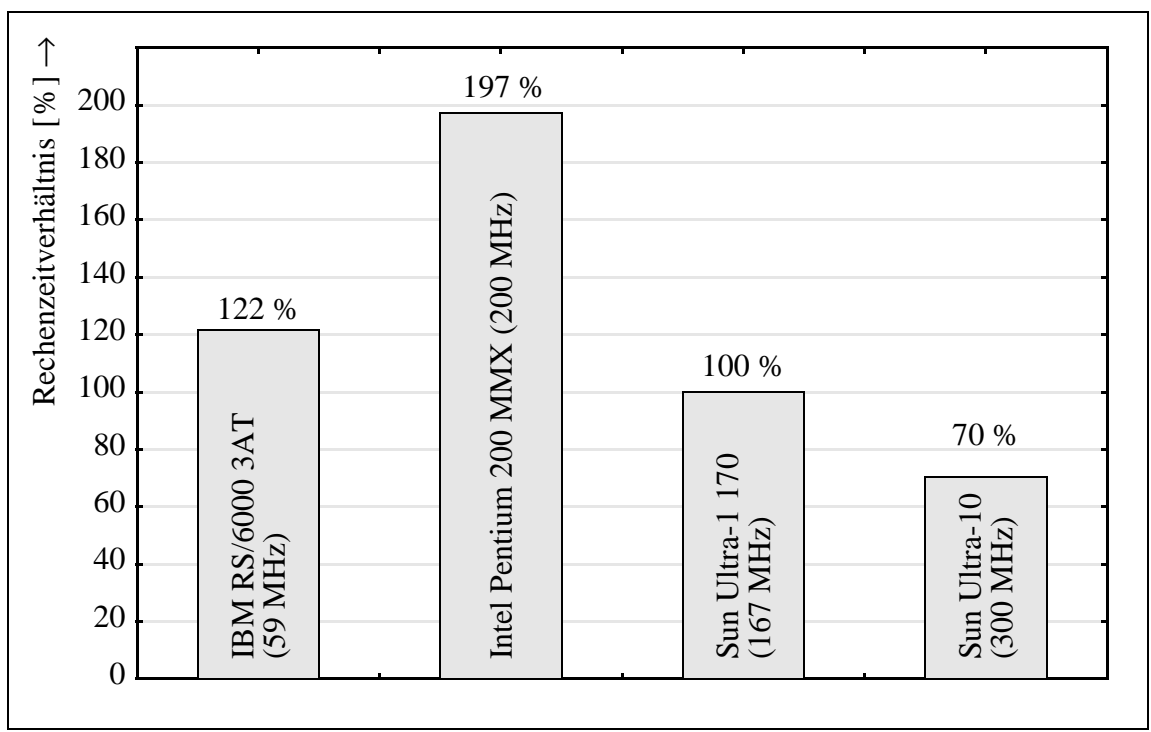

Bild 7.15: Aufgrund der `SPECfp_base95`-Kennwerte verschiedener Systeme zu erwartender Rechenzeitunterschied eines gleitkommaintensiven Programms, normiert auf eine Sun Ultra-1 170.

Als Hauptursache für die beobachtete Laufzeitdifferenz wird gesehen, daß die konsequente Eliminierung von überflüssigen Anweisungen mit Hilfe von Codelets, die zur Laufzeit an den Bedarf jeder Elementarsequenz angepaßt werden, zu dem für PARSPIN günstigen Vergleichsergebnis geführt hat. Zusätzlich könnte sich günstig ausgewirkt haben, daß in der auf ANSI-C basierenden Implementierung PARSPINs kein für objektorientierte Sprachen typischer Overhead auftritt.

Neben den Leistungsmeßergebnissen im Single-Prozessor-Betrieb ist in [69] auch die Laufzeitreduktion bei einer parallelisierten Abarbeitung auf sieben statt auf einer Workstation dokumentiert: die Laufzeit vermindert sich um den Faktor 4. Die mit PARSPIN erreichbare Steigerung des Spindurchsatzes für ein vergleichbares Experiment wurde in Abschnitt 7.3.1 gezeigt. Aus der Regressionsgeraden in Bild 7.14 läßt sich bei einer Erhöhung der Knotenzahl um den Faktor 7, also z. B. beim Übergang von 3 auf 21 Knoten, eine Erhöhung des Spindurchsatzes um den Faktor 5.8 ablesen.

Der Overhead durch die parallelisierte Abarbeitung ist offenbar bei PARSPIN geringer als in dem Werkzeug aus [69]. Ein Grund hierfür ist, daß in PARSPIN die spezialisierte, für ihre Aufgabe optimierte Bibliothek PVM eingesetzt wurde, während in [69] die Operationen für die Parallelisierung mit Standardaufrufen von Betriebssystemfunktionen realisiert wurden.

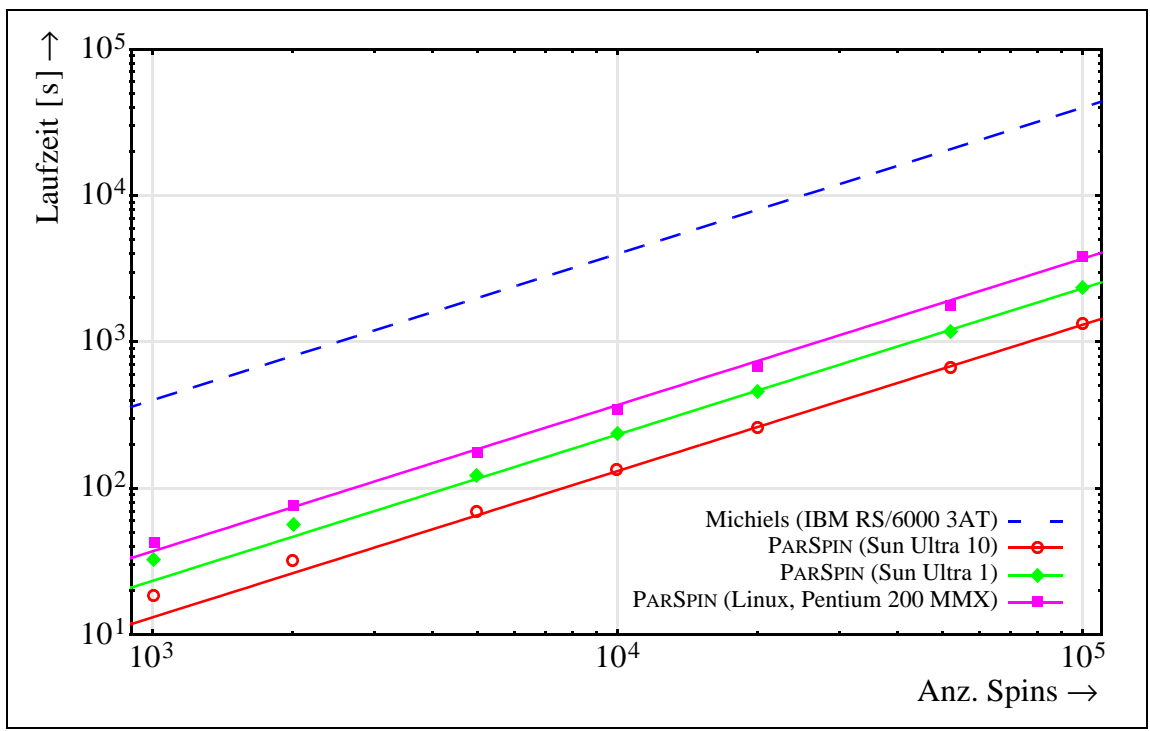

Bild 7.16: Abhängigkeit der Laufzeit von der Spin-Anzahl bei dem Simulationswerkzeug aus [69] und bei PARSPIN, jeweils gemessen im Single-Prozessor-Betrieb.

**Vergleich mit weiteren Werkzeugen**

Für einen Rechenzeitvergleich mit PARSPIN wurden einige Arbeiten neueren Datums ausgewählt, in denen Angaben zur Laufzeit einer Simulation zusammen mit dem simulierten Experiment zu finden sind. Eine Übersicht über die in den Arbeiten eingesetzten Computersysteme und die simulierten MR-Experimente ist in Tabelle 7.9 gegeben. Leider sind in den meisten der genannten Veröffentlichungen die Angaben zu den nachgestellten Sequenzen und insbesondere den simulierten Objekten ungenau.

Für einen Vergleich wurde mit PARSPIN eine Spin-Echo-Sequenz mit der selben Matrixgröße wie in dem jeweils betrachteten Aufsatz simuliert und ihre Laufzeit mit den veröffentlichten Meßdaten verglichen. Die Anzahl der Spins betrug 2120 bei einer Matrixgröße von $64^2$, 8290 bei $128^2$ und 32770 bei $256^2$. Die PARSPIN-Simulationen wurden im Single-Prozessor-Betrieb sowohl auf einer Sun-Ultra-1-170-Workstation unter Sun Solaris, als auch auf einem PC mit einem Pentium-200-MMX-Prozessor unter Linux durchgeführt.

| Arbeit | Plattform | Experiment |
|--------|-----------|------------|
| Taniguchi et al. [96] | HP Apollo 9000, 50 MHz | Spin-Echo $64^2$ |
| Olsson et al. [74] | Intel 486 50 MHz | Spin-Echo $64^2$ |
| Petersson et al. [78] | Intel 386/387 25 MHz | Spin-Echo $128^2$, 1 Gewebeart |
| Petersson et al. [79] | Pentium 120 MHz | SSFP $256^2$, 100 Pre-Pulse, 10 Gewebe-arten |
| Michiels et al. [69] | IBM RS/6000 3AT | Spin-Echo $256^2$, Spinanzahl 32000 |

Tabelle 7.9: Übersicht über die Arbeiten, die für einen Laufzeitvergleich mit PARSPIN herangezogen wurden.

Die Simulatoren von Taniguchi, Olsson und Michiels sind spinbasierte Werkzeuge, während die Arbeiten von Petersson K-t-basierte Ansätze beschreiben. Konzeptionsbedingt hängt die Laufzeit von K-t-basierten Werkzeugen nicht von einer Anzahl von Spins, sondern von der Anzahl der Kombinationen von $T_1$- und $T_2$-Relaxationskonstanten ab, also im wesentlichen von der Anzahl verschiedener Gewebearten.

Die Laufzeiten der Simulationswerkzeuge nehmen einen großen Wertebereich ein. Zum Vergleich werden sie daher in logarithmischer Darstellung aufgetragen, siehe Bild 7.17. Beim direkten Vergleich der Laufzeiten von PARSPIN mit denen seiner spinbasierten Pendants ist ein deutlicher Vorteil zugunsten PARSPINs erkennbar. Allerdings muß berücksichtigt werden, daß die in [69] [74] [96] eingesetzten Computersysteme wegen des früheren Veröffentlichungsdatums den für PARSPIN genutzten nicht ebenbürtig sind. Ein Teil des Vorsprungs ist daher der technologischen Weiterentwicklung zuzuschreiben. Dies erklärt jedoch nicht den beobachteten Rechenzeitunterschied von ein bis zwei Größenordnungen.

Deutlich ist in Bild 7.17 der ausgeprägte Laufzeitvorteil der K-t-basierten Werkzeuge von Petersson erkennbar. Sie unterbieten bei äußerst moderaten Hardwareanforderungen die mit PARSPIN erreichbaren, für einen spinbasierten Simulator vergleichsweise kurzen Rechenzeiten. Dies gilt insbesondere für das in [79] als Referenz angegebene SSFP-Experiment (vgl. Abschnitt 5.2.4) mit einer großen Anzahl von Warmlaufzyklen, die die Anzahl der Elementarsequenzen und somit die Komplexität des Experiments stark erhöht.

Dieser Laufzeitvorteil wurde bereits in Abschnitt 5.1.2 als ein wesentlicher Vorzug K-t-basierter Ansätze identifiziert. Allerdings sind die vorhandenen K-t-basierten Implementierungen nicht parallelverarbeitend ausgelegt, während PARSPIN auf mehr als einem Prozessor ablaufen kann. Durch eine Erhöhung der Anzahl von verwendeten Prozessoren kann die von PARSPIN benötigte Rechenzeit in die Nähe der von Petersson dokumentierten Werte gesenkt werden, ohne daß die konzeptionellen Einschränkungen K-t-basierter Simulatoren berücksichtigt werden müssen.

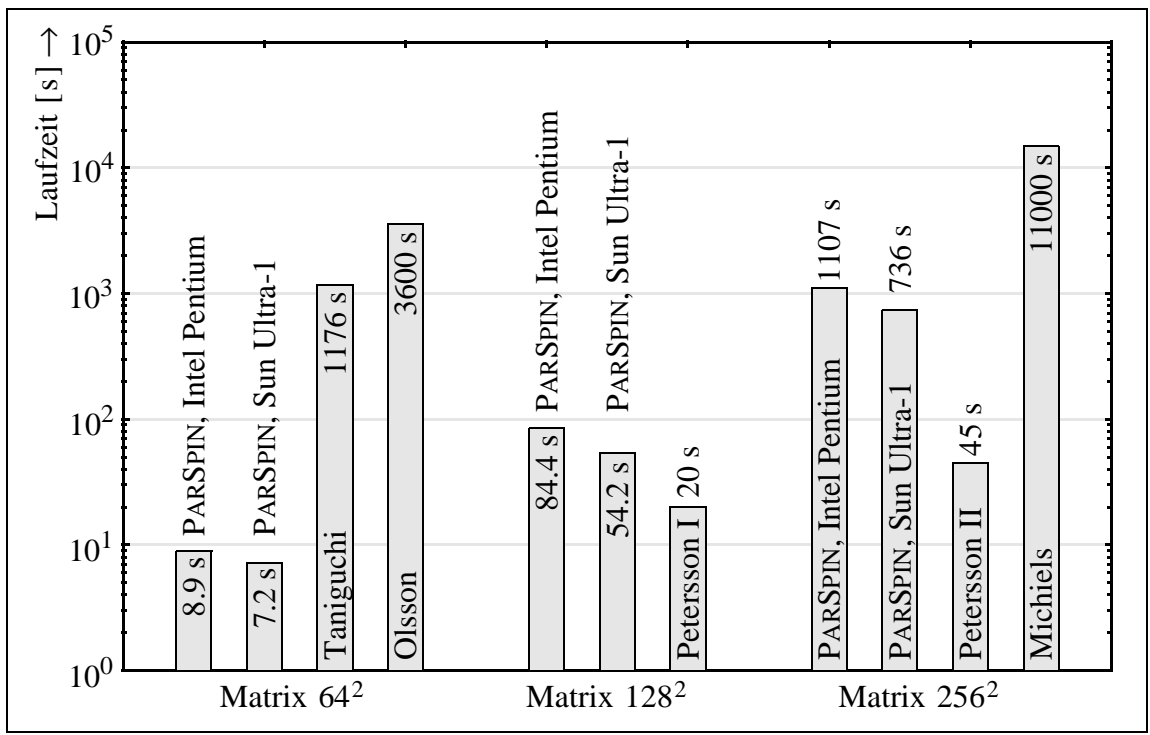

Bild 7.17: Laufzeiten verschiedener Simulatoren im Vergleich zu PARSPIN für die in den Quellen jeweils dokumentierten Sequenzen; logarithmische Unterteilung der Ordinate.

# 8 Vergleichsmessungen

Die Simulationsergebnisse des Simulators PARSPIN konnten im Rahmen von Forschungsaktivitäten des Lehrstuhls mit real gemessenen Daten verglichen werden. In [49] wurde mit Hilfe des Simulators nachgewiesen, daß bei ultraschnellen Multipuls-Bildgebungssequenzen die wesentliche Artefaktquelle die Inhomogenität des Hauptmagnetfelds ist. Aufbauend auf Simulationsstudien konnte ein optimiertes Bildrekonstruktionsverfahren entwickelt werden, das eine starke Reduktion von Verzerrungen im Ortsfrequenzraum und eine signifikante Verminderung von Artefakten im rekonstruierten Bild bewirkt. Daneben konnte eine Sequenzmodifikation entwickelt werden, die auch bei realen Experimenten Bildartefakte zusätzlich zu der durch die optimierte Rekonstruktion erreichten Verbesserung reduzierte.

Mit Hilfe von Simulationsstudien wurde daneben in [58] ein Verfahren zum Design von frequenzselektiven, hochfrequenten Pulsen vorgestellt, deren Spitzenleistung gegenüber herkömmlichen HF-Pulsen stark vermindert ist. Die beabsichtigte selektive Wirkung konnte sowohl simulativ als auch experimentell nachgewiesen werden.

Schwierigkeiten bereitet beim Vergleich von simulierten mit gemessenen Daten häufig der Umstand, daß die Eigenschaften des realen Systems nur qualitativ bekannt sind. Insbesondere ist die Homogenität des Hauptmagnetfelds meist nur durch ihr in Bruchteilen der Nennflußdichte ausgedrücktes Maximum spezifiziert. Der genaue Feldlinienverlauf des Hauptmagnetfelds hängt hingegen auf komplizierte Art von zur Homogenisierung heuristisch installierten Korrekturspulen oder korrigierenden Weicheisenbauteilen ab und läßt sich theoretisch nicht vorhersagen. Aus diesem Grunde kann bei der Untersuchung eines Bildartefakts meist durch Vergleich von Simulationsdaten ohne und mit Magnetfeldinhomogenität entschieden werden, ob diese für den Artefakt verantwortlich ist. Ein quantitativer Vergleich von simulierten mit experimentell gewonnenen Daten ist ohne genaue Kenntnis des Hauptmagnetfelds des Systems nicht möglich.

Eine weitere Hürde beim Vergleich von Bildern simulierter und gemessener Daten ist ihre automatische Nachbearbeitung auf kommerziellen MR-Systemen. So wird das Bildmaterial häufig geglättet oder auf eine größere Matrix interpoliert, um ein subjektiv besseres Bild bereitzustellen. Da ein MR-Bild zudem eine sehr hohe Dynamik aufweist, erfolgt eine adaptive Transformation des Wertebereichs auf die i. a. verfügbaren 256 Graustufen. Das Zusammenwirken aller Bildnachbearbeitungsprozeduren ist bei simuliertem Bildmaterial nicht immer gleichwertig nachstellbar.

In den folgenden Abschnitten wird das aus simulierten und experimentell gewonnenen Daten rekonstruierte Bildmaterial verschiedener Bildgebungssequenzen gegenübergestellt, und es werden die Gemeinsamkeiten und Unterschiede diskutiert. Alle eigenen Experimente wurden in der Klinik für Radiologische Diagnostik des Universitätsklinikums der RWTH Aachen durchgeführt. Beim eingesetzten MR-System handelt es sich um ein Philips ACS NT mit einer magnetischen Flußdichte des statischen Felds von 1.5 T, das für Forschungsaktivitäten im Bereich der interventionellen Radiologie verwendet wird.

## 8.1   Messung der Spin-Spin-Relaxation

Eine sehr große Bedeutung in der medizinischen Diagnostik hat das Kontrastverhalten einer Bildgebungssequenz. Mit verschiedenartigen Sequenzen kann erreicht werden, daß der Bildkontrast hauptsächlich von den ortsabhängigen Größen Spindichte, $T_1$- oder $T_2$-Relaxationszeit bestimmt wird.

Eine Sequenz zur quantitativen Bestimmung der $T_2$-Relaxationskonstanten von Gewebe und zur Ermittlung der relativen Spindichte ist beispielsweise die CPMG-Sequenz [19] [67]. Bei diesem auf einer einfachen Spin-Echo-Sequenz (siehe Bild 4.1, Seite 34) aufbauenden, einer TSE-Sequenz stark ähnelnden Verfahren werden mehrere Echos mit gleicher Phasenkodierung erzeugt, deren zeitliche Abstände zum Anregungspuls variieren.

Die aus den jeweiligen Echos generierten Bilder unterscheiden sich im $T_2$-Kontrast. Eine Auftragung der Intensität eines Pixels über der Zeit offenbart den zu erwartenden exponentiellen Zerfallsprozeß. Mit einem geeigneten Fitting-Verfahren kann aus dieser Kurve die $T_2$-Relaxationskonstante und die relative Spindichte, also die Intensität des Pixels zur Zeit des Anregungspulses, berechnet werden.

Für den Vergleich zwischen Simulation und Messung wurde eine CPMG-Sequenz mit zwölf Echos pro Anregungspuls entwickelt. Die Echozeiten $T_E$, d.h. die zeitlichen Abstände der Echomaxima vom Anregungspuls, lagen bei einer Schrittweite von 20 ms im Bereich von 20 ms bis 240 ms. Das FOV betrug 375 mm, der maximale Auslesegradient $G_x$ = 3.2 mT/m und die Matrixgröße $256^2$ Punkte. Die Schichtdicke lag bei 10 mm. Die Anregungspulse wurden im zeitlichen Abstand von $T_R$ = 1000 ms wiederholt, um nacheinander alle 256 Zeilen des K-Raums mit Daten zu füllen.

Die realen Messungen erfolgten mit einem wassergefüllten Zylinder, in dem kreisförmig vier zylinderförmige Behälter angebracht waren. In diesen Proberöhren befanden sich Substanzen mit voneinander verschiedenen Spindichten, $T_1$- und $T_2$-Relaxationszeiten. Die gemessene Schicht lag parallel zur $x$-$y$-Ebene in der Mitte des Zylinders. Aus den zwölf entstandenen Bildern wurde für jede der Proberöhren die durchschnittliche Intensität $I$ bestimmt. Es ergaben sich auf diese Weise zwölf Wertepaare $[T_E, I(T_E)]$ pro Probe. Mit einem nichtlinearen Least-Squares-Fitting-Algorithmus nach Marquardt-Levenberg [103] wurden anschließend für jede Probe die Koeffizienten $\rho$ und $T_2$ der Gleichung für die Pixelintensität

$$I(t) \,=\, \rho \cdot \exp\!\left(-\frac{t}{T_2}\right) \tag{8.1}$$

ermittelt. Dieser Fit näherte die Meßdaten sehr gut an. Die berechneten Koeffizienten wurden bei der Definition des Objekts für die Simulation verwendet.

Für die die Proberöhren umgebende Wasserfüllung ermittelte der Marquardt-Levenberg-Algorithmus eine hohe Unsicherheit für die Koeffizienten $\rho$ und $T_2$. Grund hierfür ist, daß die Wasserfüllung abhängig vom Grad der chemischen Verunreinigung eine transversale Relaxationszeit von ein bis zwei Sekunden besitzt. Der verfügbare Wertebereich von $T_E = [20\ \text{ms}, 240\ \text{ms}]$ läßt eine präzise Bestimmung derart großer Relaxationszeiten nicht zu.

Der Vergleich von Bildern aus simulierten und aus gemessenen Daten ist für alle Echozeiten in Bild 8.1 gezeigt. Bei Betrachtung der Intensitäten der vier Proberöhren zeigt sich eine ausgezeichnete Übereinstimmung zwischen simulierten und gemessenen Daten.

Das umgebende Wasser wird von der Simulation insbesondere für kleine Echozeiten $T_E < 60\ \text{ms}$ unzureichend nachgebildet. Grund hierfür ist, daß die in der Realität vorhandene Relaxationszeit größer ist als die für die Simulation gewählte. Innerhalb der Repetitionszeit $T_R$ hat sich der Zustand der Spins dem thermischen Gleichgewicht noch nicht so weit genähert, wie in der Simulation modelliert. Die vor dem Anregungspuls vorhandene longitudinale Magnetisierung ist somit kleiner als in der Simulation und führt zu einer betraglich kleineren Transversalmagnetisierung nach dem Puls und so zu einer geringeren Intensität im Bild.

Um den Grad der Übereinstimmung quantitativ zu erfassen, wurde der zeitliche Verlauf der Intensitäten aller Proberöhren im simulierten und experimentellen Fall in Diagrammen gegenübergestellt, siehe Bilder 8.2 und 8.3 (Seite 101). Während die Kurven von Proben eins bis drei sehr gute Übereinstimmung zeigen, liegt die Kurve der simulierten Daten von Probe vier etwas zu hoch. Eine Analyse der beiden Intensitätsverläufe mit dem Marquardt-Levenberg-Algorithmus zeigt, daß die Koeffizienten $T_2$ annähernd gleich und $\rho$ nur wenig verschieden sind. Insgesamt bestätigt der quantitative Vergleich den guten visuellen Eindruck.

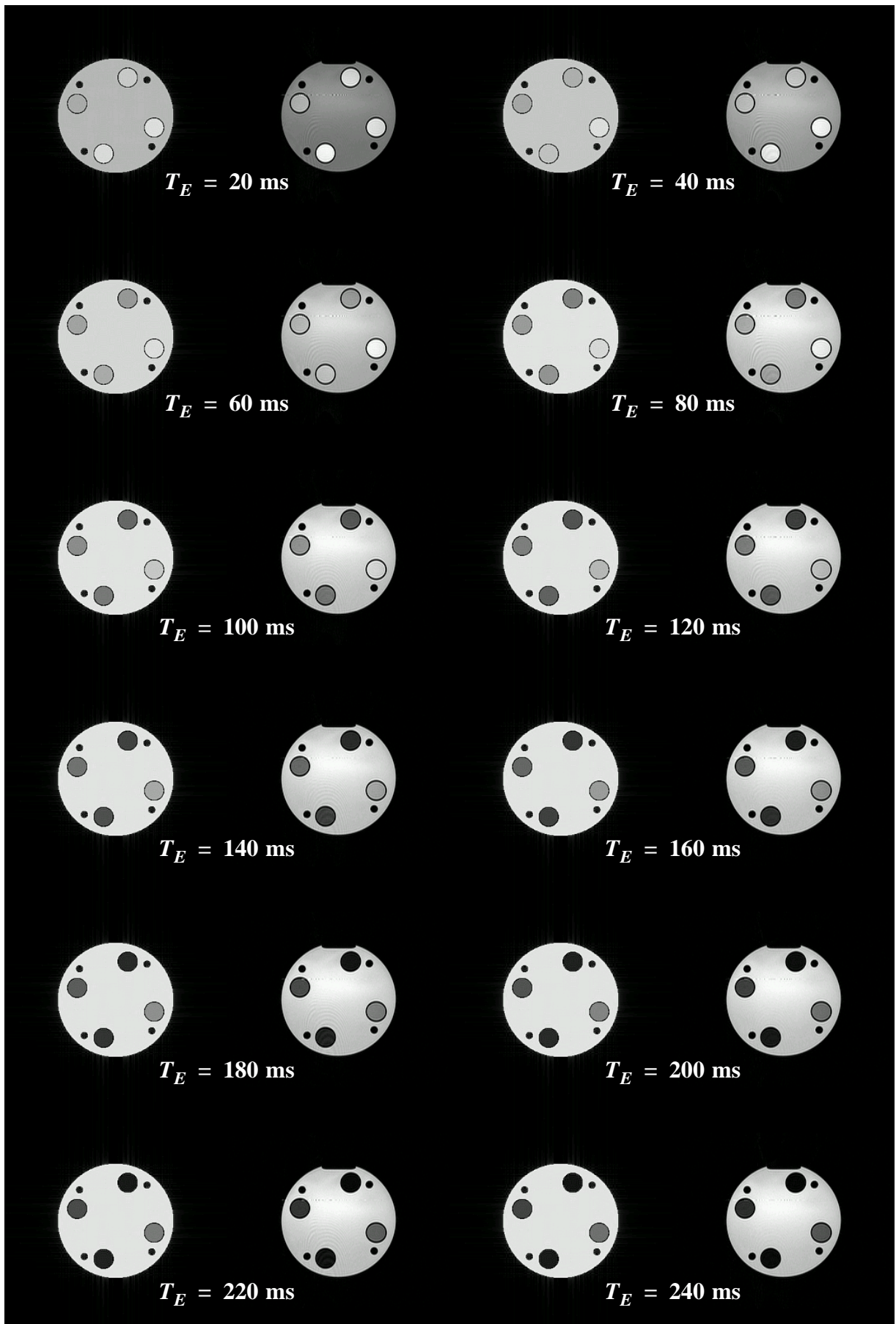

Bild 8.1: Bildvergleich von Simulation (Spalten eins und drei) und Messung (Spalten zwei und vier) für zwölf Echozeiten einer CPMG-Sequenz.

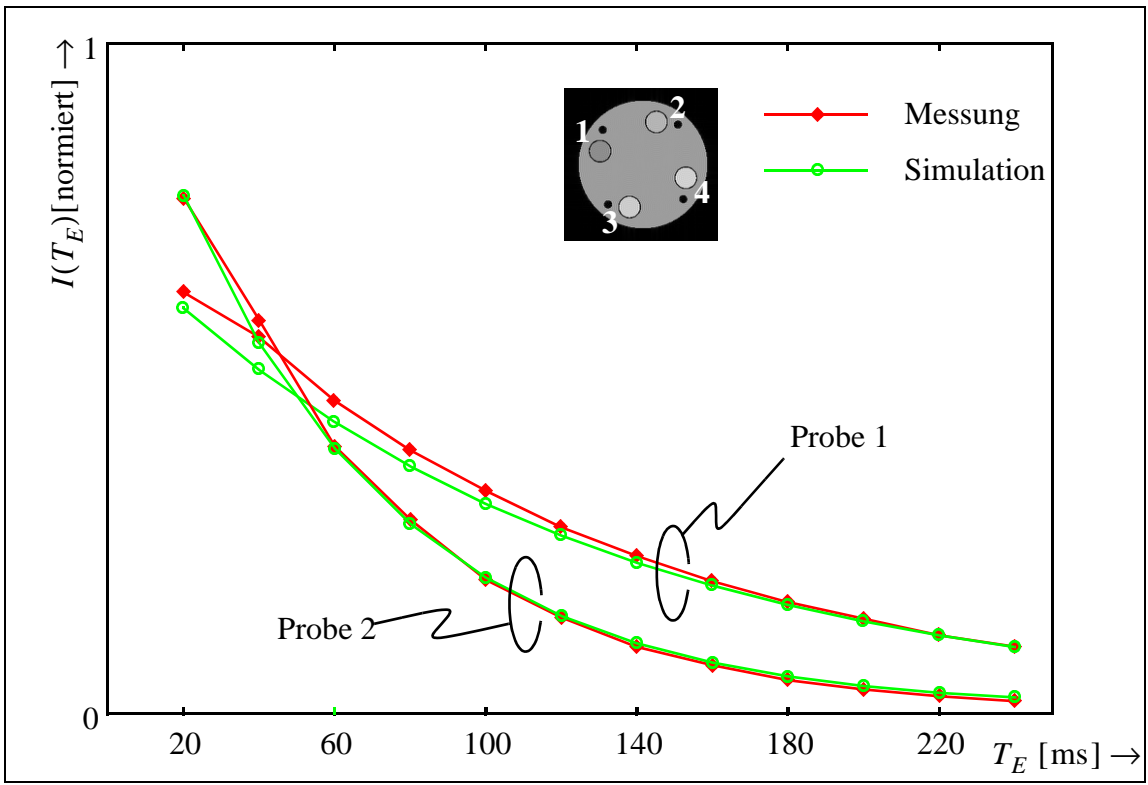

Bild 8.2: Gegenüberstellung des zeitlichen Verlaufs der mittleren Pixel-Intensitäten der Proben 1 und 2 im simulierten und gemessenen Fall.

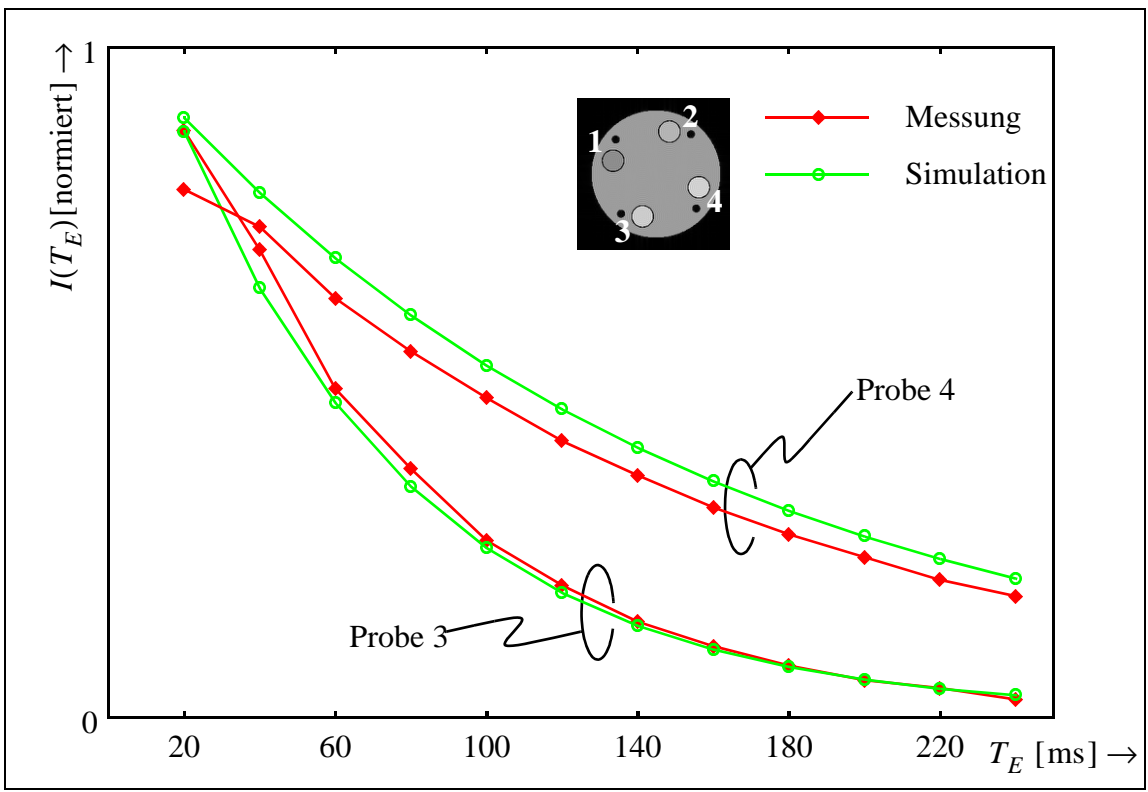

Bild 8.3: Gegenüberstellung des zeitlichen Verlaufs der mittleren Pixel-Intensitäten der Proben 3 und 4 im simulierten und gemessenen Fall.

## 8.2    Suszeptibilitätsartefakte

Suszeptibilitätsartefakte treten an Grenzflächen von Materialien mit unterschiedlicher magnetischer Suszeptibilität $\chi$ auf. Sie äußern sich im Bild in Form von Verzerrungen der Objektgeometrie und Störungen der Intensitätsverteilung. Aus dem Bereich der Anwendung in der interventionellen Radiologie liegt mit dem Aufsatz von Bakker et al. [4] eine Studie zu Suszeptibilitätsartefakten an kugelförmigen Objekten vor, in der Einflüsse auf das Bild bei Spin- und Gradienten-Echo-Sequenzen analysiert werden. Um PARSPINs Eignung in diesem Einsatzbereich nachzuweisen, wurden die Studien aus [4] nachvollzogen. Die entstandenen Simulationsergebnisse werden im folgenden mit den veröffentlichten Meßergebnissen verglichen.

Das im realen Experiment verwendete Phantom ist ein wassergefüllter Zylinder mit einer Länge von 215 mm und einem Durchmesser von 81.5 mm, in dessen Zentrum eine Stahlkugel mit einem Durchmesser von 1 mm angebracht ist. Durch die große Differenz der magnetischen Suszeptibilitäten von Wasser und Stahl ($\Delta\chi = \chi_{\text{Stahl}} - \chi_{\text{Wasser}} \approx 1 + 9.05 \cdot 10^{-6}$) kommt es zu starken Verzerrungen des statischen Magnetfelds. Mit zweidimensionalen, schichtselektiven Sequenzen werden transversale, zur $x$-$z$-Ebene parallele Schnittbilder durch den Zylinder erzeugt. Die Dicke der von der Sequenz angeregten Schichten liegt bei 10 mm. Schnitte werden in vier Abständen $y = [0\,\text{mm}, 10\,\text{mm}, 20\,\text{mm}, 30\,\text{mm}]$ von der Stahlkugel angelegt. Die Auswirkungen werden in [4] für je eine Spin-Echo- und eine Gradienten-Echo-Sequenz untersucht. Die Sequenzparameter sind in den Tabellen 8.1 und 8.2 zusammengefaßt.

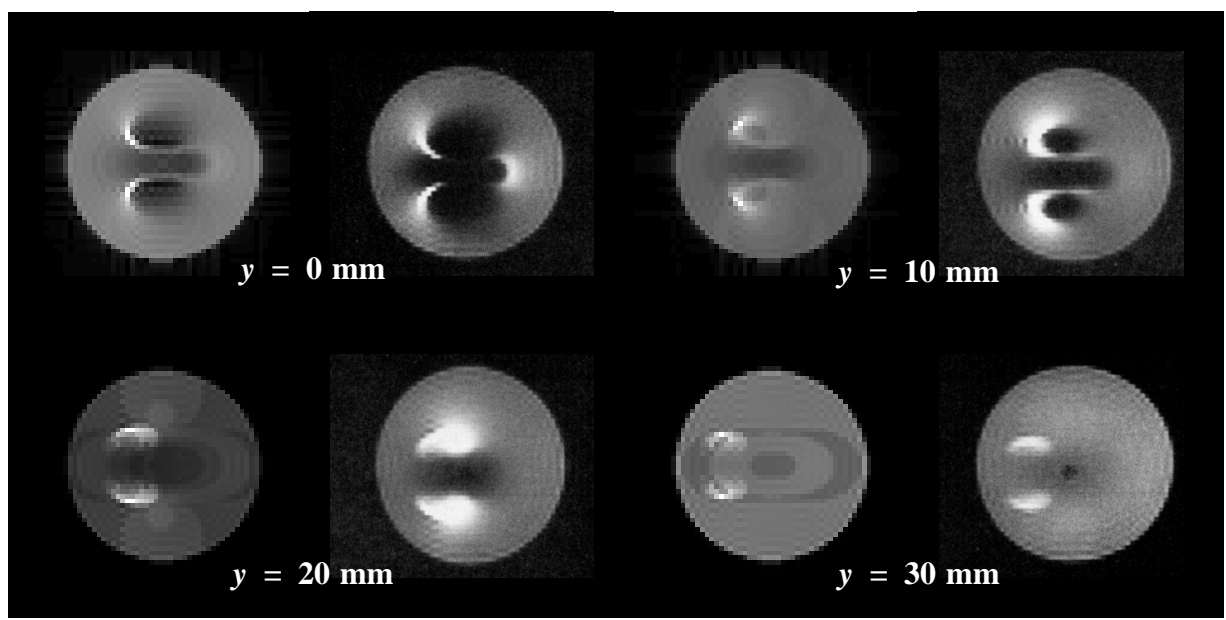

Bild 8.4: Bildvergleich von PARSPIN-Simulation (Spalten eins und drei) und Messung (Spalten zwei und vier) für verschiedene Bildgebungsschichten an den Positionen $y$ unter Verwendung einer Spin-Echo-Sequenz.

| Parameter | Wert |
|---|---|
| FOV | 128 mm |
| Matrixgröße | $64^2$ |
| maximaler Auslesegradient | $2.349\ \frac{mT}{m}$ |
| Echozeit $T_E$ | 50 ms |
| Repetitionszeit $T_R$ | 500 ms |

Tabelle 8.1: Parameter der für Bild 8.4 verwendeten Spin-Echo-Sequenz

| Parameter | Wert |
|---|---|
| FOV | 128 mm |
| Matrixgröße | $64^2$ |
| maximaler Auslesegradient | $2.349\ \frac{mT}{m}$ |
| Echozeit $T_E$ | 10 ms |
| Repetitionszeit $T_R$ | 500 ms |

Tabelle 8.2: Parameter der für Bild 8.5 verwendeten Gradienten-Echo-Sequenz

Das beschriebene Objekt und die beiden Bildgebungssequenzen wurden mit PARSPIN nachgebildet. Da PARSPIN die auftretenden Feldverzerrungen nicht eigenständig ermitteln kann, wurden sie analytisch vorgegeben und während der Simulation berücksichtigt. Bild 8.4 zeigt den Vergleich zwischen mit PARSPIN erzeugten Simulationsergebnissen und den entsprechenden Bildbeispielen eines realen Experiments aus [4][1] für eine Spin-Echo-Sequenz. Es ist eine gute Übereinstimmung zwischen Bildern aus simulierten und gemessenen Daten erkennbar. Tendenziell sind Bereiche ohne Signal – in den Bildern dunkle Gebiete innerhalb des Zylinders – in den gemessenen Bildern etwas stärker ausgeprägt.

Bild 8.5 zeigt den Vergleich zwischen mit PARSPIN erzeugten Simulationsergebnissen und den entsprechenden Bildbeispielen eines realen Experiments aus [4] für eine Gradienten-Echo-Sequenz. Im Gegensatz zum Spin-Echo-Beispiel ist der durch den Suszeptibilitätssprung verursachte Bildartefakt schärfer begrenzt. Die Wiedergabe ist daher weniger sensitiv für ungleiche Grauwertabbildungen. Die Übereinstimmung zwischen simulierten und gemessenen Bildern ist hier noch deutlicher.

Zusammenfassend kann gesagt werden, daß die Effekte von Suszeptibilitätsgrenzflächen, in diesem Fall von kugelförmiger Gestalt, mit PARSPIN sehr gut nachgebildet werden kann und der experimentellen Validierung standhält.

---

[1] Das aus gemessenen Daten rekonstruierte Bildmaterial in den Bildern 8.4 und 8.5 ist der Arbeit von Bakker et al. [4] entnommen. Die Reproduktion erfolgt mit freundlicher Genehmigung von Elsevier Science Ltd.

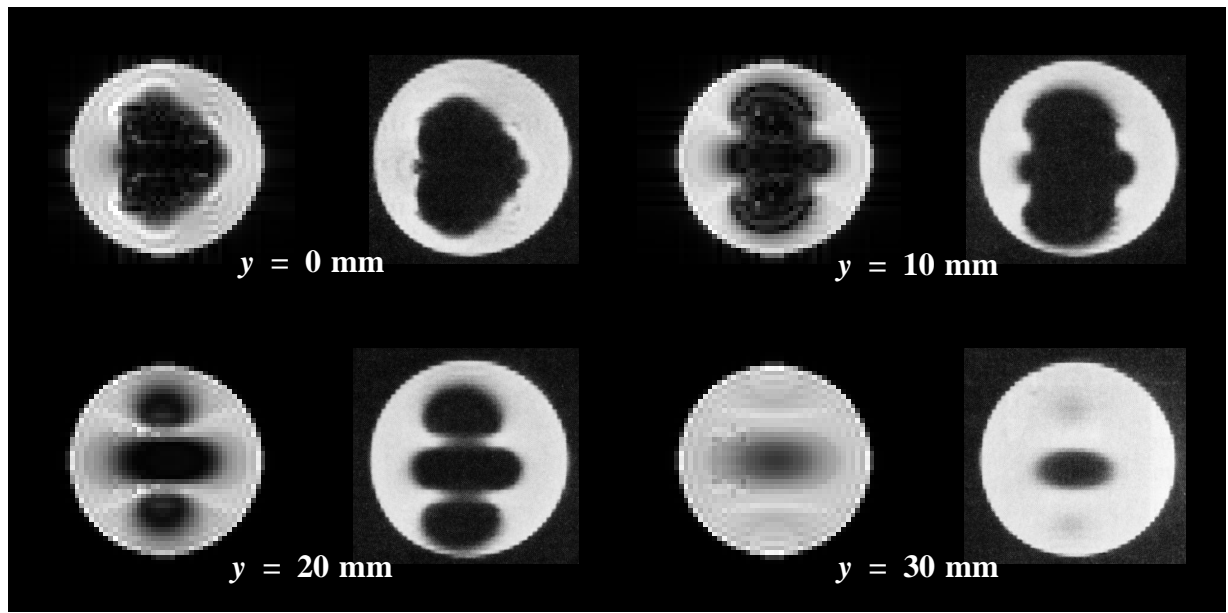

Bild 8.5: Bildvergleich von PARSPIN-Simulation (Spalten eins und drei) und Messung (Spalten zwei und vier) für verschiedene Bildgebungsschichten an den Positionen $y$ unter Verwendung einer Spin-Echo-Sequenz.

## 8.3    Einfluß der Inhomogenität des statischen Magnet-felds

Inhomogenitäten des statischen Magnetfelds haben einen großen Einfluß auf die Qualität eines Bildes. Der Grad der Beeinflussung des Bildes hängt dabei stark von der eingesetzten Bildge-bungssequenz ab. In diesem Abschnitt werden die Bilder simulierter und gemessener Daten bei einem inhomogenen statischen Magnetfeld miteinander verglichen. Der Vergleich erfolgt anhand von zwei Standardsequenzen der medizinischen MR-Bildgebung, einer Gradienten-Echo-Planar-Imaging-Sequenz und einer Turbo-Spin-Echo-Sequenz.

### 8.3.1    Echo-Planar-Imaging-Sequenzen (EPI)

Echo-Planar-Bildgebungssequenzen auf Gradientenechobasis sind generell sehr empfindlich gegenüber Inhomogenitäten des Hauptmagnetfelds. Im Gegensatz zu Spin-Echo-Sequenzen fehlt Gradientenechosequenzen die Fähigkeit, die Auswirkungen des Hauptmagnetfelds zu kompensieren, wie bereits in Abschnitt 4.3.3 diskutiert wurde. Im besonderen Maße gilt dies für Gradienten-EPI-Sequenzen, bei denen nach der HF-Anregung durch wiederholtes Umpo-len des Auslesegradienten die echobildende Konfiguration die Zeitachse im K-t-Diagramm mehrfach schneidet. Dabei ist sie der Hauptmagnetfeldinhomogenität besonders lange ausge-setzt und verbreitert sich während ihrer Lebensdauer besonders stark. Entsprechende Simula-tionsergebnisse wurden bereits in Abschnitt 7.1.4 vorgestellt.

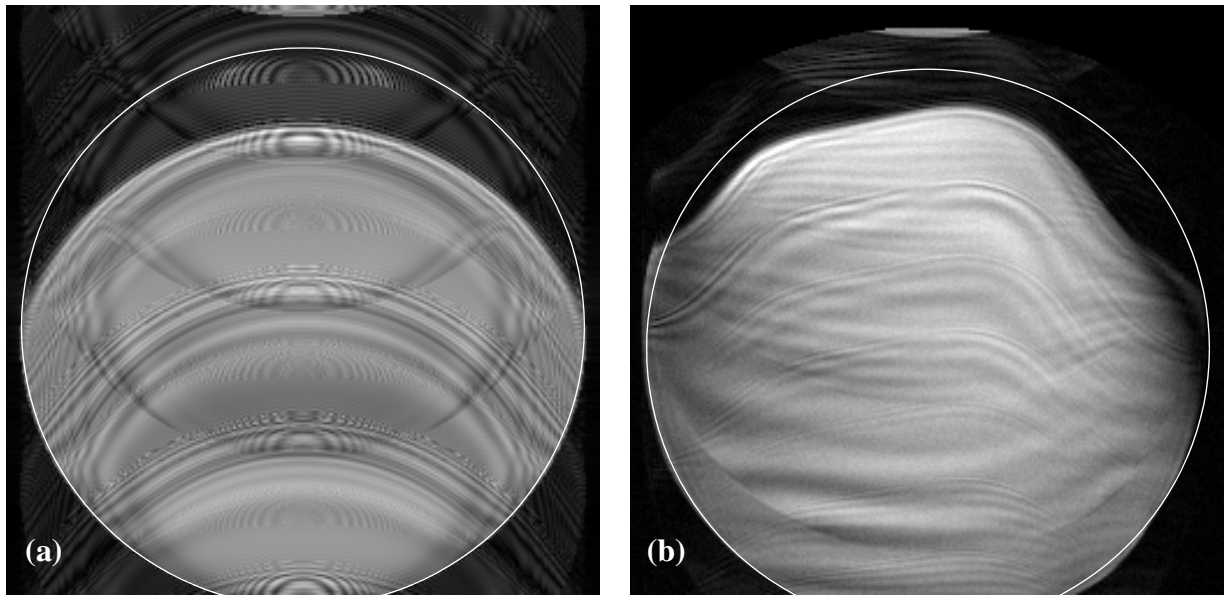

Bild 8.6: Rekonstruierte Bilder von simulierten (a) und real gemessenen Daten (b) einer Echo-Planar-Imaging-Sequenz in einem inhomogenen Magnetfeld. Der tatsächliche Umriß des Phantoms ist durch weiße Kreise gekennzeichnet.

Um Simulationsergebnisse mit realen Messungen vergleichen zu können, wurde ein zylinderförmiges, wassergefülltes Phantom mit einer Gradienten-EPI-Sequenz vermessen. Das Phantom hat einen Durchmesser von etwa 380 mm und eine Höhe von etwa 100 mm. Bei der Bildgebungssequenz handelte es sich um eine Interleaved-EPI-Sequenz, bei der der K-Raum im Zeilensprungverfahren durch mehrere Blöcke von Echos abgetastet wird. Pro Block wurden nach einer HF-Anregung 63 Echos erzeugt, so daß eine K-Raum-Matrix mit $256^2$ Abtastwerten nach vier HF-Anregungen annähernd aufgefüllt war. Der Abstand zwischen den HF-Anregungen betrug 102.9 ms, die effektive Echozeit 50.8 ms. Durch Verwendung von ortsselektiven HF-Pulsen wurde die gemessene transversale Schicht durch den Zylinder auf eine Dicke von 10 mm begrenzt.

Diese Sequenz wurde mit den von PARSPIN zur Verfügung gestellten Mitteln zur Experimentbeschreibung nachgebildet. Bild 8.6 zeigt die aus simulierten und gemessenen Daten rekonstruierten Bilder. Deutlich zu erkennen ist, daß das Innere des Phantoms stark artefaktbehaftet ist. In der Hauptsache handelt es sich bei den Artefakten um periodische Wiederholungen der Objektkanten, die auch als Geisterartefakt bezeichnet werden.

Sowohl im simulierten als auch im gemessenen Fall ist eine starke geometrische Stauchung an der Oberseite des Phantoms zu erkennen, durch die das Bild von der gewünschten Kreisform abweicht. Die Form der Abweichung unterscheidet sich in den Bildern 8.6a und 8.6b; dies kann auf die in der Simulation verwendete, mit den realen Verhältnissen nicht übereinstimmende Hauptmagnetfeldinhomogenität zurückgeführt werden. Für eine bessere Übereinstimmung hinsichtlich der geometrischen Stauchung wäre es erforderlich, das Hauptmagnetfeld des Systems zu vermessen und in numerischer Form an den Simulator weiterzugeben.Insgesamt ist jedoch die Simulation in der Lage, alle wesentlichen Artefakte zumindest qualitativ wiederzugeben. Die Aussagekraft ist daher für die Entwicklung und Verbesserung von Bildgebungssequenzen mehr als ausreichend.

## 8.3.2   Turbo-Spin-Echo-Sequenzen

Als weiteres Beispiel wird eine Turbo-Spin-Echo-Sequenz herangezogen, bei der nach einer Anregung durch einen HF-Puls 256 Echos erzeugt wurden. Die effektive Echozeit betrug 500 ms, und die Aufnahmedauer pro Echo lag bei etwa 4 ms. Die Simulation erfolgte in diesem Beispiel mit der Legendre-Inhomogenität nach Gleichung (3.6). Die Bilder 8.7a und 8.7b zeigen im Vergleich zu den Bildern 8.6a und 8.6b keine starken geometrischen Verzerrungen. Die Abfolge der in den Bildern mit A bis F bezeichneten helleren bzw. dunkleren Zonen am oberen Phantomrand ist im simulierten und im gemessenen Bild qualitativ gleich. Unterschiedlich ist jedoch die genaue Lokalisation der Zonen, ihre Ausdehnung und ihr Kontrast zu umgebenden Bildbereichen.

Auch hier zeigt sich, daß die Realisierung der Hauptmagnetfeldinhomogenität in PARSPIN nicht exakt mir derjenigen im realen System übereinstimmt. Jedoch können qualitativ die Auswirkungen des inhomogenen Hauptfelds anhand der Simulation analysiert werden. Bessere Übereinstimmungen ergäben sich auch hier, wenn das tatsächliche Hauptfeld des Systems in PARSPIN importiert und die Simulationen damit wiederholt würden. Geeignete Meßdaten und eine Schnittstelle in PARSPIN für die numerische Spezifikation von Inhomogenitäten des statischen Magnetfelds standen jedoch noch nicht zur Verfügung.

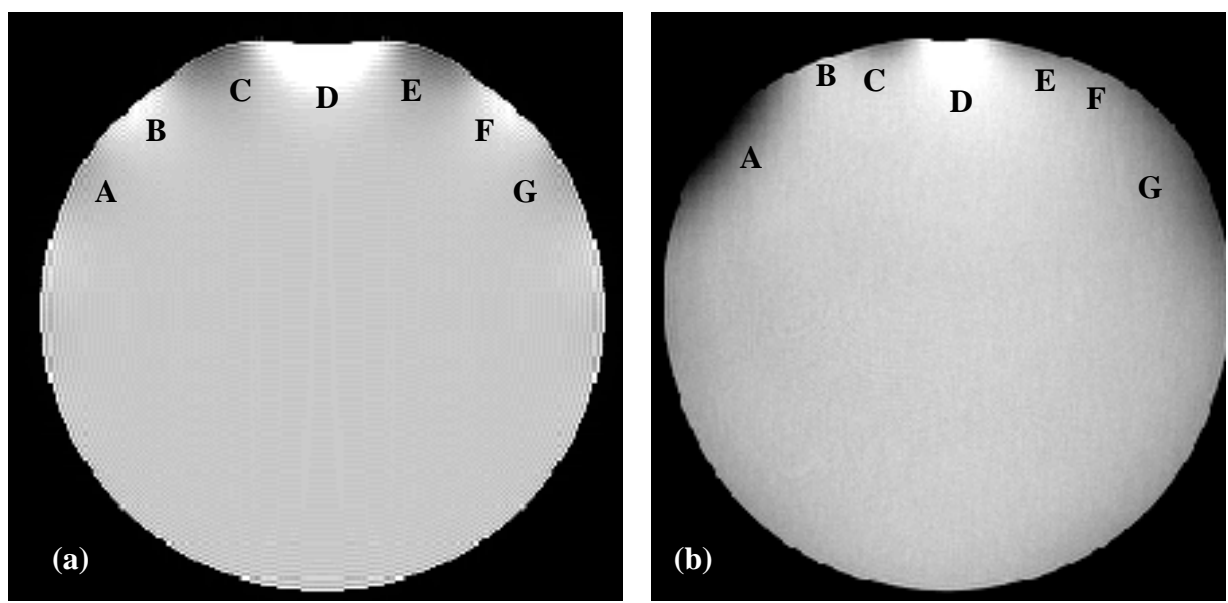

Bild 8.7: Rekonstruierte Bilder von simulierten (a) und real gemessenen Daten (b) einer Turbo-Spin-Echo-Sequenz in einem inhomogenen Magnetfeld.

# Zusammenfassung und Ausblick

Die vorliegende Arbeit beschreibt die für die Realisierung des Magnetresonanz-Simulators PARSPIN wichtigen theoretischen Grundlagen und praktischen Überlegungen. Ausgehend vom Bedarf von Forschungseinrichtungen für einen numerischen Simulator der Magnetresonanz-Bildgebung, wurde anhand einer Übersicht über die Arbeiten anderer Forschergruppen gezeigt, daß für komplexe Fragestellungen in modernen und neuartigen Bildgebungsansätzen ein universell einsetzbarer, leistungsfähiger Simulator erforderlich ist.

Mathematische Grundlage für numerische Simulationen in der Medizin ist die Blochsche Gleichung, ein gewöhnliches, inhomogenes Differentialgleichungssystem erster Ordnung. Sie beschreibt die Reaktion von ortslokalen Magnetisierungen auf das umgebende magnetische Feld. Es wurde erläutert, daß Lösungen für beliebige Zeitabhängigkeiten des Felds nur mit numerischen Methoden zur Lösung von Differentialgleichungssystemen ermittelt werden können. Für einschränkende Bedingungen lassen sich jedoch analytische Lösungen finden, deren Verwendung in einem Simulator deutliche Laufzeitvorteile gegenüber dem numerischen Lösungsvorgang verspricht. Für eine Reihe von Konstellationen von Magnetfeldern wurden analytische Lösungen angegeben.

Ausgehend von den Lösungen der Blochschen Gleichung wurde ein Formalismus zur Beschreibung von komplexen MR-Bildgebungssequenzen abgeleitet, der K-t-Formalismus. Es wurde dargelegt, daß er die Funktionsweise moderner Sequenzen effektiver beschreiben kann, als es die Betrachtung der Einzelmagnetisierungen eines Spin-Ensembles vermag. Zudem eröffnet der K-t-Formalismus unter einschränkenden Bedingungen für das Modell von Objekt und Bildgebungssystem den Zugang zu einer sehr zeiteffizienten Simulationsmethode.

Anschließend wurden verschiedene grundlegende Ansätze für die numerische Simulation des MR-Bildgebungsprozesses vorgestellt, unter ihnen ein spinbasierter Ansatz mit ortslokalen Magnetisierungsvektoren und ein auf dem K-t-Formalismus basierender Ansatz. Es wurde gezeigt, daß ein K-t-basierter Simulator in speziellen Fällen gegenüber einem spinbasierten einen beträchtlichen Laufzeitvorteil bietet. Die mit einem spinbasierten Simulator erzielbare Granularität des Modells, insbesondere im Hinblick auf nichtideale Eigenschaften des Bildgebungssystems, erreicht er jedoch nicht. Die Anforderungen bezüglich nichtidealer Systemeigenschaften wie der Inhomogenität von statischen, hochfrequenten und Gradientenmagnetfeldern gaben für diese Arbeit den Ausschlag zur Verwendung eines spinbasierten Simulationsansatzes.

Die nach dem Abtasttheorem der Systemtheorie korrekte Ortsdiskretisierung bei spinbasierter Simulation, in der Literatur nur selten thematisiert, wurde im Rahmen dieser Arbeit mit Hilfe des K-t-Formalismus allgemein behandelt. Hierzu wurden mathematische Kriterien für den Höchstabstand von Spins innerhalb eines Objekts abgeleitet, die den Einfluß beliebiger Sequenzen berücksichtigen können. Einflüsse von Inhomogenitäten des statischen Magnetfelds und von Chemical Shift konnten in die Rechnungen einbezogen werden.

Ein spinbasierter Simulator ist auch auf modernsten Rechnersystemen eine rechenzeitintensive Anwendung. Im Rahmen dieser Arbeit wurden zwei Ansätze für eine parallelverarbeitende Architektur im Hinblick auf die zu erwartende Rechenzeitreduktion bewertet. Ein Domain-Decomposition-Ansatz, bei dem Blöcke von Spins auf verschiedenen Prozessoren parallel simuliert werden, läßt eine gute Ausnutzung mehrerer Prozessoren erwarten. Er ist darüber hinaus flexibel an die Komplexität des Problems und die verfügbare Rechenleistung anpaßbar, bietet also eine sehr gute Skalierbarkeit.

Anhand von Simulationsbeispielen wurde eine Reihe von Einsatzmöglichkeiten in der Forschung und in der Bildung und Ausbildung demonstriert. Sowohl komplexe als auch einfache Experimente lassen sich schnell in einer Simulation abbilden und liefern bei Bedarf sehr anschauliche Ergebnisse. Mit Messungen der Laufzeit wurde gezeigt, daß die Ausnutzung der verfügbaren Rechenleistung in der Tat ausgesprochen gut ist. Der Vergleich mit einer ähnlichen Arbeit zeigt, daß der Simulator PARSPIN sowohl in punkto Parallelisierungsstrategie als auch in Bezug auf die Leistungsfähigkeit bei nicht paralleler Abarbeitung sehr gut abschneidet. Diese Tendenz bestätigt sich beim Vergleich mit weiteren, in der Literatur dokumentierten Arbeiten.

Der Vergleich von Bildern aus simulierten und gemessenen Daten zeigt prinzipiell gute Übereinstimmungen. Wichtige Voraussetzung für eine nicht nur qualitative, sondern quantitative Nachbildung eines realen Experiments auf einem Simulator ist die Verfügbarkeit von möglichst exakten Daten über das zu simulierende Objekt, die Sequenz und das System. Die Nachbildung der Sequenz stellt i. a. keine Schwierigkeit dar, da sämtliche Parameter der realen Sequenz aus den Einstellungen des MR-Systems abgelesen und übertragen werden können. Hingegen sind die MR-spezifischen Eigenschaften des Objekts, z. B. die Relaxationskonstanten oder die Spindichte, und die Eigenschaften der Systemkomponenten, z. B. der Feldverlauf der magnetischen Felder, i. a. unbekannt und müssen zunächst meßtechnisch bestimmt werden. In den Fällen der Demonstration von Suszeptibilitätsartefakten oder der quantitativen Bestimmung der Spin-Spin-Relaxationszeit waren die das Experiment beeinflussenden Größen mit hoher Genauigkeit bekannt. Folgerichtig konnten in diesen Beispielen sehr gute Übereinstimmungen zwischen simulierten und gemessenen Ergebnissen gezeigt werden.

Die Entscheidung zugunsten einer workstationbasierten Lösung unter dem Betriebssystem Sun Solaris hat sich als sehr vorteilhaft erwiesen. Die Arbeit konnte von der stetigen Vergrößerung des Pools von Workstations am Institut profitieren, ohne daß es zu Aufwendungen für eine Portierung gekommen ist. PARSPIN kann auch auf modernsten Workstations benutzt und erfolgreich auf anderen UNIX-Varianten wie HP-UX oder Linux übersetzt und betrieben werden.

Der Simulator PARSPIN liefert seine Simulationsergebnisse in sehr kurzer Zeit. Hinsichtlich seiner Laufzeit wird ausblickend kein Verbesserungsbedarf gesehen. Wünschenswert ist jedoch eine Vergrößerung der verfügbaren Funktionsvielfalt für die Abbildung von MR-Bildgebungsexperimenten. Ein großer Teil der in naher Zukunft z. B. für die DFG-Forschergruppe „NMR-Oberflächentomographie" benötigten Funktionen wurden in dieser Arbeit bereits im theoretischen Teil eingeführt und in das Gesamtkonzept PARSPINs eingebettet, ihre Implementierung steht jedoch teilweise noch aus.

Hierzu zählen  beispielsweise im Bereich des MR-Systems statische Magnetfelder, deren Feldvektoren nicht über dem gesamten Ort parallel zueinander liegen, und Gradientenmagnetfelder mit vom statischen Magnetfeld unabhängigen Komponenten. Gleiches gilt für die Unterstützung von inhomogenen HF-Magnetfeldern im Sendefall, die zusammen mit den bereits verfügbaren inhomogenen Empfangsfeldern den Simulator im Bereich der inhomogenen hochfrequenten Felder komplettieren würde. Wünschenswert ist in diesem Zusammenhang auch die Nachbildung von mehrkanaligem HF-Empfang, mit dem die Signale der Spulen von HF-Spulen-Arrays einzeln erfaßt und für die Auswertung bereitgestellt werden könnten.

Im Bereich des Objektmodells muß eine Erweiterung des Simulators um die Effekte von Diffusion in Betracht gezogen werden. Unter Diffusionseinfluß kommt es durch eine mikroskopische Bewegung der Spins phänomenologisch zu einer Verringerung des MR-Signals, die mit Relaxationseffekten verwandt ist. Diffusionseffekte hängen von der Stärke des Gradienten des Gesamtmagnetfelds ab und werden nicht nur vom Gradientenmagnetfeld, sondern insbesondere auch von statischen Magnetfeldern mit großen Inhomogenitäten bestimmt.

# Literatur

# A Die Blochsche Gleichung

## A.1 Relaxationseigenschaften quasi-freier Spins

Durch thermische Bewegung des Spins und seiner Umgebung findet ein Energieaustausch statt, der im wesentlichen die longitudinale Komponente der Spinmagnetisierung beeinflußt. Die longitudinale Komponente strebt mit einer exponentiellen Zeitabhängigkeit einen von der Feldstärke des äußeren Magnetfelds abhängigen Ruhewert, die Ruhemagnetisierung $M_0$, an. Die charakteristische Zeitkonstante für diesen Vorgang heißt *thermische* oder *longitudinale Relaxationszeit $T_1$*.

Ein Spinsystem, dessen Magnetisierung die Ruhemagnetisierung erreicht hat, befindet sich im *thermischen Gleichgewicht*. Der Ausgleichsvorgang wird häufig als *Spin-Gitter-Relaxation* bezeichnet, auch wenn nicht in allen interessierenden Substanzen tatsächlich ein Kristallgitter vorliegt.

Im Gegensatz zur Spin-Gitter-Relaxation steht die *Spin-Spin-Relaxation*. Sie entsteht durch lokale Magnetfeldfluktuationen, die durch das Magnetfeld entfernt benachbarter Spins hervorgerufen werden (daher die Bezeichnung Spin-Spin-Relaxation).

Dies führt zu einer ortsabhängigen Variation der Kreisfrequenz, mit der die Spins eines Spinsystems präzedieren. Hierdurch verlieren die Spins mit fortlaufender Zeit mehr und mehr an Phasenkohärenz, was sich in einer Reduktion der transversalen Magnetisierung des Spinsystems äußert. Auch diese Magnetisierungsänderung erfolgt exponentiell mit einer charakteristischen Zeit, der *transversalen Relaxationszeit $T_2$*.

## A.2 Lösungen der Blochschen Gleichung

Im Zeitbereich[1] ergibt sich aus der Blochschen Gleichung (3.4) das Differentialgleichungssystem

---

[1] Die Komponenten der Magnetisierungs- und Flußdichtevektoren sind weiterhin gemäß den Gleichungen (3.1) und (3.2) ortsabhängig.

$$\left.\begin{array}{rl} \dfrac{\mathrm{d}M_x(t)}{\mathrm{d}t} &= \gamma \cdot [M_y(t) \cdot B_z(t) - M_z(t) \cdot B_y(t)] - \dfrac{1}{T_2} \cdot M_x(t) \\[2ex] \dfrac{\mathrm{d}M_y(t)}{\mathrm{d}t} &= \gamma \cdot [M_z(t) \cdot B_x(t) - M_x(t) \cdot B_z(t)] - \dfrac{1}{T_2} \cdot M_y(t) \\[2ex] \dfrac{\mathrm{d}M_z(t)}{\mathrm{d}t} &= \gamma \cdot [M_x(t) \cdot B_y(t) - M_y(t) \cdot B_x(t)] - \dfrac{1}{T_1} \cdot [M_z(t) - M_0] \end{array}\right\}. \tag{A.1}$$

Mit einem Anfangswert $\vec{M}(t_0)$ entsteht hieraus das lineare, inhomogene Anfangswertproblem erster Ordnung in der Form

$$\left.\begin{array}{rl} \dfrac{\mathrm{d}\vec{M}(t)}{\mathrm{d}t} &= A(t) \cdot \vec{M}(t) + \vec{f}(t) \\[2ex] \vec{M}(t_0) &= (M_{x0}, M_{y0}, M_{z0})^T \end{array}\right\} \tag{A.2}$$

mit

$$A(t) = \begin{pmatrix} -\dfrac{1}{T_2} & \gamma \cdot B_z(t) & -\gamma \cdot B_y(t) \\[2ex] -\gamma \cdot B_z(t) & -\dfrac{1}{T_2} & \gamma \cdot B_x(t) \\[2ex] \gamma \cdot B_y(t) & -\gamma \cdot B_x(t) & -\dfrac{1}{T_1} \end{pmatrix} \tag{A.3}$$

und

$$\vec{f}(t) = \vec{f} = \left(0, 0, \dfrac{M_0}{T_1}\right)^T. \tag{A.4}$$

Das Anfangswertproblem (A.2) ist für beliebige zeitabhängige $A(t)$ nicht allgemein in geschlossener Form lösbar. Nach der Theorie der gewöhnlichen Differentialgleichungen [100] besitzt das System (A.2) jedoch für ein gegebenes $\vec{M}(t_0)$ genau eine Lösung $\vec{M}(t)$, sofern $A(t)$ und $\vec{f}(t)$ stetig sind.

Eine geschlossene Lösung läßt sich angeben, wenn die Matrix $A$ nicht zeitabhängig ist. Dann kann mit Hilfe ihrer Eigenwerte und Eigenvektoren ein Fundamentalsystem des homogenen Anfangswertproblems ($\vec{f}(t) = 0$) berechnet und mit der *Methode der Variation der Konstanten* eine Linearkombination der einzelnen Lösungen des Fundamentalsystems ermittelt werden, die das inhomogene System (A.2) löst.

Für spezielle Formen der Zeitabhängigkeit von $A(t)$, bei der die Zeitabhängigkeit nicht mehr wie in Gleichung (A.3) als funktionaler Zusammenhang, sondern explizit angegeben wird, kann meist auch zu System (A.2) eine geschlossene Lösung bestimmt werden. Derartige Einzelfallösungen wurden in dieser Arbeit nicht erarbeitet. Es gibt jedoch Spezialfälle für die Zeitabhängigkeit von $A(t)$, bei denen sich die Lösung noch relativ allgemein in Abhängigkeit der Elemente der Matrix angeben lassen, ohne eine explizite Zeitabhängigkeit anzunehmen.

## A.2.1 Hochfrequente Magnetfelder

Die zeitliche Abhängigkeit aller drei Komponenten des HF-Feldes führt zunächst zu maximal möglicher Komplexität bei der Aufstellung des Anfangswertproblems nach Gleichung (A.1). Eine geschlossene Lösung kann nicht in allgemeiner Form für beliebige Zeitfunktionen des HF-Feldes angegeben werden. Es ist jedoch eine Reihe von Vereinfachungen möglich, die für eine Teilmenge zeitabhängiger HF-Felder Aussagen über ihren Einfluß auf die betrachtete Magnetisierung erlauben:

1. Nach [1] kann die Komponente des HF-Felds parallel zum Hauptmagnetfeld vernachlässigt werden.

2. Eine harmonische Zeitabhängigkeit mit konstanter Amplitude ermöglicht die Definition eines rotierenden Bezugssystems, in dem die Zeitabhängigkeit nicht mehr beobachtbar ist.

Im folgenden wird zunächst hergeleitet, welche Gestalt die Blochsche Gleichung in einem rotierenden Bezugssystem annimmt. Anschließend wird skizziert, wie sich ein zeitlich kurz wirkendes magnetisches Feld, d.h. ein HF-Puls, im rotierenden Bezugssystem auf die Magnetisierung des Objekts auswirkt.

### A.2.1.1 Blochsche Gleichung im rotierenden Bezugssystem

In diesem Abschnitt werden hochfrequente Magnetfelder $\vec{B}_1$ diskutiert, die zirkular polarisiert sind, deren Feldvektor also in einer Ebene orthogonal zum Hauptmagnetfeld mit einer Kreisfrequenz $\omega_{HF}$ rotiert. Der Betrag der Flußdichte sei konstant. Für die Flußdichte eines hochfrequenten Magnetfelds, dessen Rotationsrichtung der Spinpräzessionsrichtung folgt, das also in mathematisch negativer Richtung rotiert, gilt

$$\vec{B}_1 = \begin{pmatrix} B_1 \cdot \cos(\omega_{HF}t - \phi) \\ -B_1 \cdot \sin(\omega_{HF}t - \phi) \\ 0 \end{pmatrix}. \tag{A.5}$$

Die Zeitabhängigkeit des Felds aus Gleichung (A.5) macht es schwierig, eine analytische Lösung der Blochschen Gleichung zu finden, obwohl die Existenz einer Lösung theoretisch gesi-chert ist. Wird jedoch ein *rotierendes Bezugssystem* definiert, das mit der Kreisfrequenz $\omega_{HF}$ ebenfalls in mathematisch negativer Richtung um den Vektor des Hauptmagnetfelds rotiert, so läßt sich eine Lösung für den neuen Magnetisierungsvektor $\vec{M}'$ im rotierenden Bezugssystem finden.

Im folgenden bezeichnen gestrichene Größen stets Größen im rotierenden Bezugssystem. Dreht sich das rotierende Bezugssystem mit der Kreisfrequenz $\omega_{HF}$ um die $z$-Achse des Laborsystems, so sind longitudinale Komponenten von Vektoren im rotierenden und im Laborbezugssystem identisch; es gilt also z. B. $z' = z$. Allgemein geschrieben geht ein Vektor $\vec{x}$ nach der Gleichung

$$\vec{x}'(t) \;=\; \boldsymbol{U}^{-1}(t) \cdot \vec{x}(t) \tag{A.6}$$

in den Vektor $\vec{x}'$ im rotierenden Bezugssystem über. Die Matrix $\boldsymbol{U}^{-1}$ ist die die Rotation beschreibende Transformationsmatrix; für sie gilt

$$\boldsymbol{U}^{-1} \;=\; \begin{pmatrix} \cos\omega_{HF}t & -\sin\omega_{HF}t & 0 \\ \sin\omega_{HF}t & \cos\omega_{HF}t & 0 \\ 0 & 0 & 1 \end{pmatrix}. \tag{A.7}$$

Ihre Inverse $\boldsymbol{U}$ läßt sich leicht bestimmen. Wird in Gleichung (3.4) $\vec{M}$ durch $\boldsymbol{U} \cdot \vec{M}'$ substituiert, so ergibt sich in Matrixoperatorschreibweise nach Gleichung (A.2)

$$\frac{\mathrm{d}(\boldsymbol{U} \cdot \vec{M}')}{\mathrm{d}t} \;=\; \boldsymbol{A} \cdot \boldsymbol{U} \cdot \vec{M}' + \left(0, 0, \frac{M_0}{T_1}\right)^T. \tag{A.8}$$

Bei einem äußeren Magnetfeld $\vec{B} = \vec{B}_0 + \vec{B}_1$ gilt dabei

$$\boldsymbol{A} \;=\; \begin{pmatrix} -\dfrac{1}{T_2} & \gamma \cdot B_{0z} & \gamma \cdot B_1 \cdot \sin(\omega_{HF}t - \phi) \\[2mm] -\gamma \cdot B_{0z} & -\dfrac{1}{T_2} & \gamma \cdot B_1 \cdot \cos(\omega_{HF}t - \phi) \\[2mm] -\gamma \cdot B_1 \cdot \sin(\omega_{HF}t - \phi) & -\gamma \cdot B_1 \cdot \cos(\omega_{HF}t - \phi) & -\dfrac{1}{T_1} \end{pmatrix} \tag{A.9}$$

für ein statisches Feld $\vec{B}_0 = (0, 0, B_{0z})^T$ und ein HF-Feld $\vec{B}_1$ nach Gleichung (A.5).

Mit

$$\frac{\mathrm{d}(\boldsymbol{U} \cdot \vec{M}')}{\mathrm{d}t} \;=\; \boldsymbol{U} \cdot \frac{\mathrm{d}\vec{M}'}{\mathrm{d}t} + \frac{\mathrm{d}\boldsymbol{U}}{\mathrm{d}t} \cdot \vec{M}' \tag{A.10}$$

wird Gleichung (A.8) zu

$$\frac{\mathrm{d}\vec{M}'}{\mathrm{d}t} = \boldsymbol{U}^{-1} \cdot \boldsymbol{A} \cdot \boldsymbol{U} \cdot \vec{M}' - \boldsymbol{U}^{-1} \cdot \frac{\mathrm{d}\boldsymbol{U}}{\mathrm{d}t} \cdot \vec{M}' + \left(0, 0, \frac{M_0}{T_1}\right)^T. \tag{A.11}$$

In dieser Gleichung sind die beiden auf $\vec{M}'$ wirkenden Matrixoperatoren zeitunabhängig, denn es gilt

$$\boldsymbol{U}^{-1} \cdot \boldsymbol{A} \cdot \boldsymbol{U} = \begin{pmatrix} -\dfrac{1}{T_2} & \gamma \cdot B_{0z} & -\gamma \cdot B_1 \cdot \sin\phi \\[2mm] -\gamma \cdot B_{0z} & -\dfrac{1}{T_2} & \gamma \cdot B_1 \cdot \cos\phi \\[2mm] \gamma \cdot B_1 \cdot \sin\phi & -\gamma \cdot B_1 \cdot \cos\phi & -\dfrac{1}{T_1} \end{pmatrix} \tag{A.12}$$

sowie

$$\boldsymbol{U}^{-1} \cdot \frac{\mathrm{d}\boldsymbol{U}}{\mathrm{d}t} = \begin{pmatrix} 0 & \omega_{\mathrm{HF}} & 0 \\ -\omega_{\mathrm{HF}} & 0 & 0 \\ 0 & 0 & 0 \end{pmatrix}. \tag{A.13}$$

Diese Operatoren sind äquivalent zum Vektorprodukt, so daß Gleichung (A.11) mit den Definitionen für $\boldsymbol{E}$ und $\vec{E}_0$ aus Gleichung (3.4), dem Vektor $\vec{\omega}_{\mathrm{HF}} = (0, 0, \omega_{\mathrm{HF}})^T$ sowie der Flußdichte $\vec{B}' = (B_1 \cdot \cos\phi, B_1 \cdot \sin\phi, B_{0z})^T$ umgeschrieben werden kann zu

$$\begin{aligned} \frac{\mathrm{d}\vec{M}'}{\mathrm{d}t} &= \gamma \cdot \vec{M}' \times \vec{B}' - \vec{M}' \times \vec{\omega}_{\mathrm{HF}} + \boldsymbol{E} \cdot \vec{M}' + \vec{E}_0 \\[2mm] &= \gamma \cdot \vec{M}' \times \left(\vec{B}' - \frac{\vec{\omega}_{\mathrm{HF}}}{\gamma}\right) + \boldsymbol{E} \cdot \vec{M}' + \vec{E}_0. \end{aligned} \tag{A.14}$$

Die Größe $\vec{B}' - \vec{\omega}_{\mathrm{HF}}/\gamma$ wird in der Literatur häufig als *effektive Flußdichte* $\vec{B}_{\mathrm{eff}}$ bezeichnet. Mit ihr wird Gleichung (A.14) zu

$$\frac{\mathrm{d}\vec{M}'}{\mathrm{d}t} = \gamma \cdot \vec{M}' \times \vec{B}_{\mathrm{eff}} + \boldsymbol{E} \cdot \vec{M}' + \vec{E}_0; \tag{A.15}$$

dies entspricht formal Gleichung (3.4). Die Blochsche Gleichung gilt demnach auch in einem um die Richtung des statischen Magnetfelds rotierenden Bezugssystem, wenn statt der tatsächlich vorliegenden äußeren magnetischen Flußdichte $\vec{B}$ die effektive Flußdichte $\vec{B}_{\mathrm{eff}}$ eingesetzt wird.

### A.2.1.2 Effektives Magnetfeld

Das rotierende Bezugssystem muß nicht notwendigerweise mit der Kreisfrequenz $\omega_{HF}$ rotieren. In einem solchen Fall bleibt $\vec{B}_{eff}$ in Gleichung (A.15) zeitabhängig. Der Spezialfall des Bezugssystems, das mit der Kreisfrequenz des hochfrequenten Magnetfelds rotiert und mit einem im Bezugssystem ruhenden Magnetfeldvektor einhergeht, ermöglicht jedoch erst das Auffinden einer geschlossenen analytischen Lösung der Blochschen Gleichung.

Der Feldvektor des effektiven Magnetfelds schließt mit dem des statischen Magnetfeld einen Winkel ein, der von der Differenz zwischen der Larmorfrequenz $\omega_0$ der Spins und der Kreisfrequenz des hochfrequenten Magnetfelds abhängt. Betrachtet man die $z$-Komponente des effektiven Magnetfelds $B_{0_z} - \omega_{HF}/\gamma$, so zeigt sich, daß diese verschwindet, wenn $\omega_{HF} = \omega_0 = \gamma \cdot B_{0_z}$ gilt. In diesem Fall ist das effektive Magnetfeld ein rein transversales Magnetfeld. Mit steigendem Abstand zwischen $\omega_0$ und $\omega_{HF}$ wird der Einfluß der $z$-Komponente größer.

Häufig werden hochfrequente Magnetfelder nur für kurze Zeit appliziert, es wird dann von HF-*Pulsen* gesprochen. Ist die Applikationsdauer gegenüber den Relaxationszeiten $T_1$ und $T_2$ vernachlässigbar, entfallen die Größen $E$ und $\vec{E}_0$ in Gleichung (A.15). Das entstehende Randwertproblem besitzt einen überschaubaren Lösungsvektor für die Spinmagnetisierung. Die Spinmagnetisierung beschreibt eine Präzessionsbewegung um die Achse des effektiven Magnetfelds und um den Winkel $\alpha = \gamma \cdot \left| \vec{B}_{eff} \right| \cdot \tau$, der als *Flipwinkel* bezeichnet wird, vgl. Bild A.1.

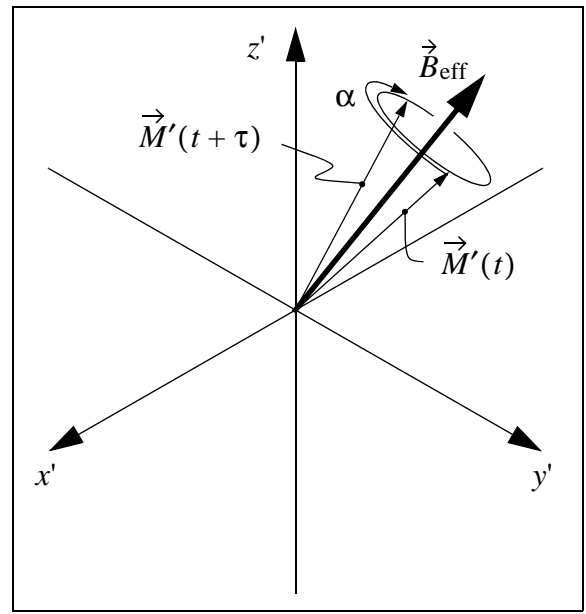

Bild A.1: Präzessionsbewegung eines Spins um die effektive Flußdichte und um den Winkel $\alpha$.

# B    Zeitliche Diskretisierung

Dieser Anhang gibt einen Überblick über die mathematischen Zusammenhänge, die für die korrekte zeitliche Diskretisierung eines MR-Experiments sowohl beim realen Experiment als auch in der Simulation erforderlich sind.

## B.1    Abtastung des Echosignals

Bei einem realen Experiment wird das empfangene Signal zunächst demoduliert und tiefpaßgefiltert. Anschließend wird das demodulierte, tiefpaßgefilterte Empfangssignal $s(t)$ abgetastet, also in eine zeitdiskrete Folge $s(n)$ umgewandelt. Dieser Vorgang erfolgt nach den bekannten Regeln der Systemtheorie [62] und erfordert die Einhaltung des Abtasttheorems. In der MR-Bildgebung wird das Empfangssignal meist nicht als Zeitsignal $s(t)$, sondern nach dem K-Raum-Konzept als Funktion $S(k_x, k_y, k_z)$ im Ortsfrequenzraum aufgefaßt. Der Übergang von Zeit- zu K-Raum-Koordinaten ist hierbei abhängig von der eingesetzten Sequenz, genauer von der Trajektorie, mit der echobildende Konfigurationen durch den K-Raum bewegt werden.

Die Abtastung des K-Raum-Signals ist im wesentlichen von der Größe des abzubildenden Ortsbereichs (dem *Field of View* oder FOV) und der gewünschten Anzahl von Abtastpunkten $N$ abhängig. Nach dem Abtasttheorem gilt im eindimensionalen Fall

$$FOV = \frac{\pi \cdot (N-1)}{2 \cdot k_{x,\,max}},\tag{B.1}$$

wobei $2 \cdot k_{x,\,max}$ der während der Datenakquisition der Länge $\tau$ von der Echokonfiguration durchlaufene K-Raum-Bereich ist; es gilt

$$2 \cdot k_{x,\,max} = \gamma \cdot \int_{t_0}^{t} G_x(\tau) \; d\tau.\tag{B.2}$$

Dieser Zusammenhang ist in den Bildern B.1 und B.2 für ein eindimensionales rechteckförmiges Objekt und seine Ortsfrequenzfunktion illustriert. Üblicherweise werden die Gleichungen (B.1) und (B.2) für die Berechnung eines zu den vorgegebenen Größen $N$, $\Delta t = t - t_0$ und $FOV$ passenden Gradienten $G_x$ verwendet. Im Fall eines zeitlich rechteckförmigen Gradientenmagnetfelds ergibt sich beispielsweise

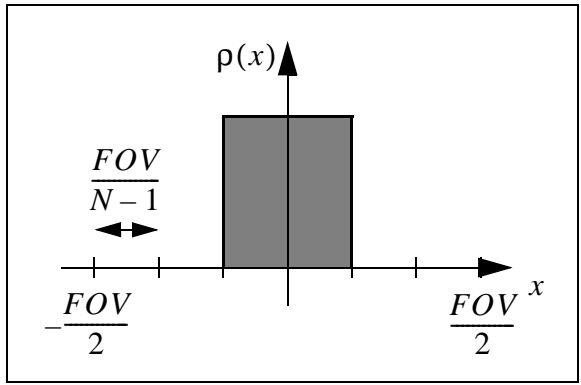

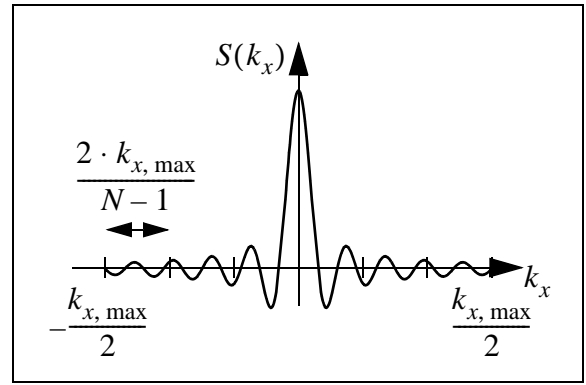

Bild B.1: Rechteckförmiges Objekt im Field Of View. Das FOV wird in $N-1$ Intervalle zerlegt.

Bild B.2: Korrespondierender Ortsfrequenz-raum, ebenfalls zerlegt in $N-1$ Intervalle, auf deren Rändern abgetastet wird.

$$G_x = \frac{2\pi \cdot (N-1)}{\gamma \cdot FOV \cdot \Delta t}. \tag{B.3}$$

Die Berechnungen für die Richtungen $y$ und $z$ in höherdimensionalen Experimenten erfolgen getrennt voneinander auf analoge Weise.

## B.2 Abtastung modulierter HF-Pulse

Die zeitliche Abtastung von modulierten HF-Pulsen muß so gewählt werden, daß es innerhalb des FOV nicht zu Aliasing-Effekten [62] kommen kann. Der Abstand der durch die Abtastung hervorgerufenen Wiederholungen des Spektrums des HF-Pulses muß demnach groß genug gewählt werden. Unter der Voraussetzung eines homogenen statischen Magnetfelds wird hier angenommen, daß der HF-Puls ortsselektiv in die $z$-Richtung wirkt. Die maximal mögliche Kreisfrequenz $\omega_{max}$ am Rand des Ortsraums, also bei $\pm FOV/2$ bei Vorhandensein eines Gradientenmagnetfelds $\vec{B}_{grad} = (0, 0, G_z \cdot z)^T$, ist

$$\pm\omega_{max} = \pm\gamma \cdot G_z \cdot \frac{FOV}{2}. \tag{B.4}$$

Um Aliasing innerhalb des FOV zu verhindern, muß der HF-Puls auf $\Delta\omega < 2 \cdot \omega_{max}$ bandbegrenzt sein. Der Abstand der HF-Puls-Spektren muß mindestens $\omega_{max} + \Delta\omega/2$ betragen. Für einen mit $N_{HF}$ Stützpunkten abgetasteten HF-Puls der Länge $\Delta t$ gilt dann nach dem Abtasttheorem die Bedingung

$$\frac{\Delta t}{N_{HF}} < \frac{\pi}{\omega_{max} + \Delta\omega/2}. \tag{B.5}$$

# Lebenslauf

| | |
|---|---|
| Name: | Jürgen Kürsch |
| Geburt: | 21. Juni 1969 in Köln |
| Eltern: | Elfriede Kürsch, geb. Bruck, Einzelhandelskauffrau |
| | Jakob Ferdinand Kürsch, Fernmeldetechniker |
| Schulbildung: | 1975 – 1979: Städtische Grundschule Bachemer Straße, Köln |
| | 1979 – 1988: Städtisches Gymnasium Elisabeth-von-Thüringen-Schule, Köln; Abschluß: Abitur |
| Wehrdienst: | 1988 – 1989: Fernmeldebataillon 6, Neumünster |
| Studium: | 1989 – 1994: Studium der Elektrotechnik mit Schwerpunkt Nachrichtentechnik an der Rheinisch-Westfälischen Technischen Hochschule Aachen; Abschluß: Diplom-Ingenieur |
| | 22. Januar 2003: Promotion zum Dr.-Ing. an der Fakultät für Elektrotechnik und Informationstechnik der Rheinisch-Westfälischen Technischen Hochschule Aachen |
| Berufliche Tätigkeit: | 1994 – 1999: Wissenschaftlicher Mitarbeiter am Lehrstuhl für Allgemeine Elektrotechnik und Datenverarbeitungssysteme der Rheinisch-Westfälischen Technischen Hochschule Aachen |
| | seit 1999: Software-Ingenieur im Bereich Enterprise Security Management bei der Beta Systems Software AG, Köln |
| E-Mail: | j.kuersch.diss@gmx.de |